\documentclass[10pt,twocolumn]{article}

\usepackage[letterpaper,margin=15.5mm,columnsep=6.0mm]{geometry}
\usepackage[T1]{fontenc}

\usepackage[scaled=0.92]{helvet}
\usepackage[scaled=0.90]{zi4}
\usepackage{amsmath,amssymb,amsthm,mathtools}
\usepackage{booktabs}
\usepackage{tabularx}
\usepackage{array}
\usepackage{xcolor}
\usepackage{microtype}
\usepackage{tikz}
\usetikzlibrary{arrows.meta,positioning,shapes.geometric,fit,calc}
\PassOptionsToPackage{hyphens}{url}
\usepackage{hyperref}

\definecolor{primary}{RGB}{15,41,66}
\definecolor{accent}{RGB}{11,87,208}
\definecolor{alert}{RGB}{179,38,30}
\definecolor{surface}{RGB}{244,247,251}

\hypersetup{
  colorlinks=true,
  linkcolor=primary,
  citecolor=accent,
  urlcolor=accent,
  pdftitle={Provenance-Isolated Epistemic Quorums: Execution-Bound Evidence Certification and Plausible-World Safety Gating in Agentic Control Planes},
  pdfauthor={Jun He and Deying Yu}
}

\makeatletter
\def\section{\@startsection{section}{1}{\z@}{-1.6ex plus -0.5ex minus -0.2ex}{0.7ex plus 0.2ex}{\reset@font\large\bfseries}}
\def\subsection{\@startsection{subsection}{2}{\z@}{-1.2ex plus -0.4ex minus -0.2ex}{0.45ex plus 0.15ex}{\reset@font\normalsize\bfseries}}
\makeatother

\newtheorem{theorem}{Theorem}
\newtheorem{lemma}[theorem]{Lemma}
\newtheorem{proposition}[theorem]{Proposition}
\newtheorem{corollary}{Corollary}
\theoremstyle{definition}
\newtheorem{definition}{Definition}
\newtheorem{assumption}{Assumption}
\theoremstyle{remark}
\newtheorem{remark}{Remark}

\newcolumntype{L}[1]{>{\raggedright\arraybackslash}p{#1}}
\newcolumntype{Y}{>{\raggedright\arraybackslash}X}

\newcommand{\Exp}{\operatorname{Exp}}
\newcommand{\Dom}{\operatorname{Dom}}
\newcommand{\Post}{\operatorname{Post}}
\newcommand{\Viab}{\operatorname{Viab}}
\newcommand{\Pred}{\operatorname{Pred}}
\newcommand{\Lin}{\operatorname{Lin}}
\newcommand{\Close}{\operatorname{Close}}
\newcommand{\Ctx}{\operatorname{Ctx}}
\newcommand{\code}[1]{\ifmmode\text{\upshape\ttfamily #1}\else{\upshape\ttfamily\def\_{\textunderscore\allowbreak}#1}\fi}

\title{\vspace{-10mm}\textbf{\Large Provenance-Isolated Epistemic Quorums: Execution-Bound Evidence Certification and Plausible-World Safety Gating in Agentic Control Planes}}

\author{
  \textbf{Jun He} \qquad \textbf{Deying Yu}\\
  OpenKedge LLC\\
  \texttt{\{junhe, deying\}@openkedge.io}
}

\date{\vspace{-4ex}}

\begin{document}

\maketitle

\begin{abstract}
Agreement among large language models is not physical evidence for mutating production infrastructure. Because generative models share pretraining priors and conform to peer rationales during deliberation, voting ensembles correlate errors instead of providing independent fault tolerance; yet attaching rigid deterministic checks to unconstrained agent outputs paralyzes incident remediation. We present Provenance-Isolated Epistemic Quorums (PIEQ), an execution-bound certification layer that keeps every language model strictly outside the trusted computing base. PIEQ unifies assumption-based truth maintenance, physical provenance isolation, control-theoretic plausible-world safety shielding, and linearizable concurrency governance around three distinct obligations: certifying diagnostic preconditions from disjoint physical failure domains, gating candidate mutations across all unrefuted fault worlds while falling back to read-only diagnostic probes under ambiguity, and binding admission certificates to target execution through snapshot barriers, revision checks, authorization freshness, trusted temporal-precondition policies, and capability-tiered enforcement.

Across formal guarantees, automated \texttt{etcd} and multi-process OS-level fault suites, frozen software benchmarks, scalability stress tests, and live language-model runs, PIEQ achieves zero within-assumption unsafe admissions and zero false-precondition commits, converting occasional agent proof omissions into recoverable false blocks rather than unsafe actions. We characterize when language models add genuine value beyond deterministic TCB rules---specifically multi-hop proof search in large rule spaces (where bounded deterministic completion saturates while targeted verification stays sub-millisecond) and unstructured incident triage grounded in authenticated telemetry---and map exact fail-closed truncation boundaries up to $M{=}64$ observations, $|\mathcal{S}_{\mathrm{cand}}|{=}256$, and Mode~C bounds. On a physical PostgreSQL/Patroni cluster, PIEQ gates supervised failovers alongside Patroni and \texttt{etcd}'s native split-brain protection and enforces a full Tier~E2 contract inside a single serializable PostgreSQL transaction for bounded capacity reservations. All benchmarks are author-labeled, our controlled-sampling direct-API confirmation run pairs a fixed local Qwen3-8B checkpoint with AWS Bedrock inference-profile aliases rather than date-stamped immutable cloud snapshots, and Tier~E2 physical failover promotion as well as multi-host hardware fault isolation remain open work.
\end{abstract}

\section{Introduction}
\label{sec:intro}

Autonomous cloud control planes increasingly invoke ensembles of large language models to diagnose incidents, shift traffic, drain nodes, and coordinate database failovers~\cite{he2026kedgeflow}. A widespread safeguard is multi-agent debate~\cite{du2023debate,choi2025debate,demystifying2026}, in which models exchange natural-language rationales and vote before issuing a mutative API call. Treating agreeing agents as a fault-tolerant quorum, however, confuses opinion consensus with physical evidence independence. Shared pretraining priors and inter-agent sycophancy correlate model errors well beyond classical fault-tolerance assumptions~\cite{lamport1982,he2026honest,hiddenbench2026,sycophancy2025}. Attempting to fix this by placing a strict deterministic verifier directly in front of free-form agent proposals creates the opposite failure mode by paralyzing remediation. In an initial 40-case five-model pilot, strict verification blocked every unsafe action (0/40) but also blocked every actionable incident (0/19 mitigations), discarding 140 of 284 submitted proofs and executing zero diagnostic probes.

Making autonomous infrastructure mutation both safe and actionable requires solving a deeper systems and control problem that agent agreement cannot address. Consider a three-node replicated database whose primary stops answering in-band SQL probes. Two incompatible fault scenarios match that symptom: the primary may have crashed, in which case promoting a replica restores availability while preserving single-primary exclusivity, or an asymmetric network partition may have isolated a healthy primary, in which case promotion induces split-brain. Physical telemetry originates from separate failure domains, such as in-band SQL probes, an out-of-band management bridge, and a consensus lease store, regardless of how many language models inspect those signals. Addressing only part of this pipeline leaves critical failure paths open. Certifying that the primary is unreachable without ruling out a network partition still admits split-brain. Conversely, blocking unsafe promotions without certifying diagnostic preconditions lets a hallucinating agent or a single faulty sensor trigger unwarranted invariant-preserving disruptions such as load shedding or standby drains. Even when both diagnostic checks succeed at observation time, physical safety still breaks if concurrent coordinators race, authorization is revoked before commit, a transient condition clears before effect, or multi-round probes mix observations from inconsistent system snapshots.

We address these coupled failures through Provenance-Isolated Epistemic Quorums (PIEQ), an execution-bound certification layer implemented in a deterministic verification proxy (\texttt{kf-proxy}) that keeps every language model strictly outside the trusted computing base. Rather than claiming individual novelty for its underlying formalisms, PIEQ composes established techniques from four disciplines across the boundary between agentic diagnosis and physical actuation. From symbolic artificial intelligence and formal verification, we adopt assumption-based truth maintenance~\cite{dekleer1986}, model-based diagnosis~\cite{dekleer1987,reiter1987}, data provenance~\cite{green2007provenance}, and proof-carrying authorization~\cite{appel1999pca,necula1997pcc}: instead of debating in prose, untrusted agents submit typed Horn-clause derivations and conflict certificates rooted in cryptographically signed telemetry, which a deterministic verifier checks while tracking minimal supporting assumptions and contradictions. From distributed fault tolerance and combinatorics, we combine Byzantine failure-domain isolation~\cite{lamport1982} with disjoint complete-proof packing, defining fault-tolerant certification by the number of independent physical telemetry domains rather than the number of agreeing agents. From control theory, we apply discrete-event supervisory control~\cite{ramadge1987}, partially observable runtime safety shielding~\cite{spf2024}, viability theory~\cite{aubin1991}, and dual control~\cite{feldbaum1960}: the proxy retains every modeled fault world not refuted by enough independent failure domains, admits a mutation only when it preserves modeled invariants across every surviving world, and shifts to read-only diagnostic probing whenever ambiguity remains. Finally, from concurrency control and distributed systems, we integrate linearizability, strict serializability, stuttering refinement, leases, fencing tokens, and escrow-style exposure accounting~\cite{herlihy1990linearizability,gray1989leases,berenson1995ansi,abadi1991refinement,oneil1986escrow} so that admission certificates remain valid, ordered, authorized, and uncontested all the way to the physical commit boundary.

This paper makes four main contributions while keeping every language model outside the trusted computing base. First, we introduce provenance-isolated epistemic certification, combining bounded truth maintenance with exact complete-proof packing over recursive physical failure-domain footprints so that fault-tolerant precondition certification depends strictly on independent telemetry domains rather than agent count or conversation context. Second, we formulate fault-tolerant plausible-world safety gating, retaining every modeled fault world not refuted by enough disjoint telemetry domains under bounded compute, admitting mutations only when they preserve state invariants across all surviving worlds, and otherwise dispatching read-only diagnostic probes or escalating deterministically. Third, we design an execution-bound target governance layer and eight-stratum conditional risk accounting framework spanning contiguous snapshot barriers, universal admission-revision compare-and-swap, physical stuttering-refinement contracts, mandatory trusted temporal-precondition policies, and capability-tiered target enforcement. Fourth, we make these gates practical for live language models without expanding the trust boundary through three supporting engineering mechanisms (typed proof submission with bounded schema-feedback retry, seven-boundary snapshot-audited active probing, and opt-in bounded deterministic completion) and evaluate the complete architecture across formal counterexamples, automated \texttt{etcd} fault suites, frozen software benchmarks, live model runs, physical PostgreSQL/Patroni failover supervision, and real-PostgreSQL transactional enforcement, while clearly delineating its formal trust boundaries and empirical scope.

The paper proceeds by formalizing the infrastructure plant, composable safety notions, and conformity collapse in agent debate, then developing provenance-isolated certification and plausible-world shielding, extending those guarantees to concurrent physical execution and risk accounting, and concluding with our multi-tier evaluation, related work, and residual limitations.

\section{System Model and Conformity Collapse in Agent Debate}
\label{sec:model}

Designing a safety layer for autonomous infrastructure requires two foundations before any certification algorithm can be written. First, we must define the physical system being controlled, the precise safety properties that hold under concurrent actuation, the physical source of observational independence, and the boundary separating trusted verification from untrusted reasoning. Second, we must examine why multi-agent deliberation fails to supply independent evidence, motivating our architectural decision to move every language model outside the trusted computing base.

\subsection{Plant, Safety Notions, Failure Domains, and Trust Boundary}
\label{sec:tcb}

An autonomous control plane supervises a partially observable infrastructure plant, such as a replicated database cluster, a routing domain, or a storage pool, whose exact physical configuration cannot be inspected instantaneously in full. To reason deterministically about risk, operators model the plant through a finite abstraction that partitions physical configurations into discrete fault regimes (such as a healthy primary, a crashed primary, or a live primary isolated by an asymmetric network partition) and specifies which regimes satisfy operational safety invariants such as single-primary exclusivity. Formally, let $\Sigma_{\mathrm{phys}}$ denote the physical state space, $\mathcal{S}_{\mathrm{cand}} \subseteq \mathcal{S}$ a finite abstract modeled universe, and $\alpha: \Sigma_{\mathrm{phys}} \to \mathcal{S}$ an abstraction map yielding the modeled state $S_\tau = \alpha(\sigma_\tau)$ at epoch $\tau$. Let $\mathcal{K} \subseteq \mathcal{S}$ denote the abstract safe state invariant corresponding to physical invariant $\mathcal{K}^{\mathrm{phys}} \subseteq \Sigma_{\mathrm{phys}}$ via the soundness implication $\alpha(\sigma) \in \mathcal{K} \implies \sigma \in \mathcal{K}^{\mathrm{phys}}$. Available control actions partition cleanly into non-mutating diagnostic probes and state-changing mutations, $\mathcal{A} = \mathcal{A}_{\mathrm{probe}} \sqcup \mathcal{A}_{\mathrm{mut}}$. For any modeled state $s \in \mathcal{S}$ and mutation $a \in \mathcal{A}_{\mathrm{mut}}$, $\Post(s, a) \subseteq \mathcal{S}$ denotes the set of modeled post-states and $A_{\mathrm{safe}}^{\mathcal{K}}(s) \subseteq \mathcal{A}_{\mathrm{mut}}$ denotes the mutations that preserve $\mathcal{K}$ from $s$.

Control theory distinguishes carefully between maintaining a state invariant at every step and maintaining long-horizon controllability, and this distinction directly shapes how concurrent safety composes. State invariant preservation requires that every intermediate and terminal state along an execution remains inside $\mathcal{K}$, whereas one-step safety ensures that from $s \in \mathcal{K}$, applying $a$ yields $\Post(s, a) \subseteq \mathcal{K}$. Controlled viability, by contrast, requires that a state belong to the viability kernel $\Viab(\mathcal{K}) \subseteq \mathcal{K}$, meaning there exists an infinite admissible control sequence staying inside $\mathcal{K}$~\cite{aubin1991}. Crucially, controlled viability does not compose by intersection under coupled invariants: in general, $\Viab(\mathcal{K}_1 \cap \mathcal{K}_2) \subsetneq \Viab(\mathcal{K}_1) \cap \Viab(\mathcal{K}_2)$ strictly. Consider a state $s_0 \in \mathcal{K}_1 \cap \mathcal{K}_2$ that admits an action $a_1$ looping forever inside $\mathcal{K}_1 \setminus \mathcal{K}_2$ and an action $a_2$ looping forever inside $\mathcal{K}_2 \setminus \mathcal{K}_1$, with no action keeping the plant inside $\mathcal{K}_1 \cap \mathcal{K}_2$. Here $s_0$ lies in both individual viability kernels, yet $\Viab(\mathcal{K}_1 \cap \mathcal{K}_2) = \varnothing$. PIEQ's compositional execution guarantees therefore target state invariant preservation, which composes soundly across closed domains by intersection ($\mathcal{K} = \bigcap_{\mathcal{D}} \mathcal{K}_{\mathcal{D}}$); achieving controlled viability requires supplying a single pre-verified global viability kernel $\mathcal{K}^\star \triangleq \Viab(\mathcal{K})$ as the target invariant.

Observational independence comes from physical infrastructure topology rather than from the number of reasoning agents inspecting a dashboard. In our database scenario, in-band SQL health checks, an out-of-band management bridge, and a consensus lease store traverse distinct physical pathways and therefore constitute separate telemetry failure domains. Formally, telemetry arrives over $M$ channels $\mathcal{O} = \{O_1, \dots, O_M\}$, each mapped by a signed topology manifest (digest $h_{\mathrm{man}}$) to one of $D$ telemetry failure domains $\mathcal{F} = \{F_1, \dots, F_D\}$ (disambiguated from closed exposure domain sets $\mathcal{D} = \Dom^\star(a)$ introduced when we analyze concurrency). At epoch $\tau$, channel $O_m$ emits a signed primitive assumption
\begin{equation}
  e_m = \langle \mathrm{id}_m, \mathrm{res}_m, \tau, d_m, v_m, h_{\mathrm{man}}, \sigma_m \rangle,
  \label{eq:prim_assump}
\end{equation}
where $\mathrm{res}_m$ scopes the target resource, $d_m \in \{1,\dots,D\}$ identifies failure domain $F_{d_m}$, $v_m = v_{\mathrm{obs}}(e_m)$ records the observed target snapshot version, and $\sigma_m$ is a collector signature covering all fields.

A collector signature $\sigma_m$ authenticates origin, payload integrity, and metadata binding, but never guarantees physical truth: a misconfigured sensor or partitioned probe still validly signs a misleading observation. PIEQ places all $N$ language-model proposers $\{A_1, \dots, A_N\}$ strictly outside the trusted computing base (TCB). The TCB comprises exactly four deterministic elements: the authenticated telemetry collectors and signed topology manifest $h_{\mathrm{man}}$; the deterministic rule and conflict verifier $\mathcal{V}_{\mathrm{rule}}$ checking derivations against the authored Horn model $\mathcal{M}_{\mathrm{sys}}$ and conflict invariants $P_{\mathrm{sys}}$; the verification proxy \texttt{kf-proxy} housing the bounded truth-maintenance engine, complete-proof packer, plausible-world safety shield, and linearizable exposure ledger $L_{\mathrm{exp}}$; and the durable target fencing and authorization adapters at the physical effect boundary (capability Tiers~E1--E2). Table~\ref{tab:notation} summarizes the core notation used throughout the paper.

\begin{table}[!t]
\centering
\caption{Core notation (disambiguating $D = |\mathcal{F}|$ telemetry failure domains from closed exposure domain sets $\mathcal{D} = \Dom^\star(a)$).}
\label{tab:notation}
\vspace{0.4mm}
\footnotesize
\setlength{\tabcolsep}{3.2pt}
\renewcommand{\arraystretch}{1.10}
\begin{tabularx}{\columnwidth}{@{}l Y@{}}
\toprule
\textbf{Symbol} & \textbf{Meaning} \\
\midrule
$\sigma_\tau, \alpha, S_\tau, \mathcal{K}$ & Physical state, abstraction map, modeled state $S_\tau{=}\alpha(\sigma_\tau)$, invariant \\
$\mathcal{F} = \{F_1,\dots,F_D\}$ & Telemetry failure domains ($D$ total; $d_m \in \{1,\dots,D\}$) \\
$\mathcal{E}_{\mathrm{prim}}, \mathfrak{L}(c), \mathcal{N}_\tau$ & Signed assumptions ($M \le 64$), ATMS label, verified nogoods \\
$\Pi, E_{\mathrm{leaf}}(\Pi), \Phi_{\mathrm{phys}}(\Pi)$ & Verified derivation DAG, primitive leaves, physical evidence footprint \\
$\mathcal{W}(a), \nu_{\mathrm{pack}}$ & Active proof DAGs for $c_a$, disjoint complete-proof packing number \\
$k_a, f = k_a - 1$ & Quorum threshold for $a$ and tolerated faulty failure domains \\
$\mathcal{B}_{\tau,v}^{(f)}, \widehat{\mathcal{B}}_{\tau,v}^{(f)}$ & Exact and bounded-compute $f$-fault-tolerant plausible world sets \\
$\mathcal{D} = \Dom^\star(a), \mathcal{K}_{\mathcal{D}}$ & Closed exposure domain set ($d \in \mathcal{D}$), scoped state invariant \\
$\theta, \mathsf{Refines}_{\mathcal{D}}$ & Physical execution trace and stuttering-refinement contract \\
$\mathcal{I}_C, \mathcal{I}_H, \mathcal{I}_U, \mathbf{r}_{\mathrm{adm}}$ & Committed, in-flight, \texttt{UNKNOWN} sets; per-domain admission revs \\
$\tau_{\mathrm{adm}}, \tau_{\mathrm{auth}}, t_{\mathrm{auth}}$ & Evidence epoch, authorization epoch, validity deadline \\
$E_1, \dots, E_8$ & Ordered residual failure events in conditional risk accounting \\
\bottomrule
\end{tabularx}
\end{table}

\subsection{Why Multi-Agent Deliberation Collapses Diversity}
\label{sec:theory}

Before replacing agent voting with physical provenance certification, we summarize why peer deliberation cannot manufacture observational independence (formal definitions, spectral proof, and counterexample appear in Appendix~\ref{app:thm1_proof}). Model $N \ge 2$ proposers evaluating a binary hypothesis through unbounded latent score residuals $\varepsilon_i^{(t)} = z_i^{(t)} - z^\star(S_\tau) \in \mathbb{R}$ with separate probability readout $y_i^{(t)} = \psi(z_i^{(t)}) \in [0,1]$, initial equicorrelation $\bar{\rho}_0 \in [0,1)$, and shared-context conformity update $\varepsilon_i^{(t+1)} = (1 - \gamma)\varepsilon_i^{(t)} + \gamma \bar{\varepsilon}^{(t)} + \eta_i^{(t)}$ ($\gamma \in (0,1]$) driven by correlated innovations $\eta^{(t)}$~\cite{choi2025debate,sycophancy2025}. Because this linear operator damps the disagreement subspace $\mathbf{1}^\perp$ by $(1-\gamma)$ at each round while leaving the consensus subspace $\operatorname{span}(\mathbf{1})$ undamped, Theorem~\ref{thm:conformity} (Appendix~\ref{app:thm1_proof}) shows that the effective sample size $N_{\mathrm{eff}}(\Sigma^{(t)}) = N/[1+(N-1)\bar{\rho}_t]$ collapses to~$1$ geometrically as $(1-\gamma)^{2t}$ without innovations and as $1 + \Theta(t^{-1})$ under persistent innovations. Three analytical subtleties accompany this collapse: the spectral participation ratio $r_{\mathrm{PR}}$ strictly overstates variance reduction relative to $N_{\mathrm{eff}}$ (Definition~\ref{def:metrics}); Popoviciu's bound ($\operatorname{Var}(y_i^{(t)}) \le 1/4$) forbids unbounded persistent additive drift directly on $[0,1]$ probabilities, necessitating the latent-score formulation; and symmetric averaging alone collapses a $5$-of-$5$ gate to a Round-0 majority rather than single-agent error (Remark~\ref{rem:counterexample_1c}), while empirical debate degrades further under shared priors ($\bar{\rho}_0 \gg 0$) and asymmetric sycophancy~\cite{hiddenbench2026,sycophancy2025,failuremodes2025}. Consequently, PIEQ never counts agreeing agents toward a safety quorum.

\subsection{When to Keep Language Models in the Loop: Synthesis vs.\ Verification Asymmetry}
\label{sec:llm_role}

Because PIEQ requires operators to author a trusted symbolic safety model $(\mathcal{S}_{\mathrm{cand}}, \mathcal{M}_{\mathrm{sys}}, P_{\mathrm{sys}}, A_{\mathrm{safe}}^{\mathcal{K}}, \widehat{\Post}, h_{\mathrm{man}}, \mathsf{TPol})$ inside the TCB, an immediate architectural question arises: once that symbolic model exists, when does keeping an untrusted language model in the control loop provide genuine value over a purely deterministic controller? Answering this candidly separates three operational regimes.

First, in \emph{small, single-hop, fully structured domains} (such as clean three-signal Patroni leader death or the small single-step confirmation incidents \code{CN10}/\code{CN12} in Section~\ref{sec:live_reliability}), a deterministic supervisor or bounded single-step TCB forward-chainer is strictly faster, cheaper, and more reliable than live LLM inference. In this regime, PIEQ is not a reason to introduce an LLM; rather, it provides a sound safety boundary when an LLM agent is already deployed for broader orchestration.

Second, in \emph{large or multi-hop rule spaces and multi-action remediation planning}, PIEQ exploits the classical \emph{synthesis-vs.-verification asymmetry} familiar from proof-carrying code and SAT/SMT witness checking~\cite{necula1997pcc,appel1999pca}. Checking a proposed Horn derivation DAG and computing its physical footprint takes sub-millisecond deterministic time in a tiny TCB, whereas exhaustive forward-chaining inside the TCB must impose tight derivation caps to remain predictable and auditable. As Section~\ref{sec:synthesis_scalability} demonstrates empirically, when background rules and derivation depth grow ($R_{\mathrm{bg}} \ge 80$, depth $d \in \{2,3,4\}$), both single-hop and bounded multi-hop TCB forward-chaining exhaust their derivation caps and fail closed ($0/8$ admitted), whereas an untrusted proposer performing goal-directed proof search submits only the relevant multi-hop DAGs, which the unchanged TCB verifies in $0.07$--$0.28\,\mathrm{ms}$ with zero TCB expansion.

Third, in \emph{unstructured incident triage feeding authenticated probing}, real operational incidents begin with messy tickets, stack traces, log snippets, change diffs, and operator runbooks that deterministic Horn engines cannot parse directly. Here the LLM serves strictly as an untrusted triage, scope-selection, and hypothesis-proposal engine: it selects which resource scope to inspect, which candidate remediation among safe alternatives best fits operational intent, and which diagnostic probes to request. PIEQ guarantees that no hallucination or prompt injection from unstructured text can ever forge physical evidence, because every admitted leaf must bind cryptographically to a collector-signed observation or read-only probe handle. Section~\ref{sec:synthesis_scalability} quantifies both the authored model footprint across all ten benchmark domains and this synthesis-vs.-completion trade-off.

\section{Complete-Proof Certification and Provenance Packing}
\label{sec:atms_packing}

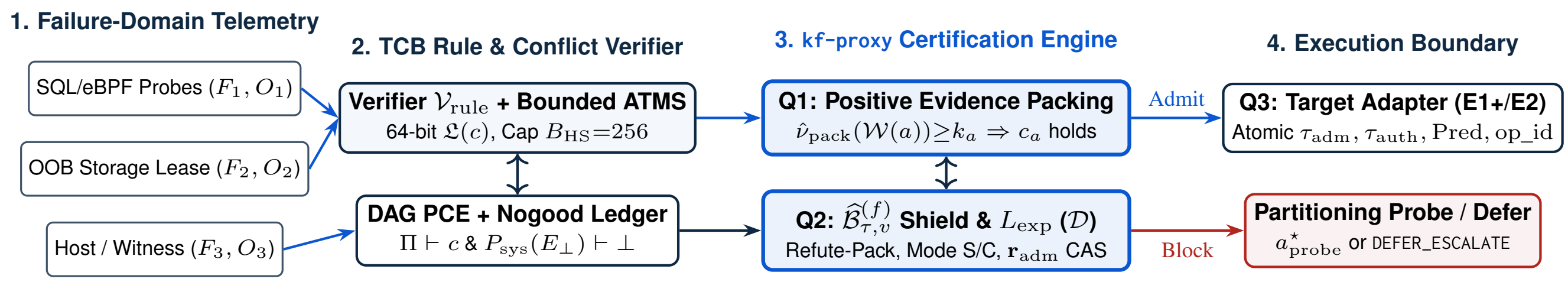
\begin{figure*}[!t]
\centering
\begin{tikzpicture}[
  font=\footnotesize\sffamily,
  box/.style={draw=primary, thick, rounded corners=3pt, fill=surface, minimum height=7.6mm, align=center, inner sep=2.8pt},
  gate/.style={draw=accent, very thick, rounded corners=3pt, fill=accent!8, minimum height=8.2mm, align=center, inner sep=3.0pt},
  sensor/.style={draw=primary!80, semithick, rounded corners=2pt, fill=white, minimum width=26mm, minimum height=6.0mm, align=center, inner sep=2.4pt},
  arr/.style={-{Stealth[length=2.0mm,width=1.4mm]}, thick, color=primary},
  bluearr/.style={-{Stealth[length=2.0mm,width=1.4mm]}, thick, color=accent}
]
\node[sensor] (s1) at (0, 1.25) {\scriptsize SQL/eBPF Probes ($F_1, O_1$)};
\node[sensor] (s2) at (0, 0.35) {\scriptsize OOB Storage Lease ($F_2, O_2$)};
\node[sensor] (s3) at (0, -0.55) {\scriptsize Host / Witness ($F_3, O_3$)};
\node[above=1.8mm of s1, font=\footnotesize\bfseries\sffamily, color=primary] {1. Failure-Domain Telemetry};

\node[box, minimum width=36mm] (atms) at (3.95, 0.92) {\textbf{Verifier $\mathcal{V}_{\mathrm{rule}}$ + Bounded ATMS}\\\scriptsize 64-bit $\mathfrak{L}(c)$, Cap $B_{\mathrm{HS}}{=}256$};
\node[box, minimum width=36mm, fill=white] (nogood) at (3.95, -0.34) {\textbf{DAG PCE + Nogood Ledger}\\\scriptsize $\Pi \vdash c$ \& $P_{\mathrm{sys}}(E_{\bot}) \vdash \bot$};
\node[above=1.8mm of atms, font=\footnotesize\bfseries\sffamily, color=primary] {2. TCB Rule \& Conflict Verifier};

\node[gate, minimum width=41mm] (cut) at (8.75, 0.92) {\textbf{Q1: Positive Evidence Packing}\\\scriptsize $\hat{\nu}_{\mathrm{pack}}(\mathcal{W}(a)){\ge}k_a \Rightarrow c_a$ holds};
\node[gate, minimum width=41mm] (diag) at (8.75, -0.34) {\textbf{Q2: $\widehat{\mathcal{B}}_{\tau,v}^{(f)}$ Shield \& $L_{\mathrm{exp}}$ ($\mathcal{D}$)}\\\scriptsize Refute-Pack, Mode S/C, $\mathbf{r}_{\mathrm{adm}}$ CAS};
\node[above=1.8mm of cut, font=\footnotesize\bfseries\sffamily, color=accent] {3. \texttt{kf-proxy} Certification Engine};

\node[box, minimum width=32mm, fill=white] (fence) at (13.75, 0.92) {\textbf{Q3: Target Adapter (E1+/E2)}\\\scriptsize Atomic $\tau_{\mathrm{adm}}, \tau_{\mathrm{auth}}, \Pred, \mathrm{op\_id}$};
\node[box, minimum width=32mm, draw=alert, fill=alert!6] (probe) at (13.75, -0.34) {\textbf{Partitioning Probe / Defer}\\\scriptsize $a_{\mathrm{probe}}^\star$ or \texttt{DEFER\_ESCALATE}};
\node[above=1.8mm of fence, font=\footnotesize\bfseries\sffamily, color=primary] {4. Execution Boundary};

\draw[bluearr] (s1.east) -- (atms.west);
\draw[bluearr] (s2.east) -- (atms.west);
\draw[bluearr] (s3.east) -- (nogood.west);
\draw[arr, <->] (atms.south) -- (nogood.north);
\draw[bluearr] (atms.east) -- (cut.west);
\draw[arr] (nogood.east) -- (diag.west);
\draw[arr, <->] (cut.south) -- (diag.north);
\draw[bluearr] (cut.east) -- node[above, font=\scriptsize] {Admit} (fence.west);
\draw[arr, color=alert] (diag.east) -- node[below, font=\scriptsize, color=alert] {Block} (probe.west);
\end{tikzpicture}
\caption{Execution-bound epistemic certification in PIEQ (\texttt{kf-proxy}). Untrusted LLMs propose candidate Horn DAGs and nogoods outside the TCB; $\mathcal{V}_{\mathrm{rule}}$ verifies derivations deterministically; \texttt{kf-proxy} separates Q1 (positive evidence certification $\hat{\nu}_{\mathrm{pack}}(\mathcal{W}(a)) \ge k_a$ for precondition $c_a$) from Q2 (plausible-world safety shielding via refutation packing plus physically refined closed-domain concurrent governance on $\mathcal{D} = \Dom^\star(a)$ with $\mathbf{r}_{\mathrm{adm}}(\mathcal{D})$ CAS); Q3 delegates durable fencing, $\Pred(a)$, and execution-time authorization freshness to target capability tiers (Table~\ref{tab:tiers}).}
\label{fig:arch}
\end{figure*}

Removing language models from the trusted computing base creates an architectural challenge: the control plane still wants to leverage agents' ability to synthesize multi-hop diagnoses across heterogeneous operational telemetry, yet it must never accept persuasive prose or double-count evidence originating from the same physical failure domain. PIEQ solves this inside \texttt{kf-proxy} (Figure~\ref{fig:arch}) by combining proof-carrying communication, assumption-based truth maintenance, and recursive physical provenance packing. As Figure~\ref{fig:arch} illustrates, signed telemetry from distinct failure domains feeds a deterministic rule and conflict verifier paired with a bounded truth-maintenance engine. Untrusted models propose candidate derivation graphs and conflict certificates outside the TCB, and \texttt{kf-proxy} enforces two separate epistemic gates before execution: positive evidence packing (Q1) certifies that the mutation's diagnostic precondition holds across independent physical domains, while plausible-world safety shielding coupled with closed-domain concurrency governance (Q2) verifies that the mutation preserves invariants across all unrefuted fault worlds and active schedules. Certified actions proceed to the target execution boundary (Q3), whereas ambiguous or under-supported proposals trigger read-only diagnostic probing or escalation.

\subsection{Derivation-DAG PCE+Nogood Channel and Bounded ATMS}
\label{sec:atms}

An assumption-based truth maintenance system (ATMS)~\cite{dekleer1986} tracks not merely which diagnostic conclusions have been derived, but exactly which minimal combinations of primitive observations support each conclusion and which combinations are mutually contradictory. By replacing free-form agent dialogue with a Proof-Carrying Evidence and Nogood (PCE+Nogood) channel~\cite{kadyrbek2026,wang2026pcaa}, PIEQ allows models to synthesize multi-hop explanations over distributed telemetry~\cite{diverse2026,hiddenbench2026} while stripping away unverified rhetoric. Specifically, let $\mathcal{E}_{\mathrm{prim}} = \{e_0, \dots, e_{M-1}\}$ ($M \le 64$) be the signed primitive assumptions bound to the active tuple $(\mathrm{res}, \tau, h_{\mathrm{man}})$. Any assumption environment $E \subseteq \mathcal{E}_{\mathrm{prim}}$ maps to a 64-bit integer mask $\mathsf{mask}(E) = \sum_{e_m \in E} 2^m$, enabling constant-time subset checks. Each ATMS node $n_c = \langle c, \mathfrak{L}(c), \mathcal{J}(c) \rangle$ stores the minimal consistent primitive environments $\mathfrak{L}(c)$ and verified derivation DAGs $\mathcal{J}(c)$ supporting proposition $c$.

\begin{definition}[Verified Derivation-DAG and Nogood Channel]
\label{def:pce_nogood}
Deterministic verifier $\mathcal{V}_{\mathrm{rule}}$ admits a message $m$ if and only if $m$ is either: (1)~a verified derivation DAG $\Pi$ rooted at proposition $c$, consisting of a finite acyclic graph whose leaves are nonempty signed assumptions $e_m \in \mathcal{E}_{\mathrm{prim}}$ matching $(\mathrm{res}, \tau, h_{\mathrm{man}})$ and whose internal nodes $v = \langle \mathrm{id}_A(v), \mathrm{ctx}(v), c_v, \mathrm{ch}(v), r_v, \sigma_v \rangle$ each apply a rule $r_v \in \mathcal{M}_{\mathrm{sys}}$ with consequent $c_v$ whose antecedent set bijectively matches the propositions of $\mathrm{ch}(v)$, with recursive primitive leaf set $E_{\mathrm{leaf}}(e_m) \triangleq \{e_m\}$ and $E_{\mathrm{leaf}}(v) \triangleq \bigcup_{u \in \mathrm{ch}(v)} E_{\mathrm{leaf}}(u) \neq \varnothing$; or (2)~a minimal nogood certificate $\mathrm{nogood}(E_{\bot}, \mathrm{inv}_{\mathrm{id}})$ where nonempty signed $E_{\bot} \subseteq \mathcal{E}_{\mathrm{prim}}$ bijectively matches the conflict set of $\mathrm{inv}_{\mathrm{id}} \in P_{\mathrm{sys}}$ ($P_{\mathrm{sys}}(E_{\bot}) \vdash \bot$).
\end{definition}

Truth maintenance preserves valid conclusions and purges contradictory derivations whenever the authored Horn rules and conflict invariants accurately describe the modeled plant.

\begin{assumption}[Trusted Deductive Rule and Invariant Soundness]
\label{assump:sound_rules}
At consistent snapshot state $S$, let $\mathcal{E}_{\mathrm{true}}(S) \subseteq \mathcal{E}_{\mathrm{prim}}$ denote primitive assumptions whose predicates physically hold in $S$. Every $r \in \mathcal{M}_{\mathrm{sys}}$ is truth-preserving in $S$, and $P_{\mathrm{sys}}$ is conflict-sound: for all $E \subseteq \mathcal{E}_{\mathrm{true}}(S)$, $P_{\mathrm{sys}}(E) \not\vdash \bot$.
\end{assumption}

\begin{theorem}[Minimal Subset Preservation and Conflict Elimination]
\label{thm:atms_soundness}
Under Definition~\ref{def:pce_nogood} and Assumption~\ref{assump:sound_rules}, let $\mathcal{N}_\tau^{(t)}$ be the verified nogoods registered by round $t$. Then: (1)~if a true assumption set $E^\star \subseteq \mathcal{E}_{\mathrm{true}}(S)$ supports $c$ via a verified DAG $\Pi_0$ with $E_0 = E_{\mathrm{leaf}}(\Pi_0) \subseteq E^\star$ ($|E_0| \le d_{\max}$), and at registration either some $\underline{E} \subseteq E_0$ already belongs to $\mathfrak{L}(c)$ or $|\mathfrak{L}(c)| < L_{\max}$, then because strict supersets are pruned before checking $L_{\max}$, $\mathfrak{L}^{(t)}(c)$ retains at least one minimal consistent subset $\underline{E} \subseteq E^\star$ for all $t \ge 1$ even if later arrivals set $\mathsf{budget\_exceeded}$; and (2)~every environment $E \supseteq E_{\bot}$ containing $E_{\bot} \in \mathcal{N}_\tau^{(t)}$ is purged from $\mathfrak{L}^{(t)}(c)$ and every DAG $\Pi$ with $E_{\bot} \subseteq E_{\mathrm{leaf}}(\Pi)$ is deactivated via $\mathsf{mask}(E) \mathbin{\&} \mathsf{mask}(E_{\bot}) = \mathsf{mask}(E_{\bot})$.
\end{theorem}

\begin{proof}
By Assumption~\ref{assump:sound_rules}, no verified nogood subsets $E_0 \subseteq \mathcal{E}_{\mathrm{true}}(S)$, so $E_0$ (or an existing $\underline{E} \subseteq E_0$) resides in $\mathfrak{L}(c)$. Any later consistent $E'$ displaces $\underline{E} \subseteq E^\star$ only if $E' \subsetneq \underline{E} \subseteq E^\star$; pruning strict supersets drops $\underline{E}$ before checking $L_{\max}$, leaving $|\mathfrak{L}_{\mathrm{pruned}}(c)| < L_{\max}$ so $E'$ inserts. Non-subsuming arrivals at capacity set $\mathsf{budget\_exceeded}$ without mutating $\mathfrak{L}(c)$, and no future verified nogood can subset $E' \subseteq \mathcal{E}_{\mathrm{true}}(S)$. Claim~(2) follows from the bitmask subset identity.
\end{proof}

Two operational properties follow from this construction. First, Theorem~\ref{thm:atms_soundness} guarantees deductive soundness rather than statistical independence among surviving observations: conditioning independent $X, Y \sim \mathrm{Bernoulli}(0.5)$ on a conflict filter enforcing $X = Y$ induces $\operatorname{Corr}(X, Y \mid X = Y) = 1$ (Berkson's paradox). Second, because general ATMS label maintenance and hitting-set enumeration can grow exponentially~\cite{provan1990}, PIEQ imposes strict online resource bounds ($M \le 64$, $|E_{\mathrm{leaf}}(\Pi)| \le d_{\max} = 6$, $|\mathfrak{L}(c)| \le L_{\max} = 64$, hitting-set cap $B_{\mathrm{HS}} = 256$, and packing cap $B_{\mathrm{pack}}$). These caps enforce dual fail-closed truncation: exhausting the positive packing cap only lowers the certified lower bound $\hat{\nu}_{\mathrm{pack}} \le \nu_{\mathrm{pack}}$ (denying admission), while exhausting belief-shield caps conservatively retains candidate fault worlds. Safety never depends on untrusted agents discovering every valid nogood.

\subsection{Recursive Dependency Footprints and Complete-Proof Packing}
\label{sec:packing}

Verifying individual Horn derivations establishes logical validity given their leaves, but it does not yet reveal how many physically independent telemetry domains back a candidate mutation. For mutation $a \in \mathcal{A}_{\mathrm{mut}}$ with diagnostic precondition $c_a$, let $\mathcal{W}(a) \triangleq \{\Pi \in \mathcal{J}(c_a) : \forall E_{\bot} \in \mathcal{N}_\tau,\, E_{\bot} \not\subseteq E_{\mathrm{leaf}}(\Pi)\}$ be its active, conflict-free verified proof DAGs. Computing fault tolerance over $\mathcal{W}(a)$ requires exact complete-proof packing rather than standard graph connectivity or shallow antecedent checks, because both simpler heuristics overcount shared dependencies. Consider three observations $\{e_1, e_2, e_3\}$ drawn from disjoint failure domains $F_1, F_2, F_3$ that assemble into three conjunctive proofs:
\begin{equation}
\begin{aligned}
  E_{\mathrm{leaf}}(\Pi_1) &= \{e_1, e_2\}, \qquad E_{\mathrm{leaf}}(\Pi_2) = \{e_2, e_3\}, \\
  E_{\mathrm{leaf}}(\Pi_3) &= \{e_1, e_3\}.
\end{aligned}
\label{eq:conj_counterexample}
\end{equation}
Unit-vertex max-flow routes three vertex-disjoint paths ($F_i \to e_i \to \Pi_i$) and reports $\operatorname{minCut} = 3$, even though every pair of proofs shares a primitive leaf so no two complete proofs are physically disjoint ($\nu_{\mathrm{pack}} = 1$). Similarly, inspecting only immediate antecedents fails on multi-hop derivations: two depth-2 proofs $d_1 \vdash c_a$ and $d_2 \vdash c_a$ derived from $\{e_1, e_2\}$ and $\{e_1, e_3\}$ look disjoint at depth~1 while sharing upstream leaf $e_1$.

Borrowing lineage propagation from data provenance~\cite{green2007provenance}, we define the physical evidence independence footprint $\Phi_{\mathrm{phys}}(\Pi)$ recursively across the derivation DAG:
\begin{equation}
  \Phi_{\mathrm{phys}}(e_m) \triangleq \{F_{d_m}, O_m, e_m\}, \quad
  \Phi_{\mathrm{phys}}(v) \triangleq \!\bigcup_{u \in \mathrm{ch}(v)}\! \Phi_{\mathrm{phys}}(u),
  \label{eq:footprint}
\end{equation}
collecting every ancestor primitive leaf, collector channel, and physical failure domain, and nothing else. Proposer context identifiers $\Ctx(\Pi)$, which tag the agent and conversation round that emitted each derivation node, are excluded deliberately. Folding $\{\mathrm{ctx}(v)\}$ into the footprint conflates four distinct notions of independence: physical telemetry failure-domain independence, authenticated primitive observation independence, language-model reasoning independence, and proof-construction context-identifier distinctness. Only the first two enter our deterministic fault-threshold guarantee. Context identifiers are a labeling artifact that a single model can mint freely and never certify independent reasoning, while model reasoning independence is empirically relevant for synthesis recall but formally unmodeled. PIEQ's fault threshold is therefore anchored strictly in independent physical evidence domains and remains completely agnostic to agent count, model family, conversation context, or submission count.

\begin{definition}[Disjoint Complete-Justification Packing Number]
\label{def:packing}
For active verified proof DAGs $\mathcal{W}$,
\begin{multline}
  \nu_{\mathrm{pack}}(\mathcal{W}) \triangleq \max\Bigl\{\, |\mathcal{J}| : \mathcal{J} \subseteq \mathcal{W}, \\
  \forall \Pi \neq \Pi' \in \mathcal{J},\; \Phi_{\mathrm{phys}}(\Pi) \cap \Phi_{\mathrm{phys}}(\Pi') = \varnothing \,\Bigr\},
  \label{eq:nupack}
\end{multline}
excluding any $\Pi$ with $\Phi_{\mathrm{phys}}(\Pi) = \varnothing$, because a derivation resting on no authenticated primitive observation certifies nothing and must not be vacuously disjoint from other proofs.
\end{definition}

An operator may optionally enforce a stricter proposer-context isolation policy requiring pairwise-distinct context tokens among packed proofs. Because augmenting the footprint with context tags yields a superset of $\Phi_{\mathrm{phys}}$, such a policy can only reduce $\nu_{\mathrm{pack}}$ and never weakens the physical fault bound; shared physical telemetry always creates footprint overlap regardless of context labeling. Under Definition~\ref{def:packing}, one agent synthesizing proofs from three independent telemetry domains legitimately reaches $\nu_{\mathrm{pack}} = 3$, whereas five agents (or five context IDs) reasoning over a single physical domain can never exceed $\nu_{\mathrm{pack}} = 1$. A single false observation from one domain is never counted twice through separate derivations.

\begin{assumption}[Failure-Domain Partition Integrity]
\label{assump:domain_partition}
The TCB topology manifest $h_{\mathrm{man}}$ accurately binds every collector $O_m$ to its true physical or logical failure domain $F_{d_m} \in \mathcal{F}$, and every verified DAG $\Pi$ has $E_{\mathrm{leaf}}(\Pi) \neq \varnothing$.
\end{assumption}

\begin{theorem}[Complete-Proof Independence and Positive Evidence Certification]
\label{thm:packing}
Let Assumptions~\ref{assump:sound_rules} and~\ref{assump:domain_partition} hold at a consistent observation snapshot $S_{v_{\mathrm{snap}}}$ (all leaves in $\mathcal{W}(a)$ observe predicates of $S_{v_{\mathrm{snap}}}$), and let $\mathcal{G}_{\mathrm{prov}}$ be the unit-vertex-capacitated physical flow graph over $\mathcal{F}_{\mathrm{src}} \to \mathcal{O}_{\mathrm{raw}} \to \mathcal{E}_{\mathrm{prim}} \to \mathcal{W}(a)$ excluding proposer contexts. Then three properties hold. First, (1)~for deterministic $k_a$-domain fault tolerance and snapshot precondition certification, $\nu_{\mathrm{pack}}(\mathcal{W}(a)) \ge k_a$ certifies $k_a$ complete verified DAGs $\mathcal{J}_a = \{\Pi_1,\dots,\Pi_{k_a}\}$ with pairwise-disjoint ancestor failure domains, collectors, and leaves regardless of proposer contexts; if $c_a$ is physically false in $S_{v_{\mathrm{snap}}}$, then at least $k_a$ distinct failure domains in $\mathcal{F}$ emitted false assumptions, so contrapositively under at most $f = k_a - 1$ faulty failure domains, $\nu_{\mathrm{pack}}(\mathcal{W}(a)) \ge k_a$ certifies $\mathsf{SnapshotPreconditionValid}(a, v_{\mathrm{snap}})$, meaning precondition $c_a$ physically held at snapshot $S_{v_{\mathrm{snap}}}$ (distinct from commit-time truth $\mathsf{CommitPreconditionValid}(a, t_{\mathrm{eff}})$, addressed in Section~\ref{sec:execution}). Second, (2)~for Menger upper bounds and single-leaf exactness, $\nu_{\mathrm{pack}}(\mathcal{W}(a)) \le \operatorname{minCut}_{\mathcal{G}_{\mathrm{prov}}}(\mathcal{F}_{\mathrm{src}}, \mathcal{W}(a))$, with polynomial-time equality on the single-leaf fragment where $|E_{\mathrm{leaf}}(\Pi)| = 1$ for every $\Pi \in \mathcal{W}(a)$. Third, (3)~for exactness and fail-closed truncation of footprint-bitmask branch-and-bound, deduplicating equal footprints and discarding any $\Pi'$ with $\Phi_{\mathrm{phys}}(\Pi) \subsetneq \Phi_{\mathrm{phys}}(\Pi')$ preserves $\nu_{\mathrm{pack}}(\mathcal{W}(a))$ exactly, and bounded branch-and-bound over arbitrary-precision footprint bitmasks returns $\nu_{\mathrm{pack}}$ when $B_{\mathrm{pack}}$ is not exhausted and a sound lower bound $\hat{\nu}_{\mathrm{pack}} \le \nu_{\mathrm{pack}}$ (with $\mathsf{trunc} = \mathsf{true}$) upon exhaustion.
\end{theorem}

\begin{proof}
For (1), structural induction on any verified DAG $\Pi$ shows that if its root proposition is false in $S_{v_{\mathrm{snap}}}$, Assumption~\ref{assump:sound_rules} implies at least one child is false in $S_{v_{\mathrm{snap}}}$, descending to a false primitive leaf $e_{m_i} \in E_{\mathrm{leaf}}(\Pi_i)$ in domain $F_{d_{m_i}} \in \Phi_{\mathrm{phys}}(\Pi_i)$. Pairwise disjointness $\Phi_{\mathrm{phys}}(\Pi_i) \cap \Phi_{\mathrm{phys}}(\Pi_j) = \varnothing$ produces $k_a$ distinct faulty domains, contradicting at most $f = k_a - 1$ faulty domains if $c_a$ were false. For (2), each disjoint packing of size $k$ induces $k$ vertex-disjoint paths in $\mathcal{G}_{\mathrm{prov}}$ by Menger's theorem; conversely, when $|E_{\mathrm{leaf}}(\Pi)| = 1$, each terminal edge $(e_m, \Pi)$ on $k$ vertex-disjoint paths is witnessed by a proof $\Pi_{e_m} \in \mathcal{W}(a)$ whose physical footprint $\{F_{d_m}, O_m, e_m\}$ is pairwise disjoint across paths. For (3), if $\Phi_{\mathrm{phys}}(\Pi) \subseteq \Phi_{\mathrm{phys}}(\Pi')$ with $\Pi \neq \Pi'$, both cannot coexist in any disjoint packing and replacing $\Pi'$ with $\Pi$ preserves validity and cardinality; branch-and-bound records only valid disjoint subsets and prunes strictly dominated branches, so termination is exact and truncation returns $\hat{\nu}_{\mathrm{pack}} \le \nu_{\mathrm{pack}}$.
\end{proof}

Certifying $\Phi_{\mathrm{phys}}(\Pi_i) \cap \Phi_{\mathrm{phys}}(\Pi_j) = \varnothing$ proves structural failure-domain disjointness under $h_{\mathrm{man}}$. It proves neither statistical independence of physical faults (handled conditionally in Section~\ref{sec:risk_bound}) nor independence of the proposers' reasoning, which PIEQ neither requires nor claims. More importantly, certifying that a diagnostic precondition holds at snapshot $v_{\mathrm{snap}}$ does not by itself guarantee that a candidate mutation preserves safety invariants when multiple fault worlds explain the same symptom, which is the control-theoretic problem addressed next.

\section{Fault-Tolerant Plausible-World Safety Shielding}
\label{sec:shield}

Even when positive proof packing satisfies $\nu_{\mathrm{pack}}(\mathcal{W}(a)) \ge k_a$, confirming a symptom cannot disambiguate incompatible underlying fault worlds. In our replicated database example, independent confirmations that SQL health checks time out still match both a crashed primary ($s_{\mathrm{dead}}$) and a live primary hidden behind an asymmetric network partition ($s_{\mathrm{part}}$). Furthermore, checking safe-action overlap across pairs of plausible worlds does not guarantee a globally safe mutation: if three plausible worlds have safe action sets
\begin{equation}
  A_{\mathrm{safe}}(s_1){=}\{a,b\},\; A_{\mathrm{safe}}(s_2){=}\{b,c\},\; A_{\mathrm{safe}}(s_3){=}\{a,c\},
  \label{eq:helly_counterexample}
\end{equation}
every pair shares a safe action, yet the global intersection $\bigcap_{i=1}^3 A_{\mathrm{safe}}(s_i)$ is empty. Conversely, allowing a single faulty sensor ($1 < k_a$) to rule out $s_{\mathrm{part}}$ would defeat the $f$-domain fault budget and expose the cluster to split-brain.

To govern actuation under both partial observability and bounded telemetry faults, PIEQ draws on discrete-event supervisory control~\cite{ramadge1987}, partially observable runtime safety shielding~\cite{spf2024,melcer2024shield}, and Feldbaum's dual control~\cite{feldbaum1960}. Rather than trusting a point diagnosis, \texttt{kf-proxy} applies complete-proof packing in reverse to refutation derivations, retaining every modeled fault world unless at least $f + 1$ physically disjoint telemetry domains refute it. A mutation is admitted only if it preserves the invariant across every surviving world simultaneously. Whenever no mutation is universally safe, the controller invokes the dual-control principle of trading immediate actuation for information gain, dispatching a read-only diagnostic probe to partition the surviving worlds or escalating deterministically to human operators.

\begin{definition}[$f$-Fault-Tolerant Plausible Belief Set]
\label{def:ft_belief}
For finite candidate universe $\mathcal{S}_{\mathrm{cand}}$, let $\mathcal{W}_{\mathrm{refute}}(s)$ be the active verified refutation DAGs incompatible with $s \in \mathcal{S}_{\mathrm{cand}}$ whose leaves are snapshot-consistent at $v_{\mathrm{snap}}$. For fault budget $f = k_a - 1$, define the exact and bounded-compute plausible belief sets as
\begin{align}
  \mathcal{B}_{\tau, v_{\mathrm{snap}}}^{(f)} &\triangleq \bigl\{ s \in \mathcal{S}_{\mathrm{cand}} : \nu_{\mathrm{pack}}(\mathcal{W}_{\mathrm{refute}}(s)) \le f \bigr\}, \label{eq:ft_belief_exact}\\
  \widehat{\mathcal{B}}_{\tau, v_{\mathrm{snap}}}^{(f)} &\triangleq \bigl\{ s \in \mathcal{S}_{\mathrm{cand}} : \mathsf{trunc}(s) \lor \hat{\nu}_{\mathrm{pack}}(\mathcal{W}_{\mathrm{refute}}(s)) \le f \bigr\}, \label{eq:ft_belief}
\end{align}
where $\mathsf{trunc}(s)$ holds if any ATMS or packing cap triggers on $s$, or if refutation leaves mix mismatched observation versions across mutating commits. By Theorem~\ref{thm:packing}(3), $\mathcal{B}_{\tau, v_{\mathrm{snap}}}^{(f)} \subseteq \widehat{\mathcal{B}}_{\tau, v_{\mathrm{snap}}}^{(f)}$ unconditionally.
\end{definition}

Retaining the true state inside this plausible belief set depends on the authored candidate universe covering the true snapshot state and refutation derivations resting on predicates that genuinely contradict the refuted world.

\begin{assumption}[Snapshot-Bound Candidate Universe and Safety Model]
\label{assump:world_model}
At consistent observation snapshot $v_{\mathrm{snap}}$ (all leaves in $\mathcal{W}_{\mathrm{refute}}(S_{v_{\mathrm{snap}}})$ satisfy $v_{\mathrm{obs}}(e_m) = v_{\mathrm{snap}}$, or span $[v_{\min}, v_{\max}]$ with no commit mutating observed predicates in $(v_{\min}, v_{\max}]$): (1)~$S_{v_{\mathrm{snap}}} \in \mathcal{S}_{\mathrm{cand}}$; (2)~every verified refutation DAG $\Pi \in \mathcal{W}_{\mathrm{refute}}(S_{v_{\mathrm{snap}}})$ cites at least one primitive leaf physically false in $S_{v_{\mathrm{snap}}}$ (automatically implied by Assumption~\ref{assump:sound_rules} whenever refutations are Horn derivations of false-in-$S_{v_{\mathrm{snap}}}$ predicates); and (3)~for every $a \in A_{\mathrm{safe}}^{\mathcal{K}}(S_{v_{\mathrm{snap}}})$, $\Post(S_{v_{\mathrm{snap}}}, a) \subseteq \mathcal{K}$.
\end{assumption}

\begin{lemma}[Fault-Tolerant Plausible-World Invariant Safety]
\label{lem:viability}
Under Assumptions~\ref{assump:domain_partition} and~\ref{assump:world_model}, suppose at most $f$ failure domains in $\mathcal{F}$ emit false telemetry assumptions at consistent snapshot $v_{\mathrm{snap}}$. Then, independently of positive proof packing $\nu_{\mathrm{pack}}(\mathcal{W}(a))$, two guarantees hold. First, (1)~for true-state retention via refutation packing, regardless of ATMS or packing truncation, $\nu_{\mathrm{pack}}(\mathcal{W}_{\mathrm{refute}}(S_{v_{\mathrm{snap}}})) \le f$ and hence $S_{v_{\mathrm{snap}}} \in \mathcal{B}_{\tau, v_{\mathrm{snap}}}^{(f)} \subseteq \widehat{\mathcal{B}}_{\tau, v_{\mathrm{snap}}}^{(f)}$. Second, (2)~for one-step invariant preservation, any mutation satisfying
\begin{equation}
  a \in \bigcap_{s \in \widehat{\mathcal{B}}_{\tau, v_{\mathrm{snap}}}^{(f)}} A_{\mathrm{safe}}^{\mathcal{K}}(s)
  \label{eq:global_intersection}
\end{equation}
guarantees $a \in A_{\mathrm{safe}}^{\mathcal{K}}(S_{v_{\mathrm{snap}}})$ and $\Post(S_{v_{\mathrm{snap}}}, a) \subseteq \mathcal{K}$, even if $\nu_{\mathrm{pack}}(\mathcal{W}(a)) = 0$. When the intersection over $\mathcal{A}_{\mathrm{mut}}$ is empty, \texttt{kf-proxy} dispatches a read-only partitioning diagnostic probe $a_{\mathrm{probe}}^\star \in \mathcal{A}_{\mathrm{probe}}$ maximizing distinct predicted outcomes across $\widehat{\mathcal{B}}_{\tau, v_{\mathrm{snap}}}^{(f)}$ (or any policy-weighted EVSI ranking over safe partitioning probes), or transitions deterministically to \texttt{DEFER\_ESCALATE}.
\end{lemma}

\begin{proof}
By Assumption~\ref{assump:world_model}(2), every $\Pi \in \mathcal{W}_{\mathrm{refute}}(S_{v_{\mathrm{snap}}})$ contains a false leaf in $S_{v_{\mathrm{snap}}}$. If $\nu_{\mathrm{pack}}(\mathcal{W}_{\mathrm{refute}}(S_{v_{\mathrm{snap}}})) \ge f + 1$, Assumption~\ref{assump:domain_partition} implies at least $f + 1$ distinct faulty failure domains, contradicting the fault budget $f$. Hence $S_{v_{\mathrm{snap}}} \in \mathcal{B}_{\tau, v_{\mathrm{snap}}}^{(f)} \subseteq \widehat{\mathcal{B}}_{\tau, v_{\mathrm{snap}}}^{(f)}$, and Eq.~\eqref{eq:global_intersection} with Assumption~\ref{assump:world_model}(3) immediately yields $a \in A_{\mathrm{safe}}^{\mathcal{K}}(S_{v_{\mathrm{snap}}})$ and $\Post(S_{v_{\mathrm{snap}}}, a) \subseteq \mathcal{K}$.
\end{proof}

Crucially, Lemma~\ref{lem:viability} never relies on an untrusted language model enumerating all candidate worlds or discovering every valid refutation. Because $\mathcal{S}_{\mathrm{cand}}$ is authored inside the TCB (Assumption~\ref{assump:world_model}(1)), omitted refutations only lower $\hat{\nu}_{\mathrm{pack}}(\mathcal{W}_{\mathrm{refute}}(s))$, retaining worlds conservatively rather than pruning $S_{v_{\mathrm{snap}}}$. Likewise, the optional bounded deterministic completion hook evaluated in Section~\ref{sec:live_reliability} only adds verifier-accepted derivations over already-authenticated leaves, so it preserves Lemma~\ref{lem:viability} whenever Assumptions~\ref{assump:sound_rules}--\ref{assump:world_model} hold while exposing any violation of Assumption~\ref{assump:sound_rules} directly.

Lemma~\ref{lem:viability} also clarifies an essential architectural separation between PIEQ's two epistemic gates. Positive proof packing ($\nu_{\mathrm{pack}}(\mathcal{W}(a)) \ge k_a$) is not logically required to preserve the modeled safety invariant $\mathcal{K}$, which refutation packing and global safe-action intersection secure on their own. Instead, positive packing supplies an independent obligation under Theorem~\ref{thm:packing}(1): certifying that the mutation's diagnostic precondition $c_a$ actually held under at most $f = k_a - 1$ faulty domains. Removing positive packing alone while keeping refutation packing inside $\widehat{\mathcal{B}}_{\tau, v_{\mathrm{snap}}}^{(f)}$ and Eq.~\eqref{eq:global_intersection} still preserves $\mathcal{K}$, yet loses precondition soundness, allowing a hallucinating agent or a single corrupted telemetry domain to trigger unwarranted false-precondition mutations ($\neg c_a$, such as unnecessary failovers, node drains, or load shedding) that happen to keep the plant inside $\mathcal{K}$. Conversely, removing or weakening refutation packing (for instance, counting raw refutations from a shared switch or using Menger path flow on Eq.~\eqref{eq:conj_counterexample} inside $\mathcal{W}_{\mathrm{refute}}(s)$) breaks true-state retention directly, allowing $f$ or fewer faulty domains to eject $S_{v_{\mathrm{snap}}}$ from the belief set and admit invariant-violating mutations such as split-brain. Both epistemic gates are therefore indispensable, yet both still evaluate a single frozen snapshot; the next section addresses what happens when multiple operations execute concurrently against a live physical target.

\section{Physically Refined Concurrency, Durability, and Risk Accounting}
\label{sec:execution}

The epistemic certification and plausible-world shield developed in Sections~\ref{sec:atms_packing} and~\ref{sec:shield} guarantee safety only at a single frozen snapshot with no competing writers. Production control planes rarely enjoy that stillness: multiple coordinators may act concurrently on overlapping resources, target state can advance between observation and actuation, authorization or policy approvals can be revoked mid-flight, and physical actuators often execute multi-phase operations rather than instantaneous transitions. To carry snapshot-level epistemic safety all the way to the physical commit boundary, this section draws on concurrency-control and distributed-systems theory, specifically linearizability, strict serializability, Abadi--Lamport stuttering refinement, escrow-style exposure accounting, leases, and fencing tokens, organizing them into closed-domain concurrent governance, verifiable physical refinement contracts, tier-classified target enforcement, and conditional risk accounting.

\subsection{Concurrent Hazards, Contiguous Snapshots, and Physical Refinement}
\label{sec:concurrent}

Because safety invariants often couple multiple infrastructure resources, concurrent governance must operate over the closed exposure domain set $\mathcal{D} = \Dom^\star(a) \triangleq \Close(\Dom(a), \mathcal{G}_{\mathrm{inv}})$, capturing every domain transitively linked to candidate mutation $a$ by shared invariants. Safe concurrent execution over $\mathcal{D}$ must close six distinct execution-layer hazards. Three arise from operation ordering and cancellation: checking only candidate $a$ after an in-flight operation $u$ misses reverse interleavings ($a \to u \to s_{\mathrm{bad}} \notin \mathcal{K}_{\mathcal{D}}$ even when $u(s_0)=s_1$ and $a(s_0)=a(s_1)=s_2 \in \mathcal{K}_{\mathcal{D}}$ yet $u(s_2)=s_{\mathrm{bad}}$); checking only terminal states overlooks unsafe predecessor prefixes masked by later recovery ($u_1(s_0) = s_{\mathrm{bad}} \notin \mathcal{K}_{\mathcal{D}}$ followed by $a(s_{\mathrm{bad}}) = s_{\mathrm{rec}} \in \mathcal{K}_{\mathcal{D}}$); and canceling or fencing $a$ (\texttt{fenced\_no\_effect}) can expose an unsafe absent-candidate schedule $(u_1, u_2)$ whenever $a \prec_{\mathrm{ord}} u_2$ was guarding $(u_1, u_2)$ only on runs where $a$ executed, because fencing $a$ still satisfies $\Pred(u_2) \subseteq \mathrm{Settled}_{\mathrm{target}}$.

Three additional hazards arise from state versioning, internal execution overlap, and concurrent coordinators. First, unversioned commits or hidden commit-version gaps break state projection: if snapshot $v_{\mathrm{snap}} = 0$ observes $x = 0$, an unversioned commit or an unacknowledged operation $u \in \mathcal{I}_U$ actually committed at $v = 1$ ($x \leftarrow 10$), and a confirmed commit $c \in \mathcal{I}_C$ committed at $v = 2$ ($x \leftarrow x + 1$), advancing only $c$ and then $u$ computes $\{1, 10\}$ and misses the true physical state $x = 11$. Second, overlapping read and write phases produce physical write-skew: starting from $(x, y) = (0, 0)$ under invariant $\neg(x=1 \land y=1)$, operation $a$ (read $y$; if $0$, set $x \leftarrow 1$) and operation $u$ (read $x$; if $0$, set $y \leftarrow 1$) are both sequentially safe ($a \to u \mapsto (1,0)$ and $u \to a \mapsto (0,1)$), yet physically overlapping phases $r_a \to r_u \to w_a \to w_u$ write $(1, 1) \notin \mathcal{K}_{\mathcal{D}}$, even in commuting double-commit variants. Third, without ledger serialization, two concurrent coordinators can independently validate against the same unresolved set $\mathcal{U}|_{\mathcal{D}}$ and mint mutually invalid stale admission certificates under either sequential or concurrent admission modes.

To neutralize version gaps and track in-flight exposure across $\mathcal{D}$, \texttt{kf-proxy} maintains an observation watermark $[v_{\min}, v_{\max}]$, a commit history $\mathcal{H}_C|_{\mathcal{D}}$ carrying commit versions $v_{\mathrm{commit}}(c) \in \mathbb{N} \cup \{\bot\}$, pairwise-disjoint exposure sets $\mathcal{I}_C(d), \mathcal{I}_H(d), \mathcal{I}_U(d)$ for each $d \in \mathcal{D}$ (with exposure totals $C(d), H(d), U(d)$ and unresolved set $\mathcal{U}_\tau|_{\mathcal{D}} \triangleq \bigcup_{d \in \mathcal{D}}(\mathcal{I}_H(d) \cup \mathcal{I}_U(d))$), and monotonic per-domain admission revisions $\mathbf{r}_{\mathrm{adm}}(\mathcal{D}) = \{(d, r_{\mathrm{adm}}(d))\}_{d \in \mathcal{D}}$. Let $\widehat{\Post}(s, o) \triangleq \Post(s, o) \cup \Post_{\mathrm{part}}(s, o)$ include declared partial effects, $\mathcal{T}_{\mathrm{det}}(S, c) \triangleq \bigcup_{s \in S} \Post(s, c)$, and $\mathcal{T}_{\mathrm{opt}}(S, c) \triangleq S \cup \mathcal{T}_{\mathrm{det}}(S, c)$, failing closed whenever any transition reaches an unmodeled world identifier outside $\mathcal{S}_{\mathrm{cand}}$.

Hidden version gaps are eliminated completely by enforcing a contiguous commit-version barrier. Let $\Delta_C^{\bot}(\mathcal{D}) \triangleq \{c \in \mathcal{H}_C|_{\mathcal{D}} : v_{\mathrm{commit}}(c) = \bot\}$ and $v_{\mathrm{head}} \triangleq \max(\{v_{\max}\} \cup \{v_{\mathrm{commit}}(c) : c \in \mathcal{H}_C|_{\mathcal{D}} \setminus \Delta_C^{\bot}(\mathcal{D})\})$. Whenever $\Delta_C^{\bot}(\mathcal{D}) \neq \varnothing$ or any integer version $v \in (v_{\min}, v_{\mathrm{head}}]$ is missing from $\{v_{\mathrm{commit}}(c) : c \in \mathcal{H}_C|_{\mathcal{D}}\}$, \texttt{kf-proxy} blocks admission on $\mathcal{D}$ until target write-ahead-log reconciliation fills the missing commit history or confirms \texttt{fenced\_no\_effect}. Once the commit sequence is contiguous, commits at or below $v_{\min}$ are skipped, commits inside the observation window $\Delta_C^{\mathrm{win}} = \{c : v_{\min} < v_{\mathrm{commit}}(c) \le v_{\max}\}$ advance optionally via $\mathcal{T}_{\mathrm{opt}}$, and strictly post-snapshot commits $\Delta_C^{> v_{\max}}$ advance deterministically via $\mathcal{T}_{\mathrm{det}}$ in ascending $v_{\mathrm{commit}}$ order:
\begin{equation}
  \mathcal{S}_\tau^{\mathrm{base}} \triangleq
  \mathcal{T}_{\mathrm{det}}^{\Delta_C^{> v_{\max}}}\!\Bigl(
    \mathcal{T}_{\mathrm{opt}}^{\Delta_C^{\mathrm{win}}}\!\bigl(\widehat{\mathcal{B}}_{\tau_{\mathrm{obs}}, [v_{\min}, v_{\max}]}^{(f)}\bigr)
  \Bigr).
  \label{eq:base_snapshot_advance}
\end{equation}

Checking abstract schedules from $\mathcal{S}_\tau^{\mathrm{base}}$ guarantees physical safety only when real target executions correspond faithfully to those abstract transitions. Let $\theta = (\sigma_0 \xrightarrow{e_1} \sigma_1 \cdots \xrightarrow{e_m} \sigma_m)$ be a physical execution trace over operations $O$ on $\mathcal{D}$. Drawing on Abadi--Lamport stuttering refinement~\cite{abadi1991refinement}, we require an explicit physical refinement contract $\mathsf{Refines}_{\mathcal{D}}(\mathcal{S}_\tau^{\mathrm{base}}, O, \prec_{\mathrm{ord}})$: every physically reachable step $\sigma_k \xrightarrow{e_{k+1}} \sigma_{k+1}$ starting from $\alpha(\sigma_0) \in \mathcal{S}_\tau^{\mathrm{base}}$ either stutters ($\alpha(\sigma_{k+1}) = \alpha(\sigma_k)$) or matches an enabled abstract step $o$ along an admissible schedule $\pi \in \Lin(O', \prec_{\mathrm{ord}}|_{O'})$ with $\alpha(\sigma_{k+1}) \in \widehat{\Post}(\alpha(\sigma_k), o)$.

PIEQ operationalizes $\mathsf{Refines}_{\mathcal{D}}$ through two attested execution regimes and a fail-closed default. Under Regime~R1 (\code{SERIALIZABLE\_ATOMIC}), the target must provide strict serializability (linearizability)~\cite{herlihy1990linearizability} at the modeled operation effect boundary rather than weak serializability or Snapshot Isolation~\cite{berenson1995ansi}: weak serializability ignores real-time precedence and can serialize post-fence or post-$\Pred(a)$ writes before an already-completed predecessor, whereas Snapshot Isolation permits write-skew ($r_a \to r_u \to w_a \to w_u \to (1,1)$). Qualifying for R1 requires a signed target capability profile bound into $h_{\mathrm{man}}$ attesting a single-boundary strictly serializable commit primitive (such as SQLite \texttt{BEGIN IMMEDIATE}, Raft/\texttt{etcd} multi-key CAS across all keys encoding $\mathcal{K}_{\mathcal{D}}$, or hardware/storage generation CAS) combined with complete writer mediation.

Under Regime~R2 (\code{PHASE\_DECOMPOSED}), where physical operations overlap internally, the verified transition relation must explicitly model six physical phenomena: internal read, decision, and write phases ordered by $\prec_{\mathrm{intra}}$; per-operation local state and cached read values embedded inside $\mathcal{S}$ (for instance, $(x, y, y_a, x_u)$); asynchronous background callbacks or webhooks represented as trailing steps remaining enabled even after client return or abort; partial effects ($\Post_{\mathrm{part}}$) and non-cancellable intra-operation prefixes that remain executable after control-plane cancellation or lease expiry; crash and recovery continuations reachable from intermediate prefixes; and retries or duplicate physical effects, either proven to be stuttering no-ops via target-native per-phase idempotency or explicitly represented as bounded retry phases. Absent attested R1 or R2 qualification, PIEQ falls back to the fail-closed default (\code{UNVERIFIED\_ASYNC}), rejecting concurrent Mode~C admission outright and rejecting Mode~S whenever two or more unordered predecessors could overlap, admitting only totally ordered non-overlapping Mode~S execution.

An important fidelity boundary governs these contracts: exhaustive exploration of a finite abstract phase graph verifies only the abstract model and never proves that a physical implementation conforms to it. Operational fault injection (Section~\ref{sec:eval}) can falsify a refinement claim and build evidence for simple single-store transactional adapters, whereas distributed asynchronous adapters require formal simulation or refinement verification (such as TLA+ or mechanized proofs). Wherever physical refinement has not been formally discharged, $\mathsf{Refines}_{\mathcal{D}}$ remains an explicit trusted target contract accounted for in stratum $E_8$ of Section~\ref{sec:risk_bound}.

\begin{definition}[Inductive Schedule Invariant, Mode~S, and Mode~C]
\label{def:concurrency_semantics}
Fix closed domain set $\mathcal{D} = \Dom^\star(a)$, invariant $\mathcal{K}_{\mathcal{D}}$, and target partial order $\prec_{\mathrm{ord}}$ encoded by $\Pred(o)$ (together with $\prec_{\mathrm{intra}}$ under R2). First, (1)~for the inductive schedule invariant, a schedule $\pi = (o_1, \dots, o_\ell)$ is sequentially safe for $\mathcal{K}_{\mathcal{D}}$ from $s_0$ if $s_0 \in \mathcal{K}_{\mathcal{D}}$ and for every $j \in \{1,\dots,\ell\}$ and every reachable $s_{j-1}$, $s_{j-1} \in \mathcal{K}_{\mathcal{D}}$, $o_j \in A_{\mathrm{safe}}^{\mathcal{K}_{\mathcal{D}}}(s_{j-1})$, and $\widehat{\Post}(s_{j-1}, o_j) \subseteq \mathcal{K}_{\mathcal{D}} \cap \mathcal{S}_{\mathrm{cand}}$. The predicate $\mathsf{Inv}_{\mathrm{sched}}(\mathcal{S}_0, \mathcal{U}, \prec_{\mathrm{ord}}; \mathcal{K}_{\mathcal{D}})$ holds if and only if every admissible subset $O \subseteq \mathcal{U}$ (allowing absent or \texttt{fenced\_no\_effect} operations, and intra-op prefix-closed under R2) and every $\pi \in \Lin(O, \prec_{\mathrm{ord}}|_O)$ is sequentially safe from every $s_0 \in \mathcal{S}_0$; let $\mathcal{R}_{\prec_{\mathrm{ord}}}(\mathcal{S}_0, \mathcal{U})$ denote all states reachable along such prefixes. Second, (2)~for exposure conservation and universal revision CAS, for every $d \in \mathcal{D}$, the sets $\mathcal{I}_C(d), \mathcal{I}_H(d), \mathcal{I}_U(d)$ remain pairwise disjoint with $C(d) + H(d) + U(d) \le B(d)$. Every admission under either Mode~S or Mode~C binds $\mathbf{r}_{\mathrm{adm}}(\mathcal{D})$ and commits atomically via linearizable CAS requiring $r_{\mathrm{adm}}^{\mathrm{cur}}(d) = r_{\mathrm{adm}}^{\mathrm{cert}}(d)$ for all $d \in \mathcal{D}$, while every reservation mutation in $L_{\mathrm{exp}}$ (admission, crash retention, timeout, commit, release, or reconciliation) increments $r_{\mathrm{adm}}(d)$ so stale concurrent coordinators must revalidate. Third, (3)~for the predecessor frontier, whenever $u \prec_{\mathrm{ord}} o$, $u.\mathrm{op\_id} \in \Pred(o)$, enforced atomically at the target as $\Pred(o) \subseteq \mathrm{Settled}_{\mathrm{target}}$ (durably \texttt{COMMITTED} or \texttt{FENCED\_NO\_EFFECT}). Fourth, (4)~for bounded Mode~C certification ($\mathsf{CompatCert}$), partition $\mathcal{U}_\tau|_{\mathcal{D}} = \mathcal{U}_S \sqcup \mathcal{U}_C$ with $\mathcal{U}_S \prec_{\mathrm{ord}} (\mathcal{U}_C \cup \{a\})$. Mode~C requires attested R1 ($|\mathcal{U}_C| \le K_{\mathrm{conc}} = 4$) or R2 ($|\mathcal{P}| \le K_{\mathrm{phase}} = 6$ phase steps) and bounded unordered $|\mathcal{U}_S| \le K_{\mathrm{conc}}$ (or totally ordered $\mathcal{U}_S$). Verifier $\mathcal{V}_{\mathrm{compat}}$ verifies $\mathsf{Inv}_{\mathrm{sched}}(\mathcal{S}_\tau^{\mathrm{base}}, \mathcal{U}_S, \prec_{\mathrm{ord}}; \mathcal{K}_{\mathcal{D}})$ yielding $\mathcal{S}_\tau^{\mathrm{pre}}$, exhaustively verifies every admissible subset $O'$ (both containing and excluding $a$, intra-op prefix-closed under R2) and every $\pi \in \Lin(O', \prec_{\mathrm{ord}}|_{O'})$ across $\widehat{\Post}$, and seals the regime and $\mathbf{r}_{\mathrm{adm}}(\mathcal{D})$ with a SHA-256 digest. (Our implementation additionally checks pairwise commutativity on unordered R1 pairs as a conservative convergence policy, though commutativity is not logically required for invariant preservation.) Outside this bounded class, Mode~C fails closed to Mode~S or \texttt{DENY}.
\end{definition}

Lifting abstract schedule safety to physical states requires mediation, refinement, durability, and isolation from disjoint domains.

\begin{assumption}[Concurrent Mediation, Refinement, Durability, and Frame]
\label{assump:concurrent}
For closed domain set $\mathcal{D} = \Dom^\star(a)$ and invariant $\mathcal{K}_{\mathcal{D}}$: (1)~all writers touching $\mathcal{D}$ mediate through linearizable $L_{\mathrm{exp}}$ with identity-checked issuance deduplication, universal $\mathbf{r}_{\mathrm{adm}}(\mathcal{D})$ CAS, and atomic multi-domain reservation across $\mathcal{D}$ (or global serialization); (2)~$0 \le \Exp_{\mathrm{act}}(u, d) \le \Exp(u, d)$ and physical execution satisfies $\mathsf{Refines}_{\mathcal{D}}$ under its attested regime (R1, R2, or non-overlapping totally ordered Mode~S); (3)~there exists $v_{\mathrm{snap}} \in [v_{\min}, v_{\max}]$ with $S_{v_{\mathrm{snap}}} \in \widehat{\mathcal{B}}_{\tau_{\mathrm{obs}}, [v_{\min}, v_{\max}]}^{(f)}$; (4)~timed-out, lost-ack, or cancellation-raced operations transfer $\mathcal{I}_H(d) \to \mathcal{I}_U(d)$ and release pre-effect only upon a verified \texttt{fenced\_no\_effect} barrier; (5)~the target durably persists epochs, $\Pred(o)$, bound at-most-once $\mathrm{op\_id}$ identities, and exposure caps across crashes; and (6)~by the closed-domain frame condition, any operation $v$ with $\Dom^\star(v) \cap \mathcal{D} = \varnothing$ leaves the physical and modeled projection governing $\mathcal{K}_{\mathcal{D}}^{\mathrm{phys}}, \mathcal{K}_{\mathcal{D}}, A_{\mathrm{safe}}^{\mathcal{K}_{\mathcal{D}}},$ and $\widehat{\Post}|_{\mathcal{D}}$ invariant.
\end{assumption}

\begin{theorem}[Closed-Domain Physical Invariant Preservation]
\label{thm:concurrent}
Let $\mathcal{D} = \Dom^\star(a)$ and $\mathcal{U}_\tau|_{\mathcal{D}} = \bigcup_{d \in \mathcal{D}}(\mathcal{I}_H(d) \cup \mathcal{I}_U(d))$ at revision vector $\mathbf{r}_{\mathrm{adm}}(\mathcal{D})$. Under Assumptions~\ref{assump:domain_partition}--\ref{assump:concurrent} with at most $f$ faulty telemetry failure domains, suppose \texttt{kf-proxy} admits $a$ only when the contiguous commit-version barrier passes (Eq.~\eqref{eq:base_snapshot_advance}), $L_{\mathrm{exp}}$ atomically commits the $\mathbf{r}_{\mathrm{adm}}(\mathcal{D})$ CAS and exposure check $C(d)+H(d)+U(d)+\Exp(a,d) \le B(d)$ across all $d \in \mathcal{D}$, and either Mode~S holds (every $u \in \mathcal{U}_\tau|_{\mathcal{D}}$ lies in $\Pred(a)$, $\mathcal{U}_\tau|_{\mathcal{D}}$ is totally ordered under \code{UNVERIFIED\_ASYNC}, $\mathsf{Inv}_{\mathrm{sched}}(\mathcal{S}_\tau^{\mathrm{base}}, \mathcal{U}_\tau|_{\mathcal{D}}, \prec_{\mathrm{ord}}; \mathcal{K}_{\mathcal{D}})$ holds, and for all $s' \in \mathcal{R}_{\prec_{\mathrm{ord}}}(\mathcal{S}_\tau^{\mathrm{base}}, \mathcal{U}_\tau|_{\mathcal{D}})$, $a \in A_{\mathrm{safe}}^{\mathcal{K}_{\mathcal{D}}}(s')$ and $\widehat{\Post}(s', a) \subseteq \mathcal{K}_{\mathcal{D}} \cap \mathcal{S}_{\mathrm{cand}}$) or Mode~C holds via a verified $\mathsf{CompatCert}(a, \mathcal{U}_C, \mathbf{r}_{\mathrm{adm}}(\mathcal{D}))$ from $\mathcal{V}_{\mathrm{compat}}$ under attested R1 or R2. Then, independently of positive proof packing: (i)~$\mathsf{Inv}_{\mathrm{sched}}(\mathcal{S}_\tau^{\mathrm{base}}, \mathcal{U}_\tau|_{\mathcal{D}} \cup \{a\}, \prec_{\mathrm{ord}}; \mathcal{K}_{\mathcal{D}})$ is preserved at the admission linearization point; (ii)~across every physically admissible trace $\theta = (\sigma_0 \to \cdots \to \sigma_m)$ of $\mathcal{U}_\tau|_{\mathcal{D}} \cup \{a\}$ on $\mathcal{D}$, every physical state satisfies $\alpha(\sigma_k) \in \mathcal{K}_{\mathcal{D}}$ and $\sigma_k \in \mathcal{K}_{\mathcal{D}}^{\mathrm{phys}}$; and (iii)~$\Exp_{\mathrm{real}}(d, t) \le C(d) + H(d) + U(d) \le B(d)$ for all $d \in \mathcal{D}$.
\end{theorem}

\begin{proof}
We establish exposure conservation, snapshot containment, abstract schedule safety, and physical refinement in sequence. For exposure conservation and revision CAS, induction on the linearizable transactions of $L_{\mathrm{exp}}$ (Assumption~\ref{assump:concurrent}(1)) shows that atomic CAS on $\mathbf{r}_{\mathrm{adm}}(\mathcal{D})$ rejects any concurrent Mode~S or Mode~C certificate evaluated against an earlier ledger state, while rollbacks, crash retentions into $\mathcal{I}_U$, deduplicated retries, timeouts ($\mathcal{I}_H \to \mathcal{I}_U$), commits ($\Exp_{\mathrm{act}} \le \Exp$), and verified \texttt{fenced\_no\_effect} releases preserve $\Exp_{\mathrm{real}}(d,t) \le C(d)+H(d)+U(d) \le B(d)$. For contiguous snapshot containment, Lemma~\ref{lem:viability}(1) and Assumption~\ref{assump:concurrent}(3) guarantee some $v_{\mathrm{snap}} \in [v_{\min}, v_{\max}]$ with $S_{v_{\mathrm{snap}}} \in \widehat{\mathcal{B}}_{\tau_{\mathrm{obs}}, [v_{\min}, v_{\max}]}^{(f)}$, and Assumption~\ref{assump:concurrent}(6) ensures only writes touching $\mathcal{D}$ alter $\mathcal{K}_{\mathcal{D}}$. Because the contiguous-version barrier confirms that every integer version in $(v_{\min}, v_{\mathrm{head}}]$ appears in $\mathcal{H}_C|_{\mathcal{D}}$ without gaps, no unacknowledged operation in $\mathcal{I}_H \cup \mathcal{I}_U$ committed inside $(v_{\min}, v_{\mathrm{head}}]$; every unacknowledged operation either already committed at or before $v_{\min}$ (captured by subsets excluding it) or takes effect after $v_{\mathrm{head}}$. Applying $\mathcal{T}_{\mathrm{opt}}^{\Delta_C^{\mathrm{win}}}$ followed by $\mathcal{T}_{\mathrm{det}}^{\Delta_C^{> v_{\max}}}$ in ascending version order therefore yields $\alpha(\sigma_0) \in \mathcal{S}_\tau^{\mathrm{base}}$. For abstract schedule safety, Assumption~\ref{assump:concurrent}(5)--(6) and $\Pred(o) \subseteq \mathrm{Settled}_{\mathrm{target}}$ ensure that every executed subset $O'$ runs along some $\pi \in \Lin(O', \prec_{\mathrm{ord}}|_{O'})$. Mode~S verifies all prefixes in $\mathcal{U}_\tau|_{\mathcal{D}}$ (covering $a \notin O'$) and $a$ from every $s' \in \mathcal{R}_{\prec_{\mathrm{ord}}}$ (covering $a \in O'$), while Mode~C exhaustively checks $\mathcal{U}_S$ and every admissible subset $O_C$ (both containing and excluding $a$, prefix-closed under R2) and every linear extension across $\widehat{\Post}$, preserving $\mathsf{Inv}_{\mathrm{sched}}$. Finally, by $\mathsf{Refines}_{\mathcal{D}}$ (Assumption~\ref{assump:concurrent}(2)), every physical state $\sigma_k$ along $\theta$ maps via $\alpha(\sigma_k)$ to a state reachable along a verified $\pi$ under $\widehat{\Post}$, yielding $\alpha(\sigma_k) \in \mathcal{K}_{\mathcal{D}}$ and $\sigma_k \in \mathcal{K}_{\mathcal{D}}^{\mathrm{phys}}$.
\end{proof}

When global safety decomposes as $\mathcal{K} = \bigcap_{\mathcal{D}} \mathcal{K}_{\mathcal{D}}$ with every invariant coupling closed inside $\mathcal{G}_{\mathrm{inv}}$, Theorem~\ref{thm:concurrent} composes by intersection to establish global physical invariant preservation (though not automatic global controlled viability $\Viab(\mathcal{K})$). Before stating the end-to-end compositional guarantee, however, two execution-semantics subtleties must be separated cleanly: how diagnostic preconditions behave across time, and when authorization revocations take physical effect across target capability tiers.

The first distinction separates snapshot truth, $\mathsf{SnapshotPreconditionValid}(a, v_{\mathrm{snap}}) \triangleq (S_{v_{\mathrm{snap}}} \models c_a)$, from actual commit-time truth, $\mathsf{CommitPreconditionValid}(a, t_{\mathrm{eff}}) \triangleq (\sigma(t_{\mathrm{eff}}) \models c_a)$. Theorem~\ref{thm:packing}(1) certifies snapshot truth, which is the exact semantics needed for historical-snapshot triggers such as revoking a compromised IAM role or certificate authority attested at $v_{\mathrm{snap}}$. For transient-state actions that require a sound system guarantee of commit-time truth, two mechanisms are individually sufficient: either (S1) interval persistence $\mathsf{Persist}(c_a, [v_{\mathrm{snap}}, t_{\mathrm{eff}}])$ paired with snapshot validity, or (S2) target-side atomic revalidation $\mathsf{AtomicReval}_a(t_{\mathrm{tx}})$ under Tier~E2, inspecting the authoritative physical state encoding $c_a$ ($\mathsf{Reval}_a(\sigma) \Rightarrow \sigma \models c_a$, never a cached mirror) inside the exact atomic transaction boundary $t_{\mathrm{tx}} = t_{\mathrm{eff}}$ as the mutation under complete writer mediation. Treating the disjunction (S1)~$\lor$~(S2) as necessary (a biconditional) is false: a transient predicate transitioning $\mathsf{true} \to \mathsf{false} \to \mathsf{true}$ over $[v_{\mathrm{snap}}, t_{\mathrm{eff}}]$ is physically true at commit even without persistence or atomic revalidation. Conversely, revalidating a predicate outside the mutation's atomic transaction admits a check-then-act time-of-check to time-of-use (TOCTOU) race after the condition clears. Furthermore, leaving revalidation to caller or proposer discretion enables silent downgrades to snapshot-only checks (demonstrated in Section~\ref{sec:live_heldout}). PIEQ therefore assigns every mutation type a trusted temporal policy $\mathsf{TPol}(a) \in \{\textsc{Snapshot-Triggered}, \textsc{Commit-Time-Sensitive}\}$ owned strictly by TCB action metadata, never by an untrusted agent proposal. In the hardened profile, unclassified mutations fail closed, \textsc{Commit-Time-Sensitive} actions mandate atomic Tier~E2 revalidation (S2) or attested persistence (S1) with trusted commit predicates injected regardless of caller omissions, the canonical policy identity is signed into the certificate and re-checked inside the atomic effect boundary so superseded policies reject, and \textsc{Snapshot-Triggered} actions avoid unnecessary revalidation and stay actionable after transient symptoms subside.

The second distinction separates Tier~E2 atomic freshness from Tier~E1+-quiescence fence-relative exclusion. Under Tier~E2 with complete writer mediation, authorization checks and physical state mutation serialize atomically at $t_{\mathrm{tx}}$, leaving zero revocation race window ($\delta_{\mathrm{fresh}}^{\mathrm{E2}} = 0$). Under Tier~E1+-quiescence, a minimal six-step trace proves that strict execution-time freshness fails: at $t=1$, certificate $C$ is minted with $\tau_{\mathrm{auth}}=5$; at $t=2$, target $r$ verifies $\tau_{\mathrm{auth}} \ge 5$, acquires $\mathsf{lock}_r$, and enters \code{prepared\_slot}; at $t=3$ ($t_{\mathrm{rev\_init}}$), the control plane initiates revocation $\tau_{\mathrm{auth}}^{\mathrm{cp}} \leftarrow 6$ and blocks on $\mathsf{lock}_r$; at $t=4$ ($t_{\mathrm{eff}}$), the already-prepared operation commits physically after revocation initiation; at $t=5$ ($t_{\mathrm{fenced}}$), $\mathsf{lock}_r$ drains, flushes $\tau_{\mathrm{auth}}^{\min}[r] \leftarrow 6$, and returns $\mathsf{FenceAck}$; and at $t=6$, subsequent $\tau_{\mathrm{auth}}=5$ operations reject deterministically. Tier~E1+-quiescence therefore guarantees fence-relative post-$\mathsf{FenceAck}$ exclusion, not zero-window revocation freshness.

\begin{corollary}[Tier-Exact Compositional Execution-Bound Admission]
\label{cor:composition}
Under Assumptions~\ref{assump:sound_rules}--\ref{assump:concurrent} with at most $f = k_a - 1$ faulty telemetry failure domains, suppose \texttt{kf-proxy} admits mutation $a$ only when: (1)~$\mathsf{EvidenceValid}$ holds via $\hat{\nu}_{\mathrm{pack}}(\mathcal{W}(a)) \ge k_a$ over consistent-snapshot DAGs; (2)~$\mathsf{SafetyValid}_{\mathcal{D}}$ satisfies the hypotheses of Theorem~\ref{thm:concurrent} on $\mathcal{D} = \Dom^\star(a)$; and (3)~$\mathsf{TargetEnforced}_T$ holds via signed certificate validation at target tier $T \in \{\mathrm{E2}, \mathrm{E1+q}\}$ enforcing $\tau_{\mathrm{adm}} \ge v_{\mathrm{fence}}$, $\tau_{\mathrm{auth}} \ge \tau_{\mathrm{auth}}^{\min}$, $t_{\mathrm{now}} \le t_{\mathrm{auth\_until}}$, and $\Pred(a) \subseteq \mathrm{Settled}_{\mathrm{target}}$ under complete writer mediation and $\mathsf{Refines}_{\mathcal{D}}$. Then four tier-exact properties follow: (a)~$\mathsf{SnapshotPreconditionValid}(a, v_{\mathrm{snap}})$ holds by Theorem~\ref{thm:packing}(1), and if either $\mathsf{Persist}(c_a, [v_{\mathrm{snap}}, t_{\mathrm{eff}}])$ holds or $T=\mathrm{E2}$ executes authoritative target-side atomic precondition revalidation $\mathsf{AtomicReval}_a(t_{\mathrm{tx}})$ sharing the physical mutation's atomic commit boundary, then the system possesses a sound guarantee of $\mathsf{CommitPreconditionValid}(a, t_{\mathrm{eff}})$ (hence $\sigma(t_{\mathrm{eff}}) \models c_a$, though the converse fails in general); (b)~every intermediate and terminal physical state satisfies $\sigma_k \in \mathcal{K}_{\mathcal{D}}^{\mathrm{phys}}$ with $\Exp_{\mathrm{real}}(d, t) \le B(d)$ by Lemma~\ref{lem:viability} and Theorem~\ref{thm:concurrent}; (c-E2)~for $T=\mathrm{E2}$, authorization and evidence epochs are fresh atomically at the physical linearization point $t_{\mathrm{tx}}$ ($\delta_{\mathrm{fresh}}^{\mathrm{E2}} = 0$); and (c-E1+)~for $T=\mathrm{E1+q}$, zero commits with revoked authorization execute after completed durable $\mathsf{FenceAck}$ ($t \ge t_{\mathrm{fenced}}$), whereas already-prepared operations holding $\mathsf{lock}_r$ may commit during the pre-$\mathsf{FenceAck}$ window $(t_{\mathrm{rev\_init}}, t_{\mathrm{fenced}})$ ($\delta_{\mathrm{fresh}}^{\mathrm{E1+q}} > 0$ relative to $t_{\mathrm{rev\_init}}$).
\end{corollary}

\subsection{Certificate Binding, Durability, and Capability Tiers}
\label{sec:fencing}

Each admission certificate cryptographically binds the twelve-tuple $(\mathrm{op\_id},\allowbreak \mathrm{res},\allowbreak \mathcal{D},\allowbreak v_{\mathrm{snap}},\allowbreak h_{\mathrm{man}},\allowbreak h_{\mathrm{proof}},\allowbreak \mathbf{r}_{\mathrm{adm}}(\mathcal{D}),\allowbreak \tau_{\mathrm{adm}},\allowbreak \tau_{\mathrm{auth}},\allowbreak t_{\mathrm{auth\_until}},\allowbreak \Pred(a),\allowbreak \sigma_{\mathrm{cp}})$. Because policy approval can be rescinded during an incident freeze or RBAC rotation even when physical telemetry remains valid, the authorization epoch $\tau_{\mathrm{auth}}$ and validity deadline $t_{\mathrm{auth\_until}}$ evolve independently of the evidence epoch $\tau_{\mathrm{adm}}$. If $t_{\mathrm{auth\_until}}$ is set and no trusted target clock is available, target validation fails closed; meanwhile, manifest rotation ($h_{\mathrm{man}}$) or late conflict nogoods advance $\tau_{\mathrm{adm}}$ atomically to fence in-flight certificates.

Handling non-cancellable external operations under crash and partition faults requires durable fence barriers rather than relying on control-plane lease expiry alone. Under Tier~\texttt{E1+-quiescence}, letting a control-plane lease expire does not stop an external RPC that has already started: if target $r$ initiates a non-cancellable call at $t = 9$ under lease deadline $t_{\mathrm{lease}} = 10$ and stalls or partitions, the call can still complete physically at $t = 11$ after lease expiry. Fence completion $t_{\mathrm{fenced}}(v)$ may therefore be declared only when target $r$ acquires $\mathsf{lock}_r$, drains active execution, and durably flushes $v_{\mathrm{fence}}[r] \ge v$ and $\tau_{\mathrm{auth}}^{\min}[r]$ via WAL/\texttt{fsync} (such as SQLite \texttt{PRAGMA synchronous=FULL} or a Raft quorum commit) before returning a signed $\mathsf{FenceAck}$, or when the external target itself enforces $t_{\mathrm{eff}}^{\mathrm{loc}} \le t_{\mathrm{lease}} - \Delta_{\mathrm{skew}}$ at commit while $t_{\mathrm{cp}} > t_{\mathrm{lease}} + \Delta_{\mathrm{skew}}$ (still retaining the operation in $\mathcal{I}_U$ until reconciled). Otherwise, \texttt{kf-proxy} keeps the operation in $\mathcal{I}_U(d)$ and blocks conflicting admissions indefinitely, prioritizing safety over liveness, until $\mathsf{FenceAck}$, reconciliation, or a hard external barrier such as STONITH or IAM revocation completes.

A related durability boundary separates local WAL deduplication from true target-native idempotency. Persisting $\mathrm{op\_id}$ bound to canonical operation identity in a local or control-plane WAL guarantees at-most-once execution only when the deduplication record and the physical state change commit in the same atomic transaction, as under Tier~E2 when the transactional store is itself the physical effect boundary. For a remote actuator across a network boundary, preventing duplicate execution after a lost acknowledgment requires genuine target-native idempotency, such as a remote idempotency key or generation CAS persisted atomically with the remote state transition. When a nontransactional remote actuator lacks target-native idempotency and its response is lost, retrying is unsafe: the control plane must never auto-retry and must hold the operation as \texttt{UNKNOWN} in $\mathcal{I}_U$ until out-of-band state reconciliation establishes whether it committed. Likewise, recording \texttt{FENCED\_NO\_EFFECT} must atomically fail if $\mathrm{op\_id}$ already committed, and post-restart recovery reconciles $\mathcal{I}_H \cup \mathcal{I}_U$ against durable target \texttt{COMMITTED} and \texttt{FENCED\_NO\_EFFECT} journals while retaining unconfirmed operations in $\mathcal{I}_U$. Table~\ref{tab:tiers} classifies target capability tiers E0--E2 so PIEQ never claims a stronger guarantee than the underlying target can physically enforce.

\begin{table*}[!t]
\centering
\caption{Target enforcement capability tiers (E0--E2) at the physical effect boundary (tier-exact authorization freshness, optional commit-time precondition revalidation, durability, and lost-response retry semantics).}
\label{tab:tiers}
\vspace{0.4mm}
\scriptsize
\setlength{\tabcolsep}{2.6pt}
\renewcommand{\arraystretch}{1.14}
\begin{tabularx}{\textwidth}{@{}>{\bfseries}L{19mm} L{29mm} L{29mm} L{22mm} L{28mm} Y@{}}
\toprule
\textbf{Tier} & \textbf{Guaranteed} & \textbf{Not Guaranteed} & \textbf{Linearization} & \textbf{Required Durable State} & \textbf{Failure, Lost-Ack \& Residual Risk} \\
\midrule
E0\newline(Proxy-Only) & Pre-dispatch check at \texttt{kf-proxy}. & Post-admission evidence/auth revocation or dedup. & $t_{\mathrm{adm}}$ only (none at target). & None at target. & Delays/revocations leak commits; lost ack on nontransactional op cannot safely retry; $\delta_{\mathrm{fresh}}^{\mathrm{E0}} > 0$. \\
\midrule
E1\newline(Async Push) & Rejects stale/revoked certs arriving after local fence. & Exclusion when cert overtakes delayed fence. & Local $v_{\mathrm{fence}}[r], \tau_{\mathrm{auth}}[r]$. & Durable async epoch + auth registers. & Delayed fence leaks commits; $\delta_{\mathrm{fresh}}^{\mathrm{E1}} > 0$. \\
\midrule
E1+-read\newline(Lock+ReadIdx) & Rejects certs revoked prior to sync \texttt{ReadIndex} $t_{\mathrm{read}}$. & Exclusion if revocation commits in $(t_{\mathrm{read}}, t_{\mathrm{eff}})$. & Split: $t_{\mathrm{read}} \neq t_{\mathrm{eff}}$. & $\mathsf{lock}_r$ + sync \texttt{ReadIndex} + durable $\mathrm{op\_id}$ log. & Revocation in $(t_{\mathrm{read}}, t_{\mathrm{eff}})$ leaks commit; $\delta_{\mathrm{fresh}}^{\mathrm{E1+-read}} > 0$. \\
\midrule
E1+-quiescence\newline(2-Phase Barrier) & Zero post-$\mathsf{FenceAck}$ commits ($t \ge t_{\mathrm{fenced}}$); blocks conflicts in $\mathcal{I}_U$ on stall. & Zero-window auth freshness in $(t_{\mathrm{rev\_init}}, t_{\mathrm{fenced}})$; CP lease expiry alone does not halt non-cancellable RPC. & $t_{\mathrm{fenced}}(v)$ on durable $\mathsf{FenceAck}$ (or target lease). & $\mathsf{lock}_r$ + \texttt{fsync}'d epochs, \texttt{RegisterPrepared}, $\mathrm{op\_id}$ log + $\mathsf{FenceBarrier}$. & Already-prepared op may commit pre-$\mathsf{FenceAck}$ ($\delta_{\mathrm{fresh}}^{\mathrm{E1+q}} > 0$ vs.\ $t_{\mathrm{rev\_init}}$; $0$ post-$\mathsf{FenceAck}$); stall holds $\mathcal{I}_U$ indefinitely until reconciled. \\
\midrule
E2\newline(Shared / Native Atomic) & Atomic $\sigma_{\mathrm{cp}}$, $h_{\mathrm{man}}$, $\tau_{\mathrm{adm}}$, $\tau_{\mathrm{auth}}$, $t_{\mathrm{auth}}$, $\Pred$, bound $\mathrm{op\_id}$, cap, and opt-in $\mathsf{CommitPreconditionValid}$ at $t_{\mathrm{tx}}$ (if store \emph{is} effect boundary). & Unmediated out-of-band external writers or remote actuators outside the atomic transaction. & Single indivisible tx / target-native CAS at $t_{\mathrm{tx}}$. & Durable atomic store (SQLite WAL \texttt{synchronous=FULL} / \texttt{etcd}) or target-native CAS. & Deterministically aborts stale/revoked certs at $t_{\mathrm{tx}}$ (survives crash/restart); idempotent retry safe on identical $\mathrm{op\_id}$; $\delta_{\mathrm{fresh}}^{\mathrm{E2}} = 0$ conditional on full mediation. \\
\bottomrule
\end{tabularx}
\end{table*}

\subsection{Conditional Eight-Risk Stratum Accounting}
\label{sec:risk_bound}

Because every formal guarantee in PIEQ is conditional on trusted domain models, accurate failure-domain topology, bounded sensor faults, and target refinement, operators need a clean accounting identity that tracks residual failure modes without double-counting. For an admitted mutation $a$ on closed domain set $\mathcal{D} = \Dom^\star(a)$ across $D = |\mathcal{F}|$ telemetry failure domains, we define eight ordered measurable failure events spanning modeling contracts, physical telemetry faults, and execution-boundary breaks. The first three events capture TCB modeling and topology errors: $E_1 \equiv E_{\mathrm{univ}}$ occurs when the candidate universe omits the true state ($S_{v_{\mathrm{snap}}} \notin \mathcal{S}_{\mathrm{cand}}$) or observations span a torn snapshot; $E_2 \equiv E_{\mathrm{rule}}$ occurs when authored rules, conflict invariants, safe-action sets, or post-state transitions in $(\mathcal{M}_{\mathrm{sys}}, P_{\mathrm{sys}}, A_{\mathrm{safe}}, \widehat{\Post})$ are unsound; and $E_3 \equiv E_{\mathrm{dom}}$ occurs when $h_{\mathrm{man}}$ misclassifies a collector's failure domain. The next two events capture physical telemetry faults: $E_4 \equiv E_{\mathrm{common}}$ denotes the occurrence of an exogenous cross-domain common-cause fault mechanism (such as a shared power, DNS, kernel, or firmware defect coupling two or more domains, defined strictly as a causal mechanism rather than the outcome $\{\sum_\ell Z_\ell \ge 2\}$ so that conditioning on $E_4^c$ does not induce negative dependence), while $E_5 \equiv E_{\mathrm{sensor}} \triangleq \{\sum_{\ell=1}^{D} Z_\ell \ge k_a\}$ captures $k_a$ or more domains emitting false assumptions, where $Z_\ell \in \{0,1\}$ indicates a false assumption in domain $F_\ell$. The final three events capture execution-layer failures: $E_6 \equiv E_{\mathrm{drift}}$ occurs upon unobserved plant drift, an unmediated external writer, or failure of both sufficient commit-time precondition mechanisms (S1 $\mathsf{Persist}$ and S2 authoritative Tier~E2 $\mathsf{AtomicReval}_a$) for a declared commit-time-truth action; $E_7 \equiv E_{\mathrm{fresh}}^T$ captures target enforcement, evidence-epoch, or tier-$T$ authorization-freshness failure (including pre-$\mathsf{FenceAck}$ commits in $(t_{\mathrm{rev\_init}}, t_{\mathrm{fenced}})$ under $T=\mathrm{E1+q}$ when evaluated against strict revocation freshness, whereas $\delta_7 = \delta_{\mathrm{fresh}}^{\mathrm{E2}} = 0$ under completely mediated $T=\mathrm{E2}$); and $E_8 \equiv E_{\mathrm{conc}}$ captures concurrent coordination failures across commit-history gaps, exposure envelopes, $\Pred$, $\mathsf{Refines}_{\mathcal{D}}$, $\mathbf{r}_{\mathrm{adm}}$ CAS, the frame condition, or $\mathsf{CompatCert}$.

\begin{proposition}[Tier-Indexed Conditional Eight-Risk Accounting]
\label{prop:risk}
Define disjoint strata $G_i \triangleq E_i \cap \bigcap_{j=1}^{i-1} E_j^c$ ($i=1,\dots,8$) and conditional risks $\delta_i \triangleq \mathbb{P}(E_i \mid \bigcap_{j=1}^{i-1} E_j^c)$, setting $\delta_i = 0$ whenever the conditioning event has probability zero. Assume only that conditioned on $\bigcap_{j=1}^4 E_j^c$, the $D$ domain-fault indicators $\{Z_\ell\}_{\ell=1}^D$ are mutually independent with $\mathbb{P}(Z_\ell = 1 \mid \bigcap_{j=1}^4 E_j^c) \le \bar{p}_{\mathrm{dom}}$. Let $\mathrm{BadCommit}_T(a)$ denote any physical execution violating: (i)~$\mathcal{K}_{\mathcal{D}}^{\mathrm{phys}}$ or exposure cap $B(d)$; (ii)~$\mathsf{SnapshotPreconditionValid}(a, v_{\mathrm{snap}})$ (and $\mathsf{CommitPreconditionValid}(a, t_{\mathrm{eff}})$ only when $a$ is declared a commit-time-truth action); or (iii)~tier-$T$'s authorization guarantee, namely execution-time freshness at $t_{\mathrm{tx}}$ for $T=\mathrm{E2}$ or post-$\mathsf{FenceAck}$ exclusion for $T=\mathrm{E1+q}$ (with pre-$\mathsf{FenceAck}$ revocation races accounted in $E_7^{\mathrm{E1+q}}$ if counted against strict freshness). Then by Corollary~\ref{cor:composition}, (1)~$\mathrm{BadCommit}_T(a) \subseteq \bigsqcup_{i=1}^8 G_i$ and $\mathbb{P}(\mathrm{BadCommit}_T(a) \mid \bigcap_{i=1}^8 E_i^c) = 0$; and (2)
\begin{align}
  \mathbb{P}\!\left(\mathrm{BadCommit}_T(a)\right)
  &\le 1 - \prod_{i=1}^{8} (1 - \delta_i) \le \sum_{i \neq 5} \delta_i \nonumber\\
  &\quad + \sum_{j=k_a}^{D} \binom{D}{j} \bar{p}_{\mathrm{dom}}^{\,j} (1 - \bar{p}_{\mathrm{dom}})^{D-j}.
  \label{eq:risk_union}
\end{align}
\end{proposition}

\begin{proof}
On $\bigcap_{i=1}^8 E_i^c$, Assumptions~\ref{assump:sound_rules}--\ref{assump:concurrent} hold, at most $f = k_a - 1$ domains fail, at least one sufficient condition (S1 or S2) holds for any declared commit-time-truth action (which by Corollary~\ref{cor:composition}(a) implies $\mathsf{CommitPreconditionValid}(a, t_{\mathrm{eff}})$), and tier-$T$ enforcement holds; hence $\mathrm{BadCommit}_T(a)$ is excluded. Disjointness gives $\mathbb{P}(\bigsqcup_{i=1}^8 G_i) = \sum_{i=1}^8 \delta_i \prod_{j=1}^{i-1}(1-\delta_j) = 1 - \prod_{i=1}^8(1-\delta_i) \le \sum_{i=1}^8 \delta_i$. Conditioning $G_5$ on $\bigcap_{j=1}^4 E_j^c$ bounds $\delta_5$ by the binomial tail via stochastic dominance.
\end{proof}

Proposition~\ref{prop:risk} is a conditional accounting identity rather than a universal empirical bound: $\delta_1, \delta_4, \delta_6,$ and $\bar{p}_{\mathrm{dom}}$ remain environment-dependent inputs. Automated regression tests verifying our execution-semantics counterexamples (the Tier~E1+-quiescence pre-$\mathsf{FenceAck}$ race, transient dip-and-return non-necessity, and non-atomic TOCTOU revalidation) are documented in Appendix~\ref{app:repro}.

\subsection{End-to-End Worked Example: Authoring and Execution Walkthrough}
\label{sec:worked_example}

To make the interaction among TCB authoring, Q1 packing, Q2 shielding, and Q3 execution concrete, we trace a compact cluster-governance scenario combining database failover with bounded capacity reservation.

\textbf{1.\ Authored TCB Model.} The operator specifies: (i)~candidate universe $\mathcal{S}_{\mathrm{cand}} = \{s_{\mathrm{dead}}, s_{\mathrm{part}}, s_{\mathrm{ok}}\}$ with safe-action map $A_{\mathrm{safe}}^{\mathcal{K}}(s_{\mathrm{dead}}) = \{\code{promote}, \code{probe}, \code{defer}\}$, $A_{\mathrm{safe}}^{\mathcal{K}}(s_{\mathrm{part}}) = \{\code{reroute}, \code{probe}, \code{defer}\}$, and $A_{\mathrm{safe}}^{\mathcal{K}}(s_{\mathrm{ok}}) = \{\code{reserve}, \code{probe}, \code{defer}\}$ (for a capacity pool capped at $B{=}10$); (ii)~signed topology manifest $h_{\mathrm{man}}$ assigning in-band SQL ($O_1 \mapsto F_1$), out-of-band BMC/REST ($O_2 \mapsto F_2$), and \texttt{etcd} lease ($O_3 \mapsto F_3$) to three disjoint telemetry failure domains ($k_a{=}3$, $f{=}2$); (iii)~Horn rules $\mathcal{M}_{\mathrm{sys}}$ mapping authenticated observations to action preconditions (\code{act:promote}, \code{act:reserve}) and world refutations (\code{refute:}$s_{\mathrm{part}}$, \code{refute:}$s_{\mathrm{ok}}$), plus conflict invariant $P_{\mathrm{sys}}$ flagging $\{\code{sql\_down}, \code{sql\_write\_ok}\} \vdash \bot$; and (iv)~trusted temporal policies $\mathsf{TPol}(\code{promote}) = \textsc{Snapshot-Triggered}$ and $\mathsf{TPol}(\code{reserve}) = \textsc{Commit-Time-Sensitive}(\code{pred:headroom})$.

\textbf{2.\ Q1 Positive Evidence Packing and Q2 Plausible-World Shielding.} Suppose the primary stops answering SQL queries at snapshot $v_{\mathrm{snap}}{=}50$. Initially only $e_1 \in F_1$ (\code{sql\_hb\_absent}) arrives. Even if five agents independently propose \code{promote} citing $e_1$ under distinct context IDs, recursive physical footprints give $\Phi_{\mathrm{phys}}(\Pi_i) = \{F_1, O_1, e_1\}$ for all $i$, capping $\hat{\nu}_{\mathrm{pack}} = 1 < 3$. Simultaneously, both $s_{\mathrm{dead}}$ and $s_{\mathrm{part}}$ survive in $\widehat{\mathcal{B}}_{\tau,50}^{(2)}$ because neither has $f{+}1{=}3$ disjoint refutations, and $A_{\mathrm{safe}}^{\mathcal{K}}(s_{\mathrm{dead}}) \cap A_{\mathrm{safe}}^{\mathcal{K}}(s_{\mathrm{part}}) = \{\code{probe}, \code{defer}\}$ contains no mutation. \texttt{kf-proxy} therefore blocks promotion and dispatches read-only partitioning probes to $F_2$ (OOB power/REST) and $F_3$ (\texttt{etcd} lease). When both probes confirm the leader is down at the same consistent snapshot $v_{\mathrm{snap}}{=}50$ and manifest $h_{\mathrm{man}}$, the agent submits three verified Horn DAGs spanning $\{F_1, F_2, F_3\}$ for \code{act:promote} ($\hat{\nu}_{\mathrm{pack}}{=}3 \ge 3$) and three disjoint refutations each for $s_{\mathrm{part}}$ and $s_{\mathrm{ok}}$, collapsing $\widehat{\mathcal{B}}_{\tau,50}^{(2)}$ to $\{s_{\mathrm{dead}}\}$ and admitting \code{promote}.

\textbf{3.\ Q3 Mode~C Compatibility and Tier~E2 Atomic Commit.} On the healthy cluster ($s_{\mathrm{ok}}$), suppose two concurrent coordinators $C_1, C_2$ each observe admission revision $r_{\mathrm{adm}}{=}1$ and propose $+4$ capacity reservations against current usage $3$ and cap $B{=}10$ ($3{+}4{+}4{=}11 > 10$). Under attested Regime~R1 (\code{SERIALIZABLE\_ATOMIC}), \texttt{kf-proxy} runs $\mathcal{V}_{\mathrm{compat}}$ over closed domain $\mathcal{D}$, verifies all subsets and linear extensions (including absent/canceled schedules), seals the certificate with $(r_{\mathrm{adm}}{=}1, \tau_{\mathrm{auth}}{=}5, \mathsf{TPol}(\code{reserve}))$, and dispatches to the Tier~E2 transactional boundary. Inside a single \texttt{SERIALIZABLE} transaction, the target recomputes the HMAC signature, locks the control row (\texttt{SELECT \dots FOR UPDATE}), checks $\tau_{\mathrm{auth}} \ge \tau_{\mathrm{auth}}^{\min}$, verifies $r_{\mathrm{adm}}^{\mathrm{cur}} = 1$, re-checks the authoritative \code{pred:headroom} row, checks $\mathrm{SUM} + 4 \le 10$, inserts the reservation, and increments $r_{\mathrm{adm}} \leftarrow 2$. Coordinator $C_1$ commits ($3{+}4{=}7 \le 10$), while $C_2$ deterministically aborts on stale $r_{\mathrm{adm}}$ (or serialization conflict) and must revalidate against $r_{\mathrm{adm}}{=}2$—where a fresh $+4$ attempt is rejected on the exposure cap, preserving $\mathrm{SUM} \le 10$ across every physical trace.

\section{Evaluation}
\label{sec:eval}

We evaluate PIEQ end-to-end using the implementation, benchmarks, traces, and test runners published in \code{research/pieq/} of the OpenKedge repository (\url{https://github.com/openkedge/openkedge}; exact commit hash and artifact paths appear in Appendix~\ref{app:repro}). Our evaluation examines six core questions: whether physical-domain evidence certification prevents false-precondition admission; whether plausible-world shielding prevents unsafe mutation under bounded faulty telemetry; whether active probing and bounded deterministic completion restore actionable remediation without weakening soundness; whether the execution boundary blocks stale, unauthorized, or conflicting commits; whether PIEQ's guarantees translate to physical PostgreSQL/Patroni failover supervision and real-PostgreSQL transactions; and whether adding language-model agents improves reliability once telemetry visibility and the execution gate are held constant.

To avoid conflating deterministic software checks with live-model behavior or physical-cluster trials, we separate our evidence strictly across six tiers and never pool them into a single headline rate. These tiers comprise the conditional formal guarantees established in Sections~\ref{sec:model}--\ref{sec:execution}; automated continuous-integration regression and fault suites spanning $249$ passing \code{pytest} tests in GitHub Actions CI (including $14$ container-backed \texttt{etcd} tests with the new Tier~E1+-quiescence barrier test), $140$ distributed three-member \texttt{etcd} fault trials in CI, $200$ baseline POSIX/SQLite WAL fault trials, $80$ multi-process OS-level Section~5 concurrency and revocation fault trials, and containerized PostgreSQL~16 Tier~E2 concurrent-coordinator and mid-flight revocation races (Appendix~\ref{app:repro}); frozen software benchmarks run with deterministic proposers against on-disk Tier~E2 SQLite WAL alongside our new synthesis-vs.-completion and scalability stress suites; audited offline counterfactual replays of our initial five-model pilot; live language-model evaluations across single-run and repeated studies; and physical PostgreSQL/Patroni failover supervision paired with real-PostgreSQL Tier~E2 transactional enforcement. Throughout every tier, within-assumption and out-of-assumption outcomes are reported separately, i.i.d.\ confidence intervals over fixed deterministic fixtures are omitted, and local verifier latency is separated from end-to-end recovery time. To keep empirical tables scannable, Tables~\ref{tab:agent_and_factorial}--\ref{tab:rev11_rep} use standardized outcome columns: \emph{Mitig.}\ (actionable safe mitigations), \emph{False Blk}\ (actionable incidents unnecessarily blocked), \emph{Mand.\ Defer}\ (mandatory deferrals correctly blocked), \emph{Within Unsafe}\ (within-assumption invariant violations), \emph{False Precond.}\ (invariant-safe but diagnostically unjustified mutations), \emph{UBSA}\ (epistemically admitted enforcement-deferral cases caught only by the execution gate), and \emph{OOA Unsafe}\ (intentional out-of-assumption controls admitted outside Proposition~\ref{prop:risk}'s trust boundary). Three methodological boundaries apply across the study: every benchmark is author-labeled rather than externally annotated, earlier CLI-tier live runs invoke unpinned locally authenticated snapshots under uncontrolled sampling (supplemented in Section~\ref{sec:live_reliability} by a controlled-sampling direct-API three-family confirmation run with complete billing over Bedrock inference-profile aliases and local SGLang Qwen3-8B), and the physical testbed containers share a single VM kernel.

\subsection{Supporting Engineering Mechanisms}
\label{sec:enabling_mech}

Turning strict deterministic verification into a practical control plane for untrusted language models requires bridging syntax, ambiguity, and proof-omission friction without expanding the TCB. PIEQ pairs its unchanged verifier with three supporting engineering mechanisms.

First, a typed proof-submission interface exposes a finite verifier-known vocabulary covering permissible actions, candidate fault worlds, authored rules $\mathcal{M}_{\mathrm{sys}}$, and authorized evidence handles mapped to primitive propositions. Proposers emit decoupled action proposals alongside typed \code{(rule\_id, evidence\_handles)} claims, while any proposer-supplied TCB metadata such as signatures, failure domains, bitmasks, epochs, or manifest hashes is rejected fail-closed. The TCB resolves handles deterministically and runs bounded schema-feedback retry (at most two retries) without leaking cross-agent observations.

Second, a bounded read-only active probing loop operationalizes dual control whenever the plausible-world shield blocks immediate actuation. Candidate probes are filtered fail-closed on verified read-only safety contracts and cost or load caps, ranked by worst-case plausible-world reduction and domain novelty (or calibrated EVSI), and audited across rounds along seven snapshot-consistency boundaries enforcing Assumptions~\ref{assump:sound_rules}--\ref{assump:world_model}: plant version drift between passive telemetry and a probe, concurrent target commits during probe execution, stale observation handles barred from cross-version conjunctive rules, collectors observing inconsistent snapshots that must never be packed into one quorum, probe retries returning a different state epoch, mid-probe manifest rotation or failure-domain mismatch, and previously verified proofs invalidated by newly observed $P_{\mathrm{sys}}$ conflicts (triggering automatic nogood registration and exclusion of contested leaves, alongside same-domain non-inflation). Whenever snapshot consistency cannot be verified across any boundary, PIEQ fails closed by rejecting positive certificates and retaining all candidate worlds.

Third, an opt-in bounded deterministic completion helper, implemented as a default-off $\sim$70-line TCB forward-chaining routine, completes verifier-accepted derivations strictly over already-authenticated observations and $\mathcal{M}_{\mathrm{sys}}$ under fail-closed resource caps. Section~\ref{sec:live_reliability} measures both the liveness it recovers and the explicit out-of-assumption risk trade-off it introduces.

\subsection{Pilot Forensics and Counterfactual Replay}
\label{sec:r4_forensics}

Before building the full evaluation suite, we diagnosed why our initial 40-case five-model pilot ($800$ calls) paralyzed remediation. Although that pilot achieved $0/40$ unsafe authorizations (compared to $3/40$ under debate and $5/40$ over raw telemetry channels), it achieved $0/19$ actionable mitigations ($19/19$ false blocks) and executed $0/549$ requested probes. Classifying all $140/284$ rejected Round-0 proofs (detailed in the companion supplement) revealed that $95/140$ ($67.9\%$) were completely verifier-valid derivations discarded solely because the prototype coupled proof acceptance to final action votes whenever cautious agents voted to probe or defer; the remaining rejections stemmed from $33/140$ ($23.6\%$) bitmask or antecedent mapping mistakes and $12/140$ ($8.6\%$) schema errors.

Replaying those frozen Round-0 pilot replies under controlled mechanism toggles (Table~\ref{tab:r4_audit}) exposed two methodological lessons. Decoupling proof acceptance from action votes (\code{R4-B}) restored valid proofs from $144/284$ to $239/284$ and raised mitigations from $0/19$ to $5/19$ at $0/40$ unsafe admissions, whereas dropping positive packing (\code{R4-D}) immediately admitted two uncertified-positive mutations (Cases~72 and 74). In \code{R4-E}, all $43$ probes re-queried passive observations already present in the fixture where pilot agents had omitted valid Horn rules ($4/14$ rescued cases missed refutation rules only, $2/14$ missed positive rules only, and $8/14$ missed both); because that prototype loop forward-chained $\mathcal{M}_{\mathrm{sys}}$ deterministically without live models and built its probe menu from realized fixture observations, it indirectly leaked the realized support set. Auditing the seventh temporal boundary also uncovered a missed conflict nogood on reprobed leaves in Cases~1, 39, and 71, after which repairing that conflict check ran $46$ probes ($30$ effective) while preserving $19/19$ mitigations and $0/40$ unsafe admissions.

Constructing the probe menu strictly from the public incident-class schema (\code{R4-E-strict}, returning absent markers for unobserved predicates) eliminated menu leakage while retaining $19/19$ mitigations and $0/40$ unsafe admissions, yet increased probe volume $2.5\times$ from $43$ to $108$ ($28$ effective). This confirmed that \code{R4-E}'s lower probe count had benefited from menu leakage.

\begin{table}[!t]
\centering
\caption{Audited offline counterfactual replays on frozen pilot Round-0 replies ($40$ cases: $19$ actionable, $21$ defer; zero TCB weakening).}
\label{tab:r4_audit}
\vspace{0.3mm}
\scriptsize
\setlength{\tabcolsep}{2.4pt}
\renewcommand{\arraystretch}{1.10}
\begin{tabularx}{\columnwidth}{@{}L{29mm} r r r r Y@{}}
\toprule
\textbf{Replay} & \textbf{Mit.(19)} & \textbf{Def.(21)} & \textbf{Uns.} & \textbf{Prb.(Eff)} & \textbf{Audited Provenance Label} \\
\midrule
\code{R4-A} (Archived) & 0/19 & 21/21 & 0 & 0 (0) & Live pilot (strict coupling; 144/284) \\
\code{R4-B} (Decoupled) & 5/19 & 21/21 & 0 & 0 (0) & Offline reply replay (239/284) \\
\code{R4-D} (No Pos.\ Pack) & 7/19 & 21/21 & 0 & 0 (0) & Offline replay (2 uncert.\ pos.) \\
\code{R4-E} (Baseline) & 19/19 & 21/21 & 0 & 43 (33)\textsuperscript{*} & Fixture-backed + TCB rule compl. \\
\code{R4-E-strict} & 19/19 & 21/21 & 0 & 108 (28) & Leakage-free public-schema menu \\
\bottomrule
\multicolumn{6}{@{}p{\columnwidth}@{}}{\scriptsize \textsuperscript{*}After repairing the \code{P\_sys} nogood check: 46 (30) probes (Cases~1, 39, 71); same 19/19, 0 unsafe.}
\end{tabularx}
\end{table}

\subsection{Held-Out Software Benchmark, Ablations, and Confirmation}
\label{sec:heldout_fair}

After freezing implementation hashes, we authored a disjoint 36-case held-out benchmark (\code{H01}--\code{H36}) with zero overlap against the original 40 pilot cases across four new infrastructure domains: \code{bgp\_route\_leak}, \code{storage\_quorum\_fence}, \code{k8s\_control\_plane\_drain}, and \code{tls\_ca\_rotation}. The suite spans nine failure archetypes of four cases each (passive recoverable, probe-recoverable, unresolvable uncertainty, correlated or misclassified telemetry, valid-evidence unjustified mutations, insufficient execution authorization, concurrent or stale certificates, transient preconditions with mid-probe drift, and trusted-model-defect controls), pairing each incident with a leakage-free domain-level probe catalog and sealed ground truth in four classes: $9$ actionable safe mitigations, $5$ safe-but-diagnostically-unjustified mutations, $17$ mandatory deferrals, and $5$ out-of-assumption negative controls (\code{H16}, \code{H33}--\code{H36}). All software runs in this subsection use deterministic proposers paired with real on-disk Tier~E2 SQLite WAL (\texttt{PRAGMA synchronous=FULL}) and zero retuning.

Evaluating our initial frozen implementation (prior to adding cross-round probe snapshot auditing) exposed three sharp behavioral boundaries. First, on \code{H31} (a mid-probe plant version jump from $v=50$ to $v=51$), the initial code committed $1/31$ within-assumption unsafe invariant violation because probing did not audit cross-round snapshot consistency; enabling the seven-boundary temporal audit with zero retuning flipped \code{H31} to a snapshot-inconsistency deferral, achieving $0/31$ within-assumption unsafe admissions, $9/9$ mitigations ($4$ via leakage-free probes \code{H05}--\code{H08} and $1$ historical-snapshot CA revocation \code{H32}), and $17/17$ mandatory deferrals. Second, on \code{H30} (a transient I/O spike clearing before commit under caller-discretion revalidation), PIEQ blocked all $4/4$ passive false-precondition cases (\code{H17}--\code{H20}) yet committed one stale-precondition mutation with zero invariant violations, directly motivating the mandatory temporal-policy registry evaluated live in Section~\ref{sec:live_heldout}. Third, on all $5/5$ out-of-assumption controls (\code{H16} in $E_3$, \code{H33} in $E_1$, \code{H34} in $E_2$, \code{H35} in $E_5$, and \code{H36} in $E_6$), PIEQ admitted unsafe mutations exactly as modeled outside Proposition~\ref{prop:risk}'s trusted boundary.

Because PIEQ's fault threshold counts physical failure domains rather than agent count, we next ran a causally fair agent-count isolation study holding the verifier, vocabulary, probe budget, and Tier~E2 SQLite WAL constant while varying agent count and telemetry visibility (Table~\ref{tab:agent_and_factorial}, top). This exposed an earlier shared-context confound: \code{1A\_LocalTel} ($0/9$ passive on \code{heldout36}, $0/19$ on the 40-case suite) had received only one-fifth of telemetry, and even with full telemetry handles, \code{1A\_FullTel\_SharedCtx} stayed at $0/9$ passive because folding a single proposer context ID into proof packing capped $\nu_{\mathrm{pack}} \le 1$. Defining $\nu_{\mathrm{pack}}$ solely over physical footprints $\Phi_{\mathrm{phys}}$ (Eq.~\eqref{eq:footprint}) removed that artifact. With full telemetry and physical-footprint packing, \code{1A\_FullTel\_DomainIsolated} matched five-agent \code{5A\_PartTel} and \code{5A\_FullTel} identically in both safety and liveness across \code{heldout36} ($5/9$ passive, $9/9$ with probes, $0/31$ unsafe, $17/17$ deferrals) and \code{frozen40} ($19/19$ mitigations, $0/40$ unsafe), while issuing $6\times$ fewer proposer calls ($36$ vs.\ $216$ and $40$ vs.\ $240$). By contrast, matched unshielded single-agent and 3-of-5 debate baselines given identical telemetry and Tier~E2 fencing both incurred $6/31$ unsafe invariant violations and $5/5$ false-precondition admissions. Section~\ref{sec:live_heldout} tests whether stochastic live language models diverge from this deterministic equivalence under repeated sampling.

To isolate the causal role of each epistemic mechanism under identical telemetry and Tier~E2 SQLite WAL, Table~\ref{tab:agent_and_factorial} (bottom; full 17-row matrices in the companion supplement) summarizes a controlled mechanism-isolation factorial. Disabling positive packing alone left invariant violations at $0/31$ yet raised false-precondition admissions from $1$ to $5/5$ by admitting all four passive traps (\code{H17}--\code{H20}), confirming that physical-domain positive certification governs precondition soundness. Disabling fault-tolerant refutation packing ($f=0$) let a single faulty domain eject the true world, collapsing passive mitigations from $5/9$ to $1/9$ on \code{heldout36} and causing $6/40$ unsafe admissions ($0/19$ mitigations) on \code{frozen40}; disabling both packings produced $4/31$ unsafe admissions and $5/5$ false-precondition admissions on \code{heldout36}. Removing the plausible-world shield (\code{NO\_SHIELD}) admitted the \code{H09} Helly overlap trap ($1/31$ unsafe on \code{heldout36}; $7/40$ on \code{frozen40}), confirming the necessity of global safe-action intersection. On the liveness side, typed decoupling lifted passive mitigations from $0/9$ to $5/9$ ($0/19 \to 14/19$ on \code{frozen40}), and adding active probing with the seven-boundary temporal audit lifted mitigations further from $5/9$ to $9/9$ ($14/19 \to 19/19$ on \code{frozen40}) with zero within-assumption unsafe admissions.

Because \code{H31} prompted adding the seven-boundary temporal audit, re-evaluating \code{heldout36} serves only as regression verification rather than fresh out-of-sample testing. After freezing the updated implementation hashes, we therefore authored a disjoint 12-case software confirmation set in a new fifth domain (\code{dns\_anycast\_cluster}) targeting cross-round snapshot drift, mixed snapshots, mid-probe commits, manifest rotation, same-domain handles, late probe-derived nogoods, stale retries, concurrent ledger CAS races, two actionable controls, and two temporal-policy checks. First execution achieved $9/9$ mandatory deferrals, $3/3$ mitigations, $0$ unsafe admissions, and $0$ false blocks.

\begin{table*}[!t]
\centering
\caption{Frozen 36-case held-out benchmark (\code{H01}--\code{H36}: $31$ within-assumption [$9$ actionable, $17$ mandatory-deferral, $5$ false-precondition] + $5$ out-of-assumption controls; deterministic software proposers, Tier~E2 SQLite WAL \texttt{synchronous=FULL}). Top: causally fair agent-count isolation. Bottom: key cells of the $2^4$ mechanism-isolation factorial (T=Typed, P=PosPack, R=RefutePack, B=Probes; full 17-row matrices in the companion supplement).}
\label{tab:agent_and_factorial}
\vspace{0.3mm}
\scriptsize
\setlength{\tabcolsep}{3.5pt}
\renewcommand{\arraystretch}{1.10}
\begin{tabularx}{\textwidth}{@{}Y r r r r r r c r@{}}
\toprule
\textbf{Configuration (Held-Out 36: 31 Within + 5 OOA)} & \textbf{Mitig.\,(9)} & \textbf{False Blk} & \textbf{Mand.\,Def.\,(17)} & \textbf{Within Uns.} & \textbf{False Pre.\,(5)} & \textbf{OOA Uns.\,(5)} & \textbf{Probes (Eff)} & \textbf{Calls} \\
\midrule
\multicolumn{9}{@{}l}{\textit{Agent-Count Isolation (Leak-Free Probes, Tier~E2 SQLite WAL; Initial Frozen Baseline shown for \code{5A\_PartTel})}} \\
\code{5A\_PartTel} (Initial frozen baseline) & 9/9 & 0 & 16/17 & 1\rlap{\textsuperscript{a}} & 1\rlap{\textsuperscript{b}} & 5/5 & 17 (7) & 216 \\
\code{1A\_LocalTel} + Shield (1/5 telemetry) & 8/9 & 1 & 17/17 & 0 & 0 & 0/5 & 31 (16) & 36 \\
\code{1A\_FullTel\_SharedCtx} (shared ctx cap) & 8/9 & 1 & 17/17 & 0 & 0 & 0/5 & 27 (12) & 36 \\
\code{1A\_FullTel\_DomainIsolated} (Fair 1A) & 9/9 & 0 & 17/17 & 0 & 1\rlap{\textsuperscript{b}} & 5/5 & 17 (6) & 36 \\
\code{5A\_PartTel} (Full PIEQ, temporal audit) & 9/9 & 0 & 17/17 & 0 & 1\rlap{\textsuperscript{b}} & 5/5 & 17 (6) & 216 \\
\code{5A\_FullTel} (5 Agents, Full Telemetry) & 9/9 & 0 & 17/17 & 0 & 1\rlap{\textsuperscript{b}} & 5/5 & 17 (6) & 216 \\
\code{1A\_FullTel\_Unshielded} (+ Matched E2) & 9/9 & 0 & 11/17 & 6 & 5 & 5/5 & 0 (0) & 36 \\
\code{5A\_FullTel\_Debate} (3/5 Maj.\ + Matched E2) & 9/9 & 0 & 11/17 & 6 & 5 & 5/5 & 0 (0) & 180 \\
\midrule
\multicolumn{9}{@{}l}{\textit{Mechanism-Isolation Factorial (\code{5A\_PartTel}, Temporal Audit, Matched Tier~E2 SQLite WAL)}} \\
Strict Action Coupling (\code{T0\_P1\_R1\_B0}) & 0/9 & 9 & 17/17 & 0 & 0 & 0/5 & 0 (0) & 180 \\
Typed, No Probes (\code{T1\_P1\_R1\_B0}) & 5/9 & 4 & 17/17 & 0 & 1\rlap{\textsuperscript{b}} & 5/5 & 0 (0) & 216 \\
Drop Positive Packing (\code{T1\_P0\_R1\_B0}) & 5/9 & 4 & 17/17 & 0 & 5\rlap{\textsuperscript{c}} & 5/5 & 0 (0) & 216 \\
Drop Refutation Packing, $f{=}0$ (\code{T1\_P1\_R0\_B0}) & 1/9 & 8 & 17/17 & 0\rlap{\textsuperscript{d}} & 1\rlap{\textsuperscript{b}} & 5/5 & 0 (0) & 216 \\
Drop Both Packings (\code{T1\_P0\_R0\_B0}) & 5/9 & 4 & 13/17 & 4 & 5 & 5/5 & 0 (0) & 216 \\
No Plausible-World Shield (\code{T1\_P1\_NO\_SHIELD\_B0}) & 5/9 & 4 & 16/17 & 1\rlap{\textsuperscript{e}} & 1\rlap{\textsuperscript{b}} & 5/5 & 0 (0) & 216 \\
\textbf{Full PIEQ + Probes (\code{T1\_P1\_R1\_B1})} & \textbf{9/9} & \textbf{0} & \textbf{17/17} & \textbf{0} & 1\rlap{\textsuperscript{b}} & 5/5 & 17 (6) & 216 \\
\bottomrule
\multicolumn{9}{@{}p{\textwidth}@{}}{\scriptsize \textsuperscript{a}\code{H31} cross-snapshot bug caught prior to seven-boundary temporal audit ($v{=}50/51$). \textsuperscript{b}\code{H30} un-revalidated transient drop under caller-discretion mode ($0/4$ passive false preconditions on \code{H17}--\code{H20}); blocked by mandatory registry (Sec.~\ref{sec:live_heldout}). \textsuperscript{c}Admits all 4/4 passive false preconditions (\code{H17}--\code{H20}) + \code{H30}. \textsuperscript{d}On \code{frozen40}, \code{T1\_P1\_R0\_B0} incurs 6/40 unsafe ($0/19$ mitig.). \textsuperscript{e}Admits \code{H09} Helly trap ($7/40$ unsafe on \code{frozen40}). OOA = out-of-assumption controls (\code{H16}, \code{H33}--\code{H36}).}
\end{tabularx}
\end{table*}

\subsection{Verification Funnel, Latency, and Evidence Attribution}
\label{sec:costs_attribution}

Tracking proposals through the verification pipeline clarifies where deterministic checks filter malformed submissions and how little latency the TCB adds relative to external I/O and inference. On \code{heldout36} under \code{5A\_PartTel}, $216$ total proposals ($180$ first-attempt plus $36$ retries, with zero retry exhaustions) yielded $144/180$ ($80.0\%$) first-attempt validity. First-attempt derivations progressed from $277$ submitted to $241$ ($87.0\%$) schema-valid and $205$ ($74.0\%$) Horn- and provenance-valid, reaching $205/205$ accepted certificates after retry. Cases reaching quorum $k_a$ rose from $21$ before probing to $25$ after probing, with $4$ admitted via probes, $12$ rejected for insufficient evidence or snapshot drift, and $21$ deferred. Under \code{1A\_FullTel\_DomainIsolated}, all $36/36$ proposals and $204/204$ derivations passed on attempt~1 with zero retries.

Timing measurements (Apple M-series, Python~3.13.9, pre-initialized SQLite WAL with \texttt{PRAGMA synchronous=FULL}; extended breakdown in the companion supplement) show that local TCB verification runs four to five orders of magnitude faster than language-model inference and physical failover. Within the local TCB, typed build and verification takes $0.088\,\mathrm{ms}$ p50 ($0.189\,\mathrm{ms}$ p95), retry verification $0.002\,\mathrm{ms}$ ($0.013\,\mathrm{ms}$), ATMS packing and shielding $0.031\,\mathrm{ms}$ ($0.059\,\mathrm{ms}$), probe-loop temporal auditing $0.185\,\mathrm{ms}$ ($0.304\,\mathrm{ms}$), and Tier~E2 SQLite WAL commit $0.101\,\mathrm{ms}$ ($0.118\,\mathrm{ms}$; atomic rejection $0.097\,\mathrm{ms}$), yielding a total local TCB time to an authorized action across the nine actionable held-out cases of $0.531\,\mathrm{ms}$ p50 ($0.576\,\mathrm{ms}$ p95). By comparison, simulated out-of-band probe round-trips add $20.0\,\mathrm{ms}$ p50 ($38.0\,\mathrm{ms}$ p95), real three-member \texttt{etcd}~3.7.3 Raft consensus spans $45$--$2{,}467\,\mathrm{ms}$ schedule medians across $140$ CI fault trials (Appendix~\ref{app:repro}), and live pilot LLM inference required $15.72\,\mathrm{s}$ p50 ($40.75\,\mathrm{s}$ p95) per call and $24.25\,\mathrm{s}$ p50 ($97.61\,\mathrm{s}$ p95) parallel Round-0 wall-clock time per incident ($82.74\,\mathrm{s}$ sequential; $19.78\mathrm{M}$ input and $607\mathrm{k}$ output tokens, \$$5.13$ reported cost). Physical PostgreSQL failover (Section~\ref{sec:physical_eval}) took $9.9$--$10.2$\,s for native auto-failover versus $57$--$67$\,s for paused live-LLM supervised promotion, reflecting distinct availability policies rather than directly comparable latency targets. Table~\ref{tab:attribution} summarizes how each PIEQ contribution maps to its formal basis, empirical evidence, and remaining gaps.

\begin{table}[!t]
\centering
\caption{Contributions-to-evidence attribution map separating formal guarantees from empirical software, live-LLM, and physical evidence alongside explicit open gaps.}
\label{tab:attribution}
\vspace{0.3mm}
\scriptsize
\setlength{\tabcolsep}{2.2pt}
\renewcommand{\arraystretch}{1.10}
\begin{tabularx}{\columnwidth}{@{}L{22mm} L{26mm} Y@{}}
\toprule
\textbf{Contribution} & \textbf{Formal Basis} & \textbf{Empirical Support \& Open Gaps} \\
\midrule
1. Epistemic provenance cert. & Thm.~\ref{thm:atms_soundness}, Thm.~\ref{thm:packing} & Factorial \code{P0} vs.\ \code{P1}: cuts passive false preconditions $4/4{\to}0/4$ (\code{heldout36}); \code{H13}--\code{H15} blocked; $M{\to}64$ packing stress \& synthesis-vs.-completion (Sec.~\ref{sec:synthesis_scalability}); \code{H16} ($E_3$) shows $h_{\mathrm{man}}$ boundary. \\
2. Plausible-world gating & Lem.~\ref{lem:viability} & \code{R0}/\code{NO\_SHIELD}: cuts unsafe $6{\to}0$ (\code{frozen40}), $4{\to}0$ (\code{heldout36}); $|\mathcal{S}_{\mathrm{cand}}|{\to}256$ shield stress ($2.23$\,ms); \code{H33} ($E_1$) shows world-completeness boundary. \\
3. Execution-bound governance & Thm.~\ref{thm:concurrent}, Cor.~\ref{cor:composition}, Prop.~\ref{prop:risk} & \code{H21}--\code{H29} blocked; $249$ CI tests + $200$ POSIX/SQLite + $80$ Sec.~5 multi-proc.\ races + $140$ \texttt{etcd} trials + CI \texttt{etcd} quiescence \& PG-16 race suite pass; PG/Patroni failover gated at Tier~E0+native and real-PG Tier~E2 enforced for bounded reservations (Sec.~\ref{sec:physical_eval}). Scope: Patroni promotion remains Tier~E0+native; multi-host cluster faults open. \\
Supporting: Typed, Probes, Completion & Assump.~\ref{assump:sound_rules}--\ref{assump:world_model} (7-boundary audit) & Lifts software mitig.\ $0/9{\to}5/9{\to}9/9$; live CLI runs ($9/9$ held-out, $20\times$ reliability, $5\times$ agent-count) + direct-API 3-family \code{confirmation12} ($4/4$ mitig., $0/11$ unsafe, \$0.14 complete cost); $16/16$ targeted vs.\ $1/16$ 1-hop completion. Scope: Bedrock cloud endpoints use profile aliases; completion exposes $E_2$ defect (\code{CN11}). \\
\bottomrule
\end{tabularx}
\end{table}

\subsection{Live LLM Held-Out Evaluation and Agent-Count Scaling}
\label{sec:live_heldout}

To test PIEQ with stochastic reasoning rather than deterministic proposers, we evaluated live language-model inference across all 36 held-out incidents with zero replay, zero oracle access, zero deterministic completion, and a retry budget of at most two. Five-agent heterogeneous ensembles paired three Claude models (\code{claude-sonnet-5-5}, \code{claude-haiku-5-5}, \code{claude-opus-5-5}) with two GPT instances (\code{openai.gpt-5.6-sol}), while single-agent and homogeneous runs used Claude Sonnet. Locally authenticated agent runners were used throughout; GPT token billing is uncaptured (reported dollar costs cover Claude usage only), and runner harness overhead adds roughly $18$--$21\mathrm{k}$ input tokens per call. Tables~\ref{tab:rev9_live} and~\ref{tab:rev10_live_ops} report single-run safety, liveness, and operational metrics across eight configurations (Configs~A--E0, E1, S1, and H1).

\begin{table*}[!t]
\centering
\caption{Live LLM evaluation on all 36 frozen held-out incidents ($31$ within-assumption [$9$ actionable, $17$ mandatory-deferral, $5$ false-precondition] + $5$ out-of-assumption controls). Bottom three rows enable the mandatory temporal-policy registry; E0 uses caller-discretion validation. Operational metrics appear in Table~\ref{tab:rev10_live_ops}.}
\label{tab:rev9_live}
\footnotesize
\setlength{\tabcolsep}{4.5pt}
\renewcommand{\arraystretch}{1.12}
\begin{tabular}{@{}lccccccc@{}}
\toprule
\textbf{Configuration} & \textbf{Mitig.\,(9)} & \textbf{False Blk} & \textbf{Within Uns.\,(31)} & \textbf{Mand.\,Def.\,(17)} & \textbf{False Pre.\,(5)} & \textbf{UBSA} & \textbf{OOA Uns.\,(5)} \\
\midrule
A: Free-form (no exec gate) & 4/9 & 5 & 0/31 & 13/17 & 1/5 & 4 & 2/5 \\
B: Typed (no exec gate) & 4/9 & 5 & 0/31 & 9/17 & 1/5 & 8 & 4/5 \\
C: Typed + probes (no exec gate) & 9/9 & 0 & 0/31 & 9/17 & 1/5 & 8 & 4/5 \\
D: 1-model full-tel (no exec gate) & 9/9 & 0 & 0/31 & 12/17 & 0/5 & 5 & 3/5 \\
E0: Full PIEQ, caller-discretion & 9/9 & 0 & 0/31 & 17/17 & 1/5 (\code{H30} committed) & 0 & 2/5 \\
\midrule
E1: Full PIEQ, mandatory registry & 9/9 & 0 & 0/31 & 17/17 & 0/5 (\code{H30} blocked) & 0 & 3/5 \\
S1: One Sonnet, full tel, full gate & 9/9 & 0 & 0/31 & 17/17 & 0/5 & 0 & 3/5 \\
H1: Five homogeneous Sonnet & 9/9 & 0 & 0/31 & 17/17 & 0/5 & 0 & 4/5 \\
\bottomrule
\end{tabular}
\end{table*}

\begin{table*}[!t]
\centering
\caption{Live LLM operational metrics on all 36 held-out incidents (single run per configuration; multi-repetition results in Table~\ref{tab:rev11_rep}). Valid Proposals = replies carrying $\ge 1$ verifier-accepted derivation, first-attempt$\to$post-retry (not comparable across telemetry splits because partitioned agents may lack sufficient evidence). Reported cost covers Claude usage only (GPT usage unbilled); runner harness overhead adds $\approx$18--21k input tokens per call.}
\label{tab:rev10_live_ops}
\footnotesize
\setlength{\tabcolsep}{4.8pt}
\renewcommand{\arraystretch}{1.12}
\begin{tabular}{@{}lcccccc@{}}
\toprule
\textbf{Configuration} & \textbf{Valid Proposals} & \textbf{Probes (Eff)} & \textbf{LLM Calls} & \textbf{Tokens (In/Out)} & \textbf{p50 Latency/Case} & \textbf{Reported Cost} \\
\midrule
A: Free-form & 62$\to$62/180 & 0 & 180 & 3.9M / 71k & 43.2\,s & \$2.80 \\
B: Typed & 73$\to$74/180 & 0 & 182 & 3.9M / 76k & 38.7\,s & \$2.68 \\
C: Typed + probes & 69$\to$74/180 & 17 (6) & 189 & 4.1M / 85k & 46.3\,s & \$2.85 \\
D: 1-model full-tel & 30$\to$33/36 & 19 (8) & 39 & 0.9M / 18k & 6.4\,s & \$1.00 \\
E0: Caller-discretion full PIEQ & 73$\to$74/180 & 17 (6) & 181 & 3.9M / 76k & 44.4\,s & \$2.81 \\
E1: Mandatory registry & 73$\to$74/180 & 18 (7) & 183 & 3.9M / 77k & 52.9\,s & \$2.77 \\
S1: One Sonnet full gate & 30$\to$33/36 & 19 (8) & 39 & 0.9M / 18k & 6.3\,s & \$1.00 \\
H1: Five Sonnet\rlap{\textsuperscript{*}} & 68$\to$72/180 & 17 (6) & 188 & 4.2M / 62k & 30.2\,s & \$4.58 \\
\bottomrule
\multicolumn{7}{@{}l@{}}{\scriptsize \textsuperscript{*}Includes $1$ transient credential-export failure recovered on retry; $0$ provider failures after retry across all other runs.}
\end{tabular}
\end{table*}

Comparing Configs~A through E1 isolates four live effects. Active probing restores live remediation without weakening safety: free-form (A) and typed-only (B) configurations each resolved $4/9$ actionable incidents with $5$ false blocks (mainly probe-recoverable cases \code{H04}--\code{H08}), whereas adding bounded leakage-free probing (C) lifted live models to $9/9$ mitigations and $0$ false blocks while holding within-assumption unsafe admissions at $0/31$. Execution binding is equally indispensable: without the execution gate (A--D), live models admitted $4$--$8$ unjustified mutations on authorization, stale-certificate, and concurrency cases (\code{H21}--\code{H29}), whereas full PIEQ's Tier~E2 boundary blocked every one ($17/17$ mandatory deferrals, $0$ UBSA). Enabling the mandatory trusted temporal-policy registry closed the remaining transient-precondition gap: under caller-discretion validation (E0), transient case \code{H30} still committed ($1/5$ false-precondition; Config~D avoided it incidentally by deferring after probing), whereas turning on the mandatory registry (E1) flipped \code{H30} to blocked ($0/5$). Trace inspection confirmed that the live model proposed the mutation, passed packing and shielding, and received a signed commit-time policy before the Tier~E2 transaction atomically rejected the mutation because the trusted predicate was false at commit, while snapshot-triggered actions including historical \code{H32} remained actionable with zero new false blocks. Across Configs~A--E0 ($771$ calls, \$$12.13$ reported cost), retry rescued $1$--$5$ valid proposals per configuration with zero post-retry provider failures, and out-of-assumption controls admitted unsafe mutations in $2$--$4$ of $5$ cases as designed while live agents deferred the rest.

Holding the full hardened gate constant in a single run, one Sonnet with full telemetry (S1) matched both the five-agent heterogeneous ensemble (E1) and five homogeneous Sonnets (H1) across all $31$ within-assumption cases and all $5$ false-precondition controls (zero within-assumption disagreements and two opposite-direction divergences on out-of-assumption \code{H33}/\code{H35}), while consuming far fewer resources ($39$ vs.\ $183$--$188$ calls, \$$1.00$ vs.\ \$$2.77$--\$$4.58$, and $6.3$\,s vs.\ $30.2$--$52.9$\,s p50 per incident). To test whether this single-run equivalence persists under stochastic repetition, we conducted a preregistered five-repetition paired comparison of S1, H1, and E1 with condition order randomized per repetition ($2{,}030$ calls, \$$41.28$ reported cost). The $540$ case executions split into $465$ within-assumption ($155$ per condition) and $75$ out-of-assumption ($25$ per condition; Table~\ref{tab:rev11_rep}).

Consistent with Theorem~\ref{thm:packing} and Lemma~\ref{lem:viability}, within-assumption safety remained invariant to agent count across all five repetitions: PIEQ recorded $0/465$ unsafe admissions ($0/155$ per condition), $0/75$ false-precondition commits ($0/25$ each), and $255/255$ mandatory deferrals ($85/85$ each). On the five intentional trusted-model-defect controls (\code{H16}, \code{H33}--\code{H36}), PIEQ recorded $47/75$ out-of-assumption unsafe admissions ($11/25$ S1, $20/25$ H1, $16/25$ E1) while model reticence deferred the other $28/75$ (including \code{H34}, deferred $15/15$), reflecting designed assumption violations outside PIEQ's trusted boundary. For liveness, single-run equivalence did not completely survive repetition: S1 false-blocked the historical CA-revocation case \code{H32} in $2/5$ repetitions ($43/45$ mitigations; Wilson 95\% $[0.85, 0.99]$, clustered by $9$ incidents), whereas both five-agent conditions reached $45/45$ $[0.92, 1.00]$ with zero H1-vs-E1 within-assumption discordances, at $4$--$6\times$ latency ($26.4$--$43.0$\,s vs.\ $6.8$\,s p50) and $2.7$--$4.3\times$ reported cost (\$$13.81$--\$$22.27$ vs.\ \$$5.20$). Because this difference concentrated in a single incident ($2/45$ discordant pairs) over $n=5$ repetitions, it reflects bounded single-shot liveness redundancy rather than statistical proof of multi-agent superiority; model heterogeneity conferred no within-assumption benefit, and deterministic completion (examined next) targets the exact same proof-omission mode without multiplying inference cost.

\subsection{Live Reliability and Bounded Deterministic Completion}
\label{sec:live_reliability}

After freezing pipeline hashes, we authored a preregistered 12-case live confirmation set (\code{CN01}--\code{CN12}) in two previously unseen domains (\code{cdn\_tls\_edge} and \code{mq\_broker\_cluster}), comprising $4$ actionable mitigations, $6$ mandatory deferrals, $1$ commit-time false-precondition control (\code{CN09}), and $1$ out-of-assumption unsound-rule control (\code{CN11}). On the initial frozen run, every safety criterion passed ($0$ unsafe, \code{CN09} blocked, $6/6$ deferrals), yet the liveness criterion failed ($2/4$ actionable) because the model submitted both positive proofs while omitting one of two required refutations per world (an exploratory rerun reached $4/4$).

To measure this single-shot variance precisely, we ran $20$ unmodified live repetitions of the 12-case set ($240$ executions: $220$ within-assumption plus $20$ out-of-assumption \code{CN11}; $248$ calls, \$$5.47$ reported cost, zero provider failures; Table~\ref{tab:rev11_rep}). Within-assumption safety held on every execution: $0/220$ unsafe admissions, $0/20$ false-precondition commits on \code{CN09} ($11$ rejected atomically at Tier~E2 after certification and $9$ deferred earlier by epistemic ambiguity), $120/120$ mandatory deferrals, and $0/20$ admissions on \code{CN11} under baseline model reticence. Actionable liveness was strongly bimodal: \code{CN01} and \code{CN02} mitigated $20/20$ (Wilson 95\% $[0.84, 1.00]$), whereas \code{CN10} reached $11/20$ $[0.34, 0.74]$ ($9$ false blocks: $5$ missing refutations and $4$ missing positives) and \code{CN12} reached $9/20$ $[0.26, 0.66]$ ($11$ false blocks, all missing refutations), totaling $60/80$ mitigations and $20$ false blocks. Every false block arose from omitted verifier-acceptable derivations, with zero invalid-Horn, provenance, or probe-budget failures, which schema-driven retry cannot remedy because the submissions are syntactically valid. Fail-closed verification thus converts model under-submission into recoverable false blocks rather than unsafe mutations.

We then evaluated the bounded deterministic completion helper (verified by $13$ adversarial unit tests covering shared domains, shared leaves, nogoods, truncation, and missing sensors) through paired case-by-case replays against all $20$ baseline runs ($240$ replays each without and with in-probe completion, $480$ total; Table~\ref{tab:rev11_rep}). In both modes, completion recovered all $20$ actionable false blocks: \code{CN10} rose from $11/20$ to $20/20$ ($+9$ recoveries, $20$ derivations supplied) and \code{CN12} rose from $9/20$ to $20/20$ ($+11$ recoveries, $22$ derivations supplied), lifting mitigations from $60/80$ to $80/80$. Within assumptions, completion incurred $0/220$ unsafe admissions, $0/20$ false-precondition commits (\code{CN09} remained blocked $20/20$, with $9/20$ shifting from epistemic deferral to atomic Tier~E2 rejection), $120/120$ unchanged mandatory deferrals, and zero true-world refutations. A live type-level missing-proof feedback loop ($n=3$ runs, $46$ calls, \$$0.43$) reproduced these gains ($12/12$ mitigations, zero within-assumption regressions).

This experiment also exposed an important negative finding. Because deterministic completion systematically derives every rule-valid conclusion, it flipped the out-of-assumption unsound-rule control (\code{CN11}, stratum $E_2$) from deferred $20/20$ under baseline model hesitation to admitted-unsafe $20/20$ under completion (and $3/3$ under live feedback). Completion strips away the accidental protection that language-model reticence provides against trusted-rule defects (protection no PIEQ theorem ever relied upon) and places soundness squarely in the authored domain model (Proposition~\ref{prop:risk}, stratum $E_2$). We therefore keep completion disabled by default as an explicit operator trade-off between single-shot liveness and exposure to rule-authoring defects rather than an unconditional safety gain.

\textbf{Direct-API Controlled-Sampling Three-Family Confirmation Run.} To close the uncontrolled-sampling, CLI-harness-bloat, and partial-billing gaps on the same frozen 12-case confirmation suite (\code{CN01}--\code{CN12}), we executed a direct-HTTPS five-agent run (\code{full\_pieq} with the mandatory temporal-policy registry, \emph{zero} deterministic completion, $\mathrm{temperature}{=}0.0$, $\mathrm{top\_p}{=}1.0$, $\mathrm{seed}{=}17$) spanning three independent model families: two AWS Bedrock Anthropic models (\code{claude-sonnet-5-5}, \code{claude-haiku-5-5}), two AWS Bedrock OpenAI models (\code{us.openai.gpt-5.6-sol}), and one local SGLang open-weight model (\code{Qwen/Qwen3-8B}; Table~\ref{tab:rev11_rep}). Every single API call succeeded on attempt~1 with zero retries and zero malformed outputs ($60/60$ valid proposals, $173/173$ accepted derivations, $9$ probes [$3$ effective], $19.2$\,s p50 per incident). Removing the CLI wrapper cut token usage by over an order of magnitude to $45.1\mathrm{k}$ input / $17.6\mathrm{k}$ output tokens ($\approx 751$ input tokens per call) at a complete multi-provider cost of \$$0.139$ (\$$0.0765$ Bedrock Anthropic, \$$0.0625$ Bedrock OpenAI, \$$0.00$ local SGLang). Within assumptions, PIEQ achieved $11/11$ exact outcomes: $4/4$ actionable mitigations ($0$ false blocks, including \code{CN02} via $2$ effective probes), $6/6$ mandatory deferrals, $0/1$ false-precondition commits (\code{CN09} atomically rejected at Tier~E2 via \code{REJECTED\_COMMIT\_PRECONDITION\_FALSE}), and $0/11$ unsafe admissions. On the $E_2$ unsound-rule control (\code{CN11}), per-agent traces directly corroborated our EXP-06 finding without deterministic completion: both Claude agents still omitted the unsound rules out of model reticence, whereas both Bedrock GPT agents and local \code{Qwen/Qwen3-8B} faithfully applied the trusted rules, yielding $1/1$ out-of-assumption admission and demonstrating empirically that reticence against $E_2$ defects is model-specific luck rather than an architectural guarantee. To preserve strict identity disclosure, while \code{Qwen/Qwen3-8B} is a fixed local open-weight checkpoint and all $60$ calls returned server-reported model IDs, the AWS Bedrock endpoints use cross-region US inference-profile aliases rather than date-stamped immutable snapshots.

\begin{table*}[!t]
\centering
\caption{Repeated CLI live evaluations and direct-API controlled-sampling confirmation on the frozen 12-case confirmation set ($20\times$ reliability + paired completion replays + $1\times$ direct-API 3-family run) and frozen 36-case held-out suite ($5\times$ paired agent-count repetitions). Within-assumption and out-of-assumption (OOA) denominators are always reported separately; repeated runs of the same incident are clustered by case, not i.i.d.\ scenario samples.}
\label{tab:rev11_rep}
\vspace{0.2mm}
\scriptsize
\setlength{\tabcolsep}{2.8pt}
\renewcommand{\arraystretch}{1.10}
\begin{tabularx}{\textwidth}{@{}L{43mm} c c c c c c Y@{}}
\toprule
\textbf{Study / Condition} & \textbf{Mitig.} & \textbf{False Blk} & \textbf{Mand.\,Def.} & \textbf{False Pre.} & \textbf{Within Uns.} & \textbf{OOA Uns.} & \textbf{Operational / Notes} \\
\midrule
\multicolumn{8}{@{}l}{\textit{Frozen 12-Case Confirmation Set: $20\times$ CLI Runs ($220$ within + $20$ OOA) \& $1\times$ Direct-API 3-Family Run ($11$ within + $1$ OOA)}} \\
Baseline live Sonnet (CLI, $20\times$) & 60/80 & 20\rlap{\textsuperscript{a}} & 120/120 & 0/20 & 0/220 & 0/20 & $248$ calls, \$5.47; \code{CN10} 11/20, \code{CN12} 9/20 \\
Deterministic Completion (paired) & 80/80 & 0 & 120/120 & 0/20\rlap{\textsuperscript{b}} & 0/220 & 20/20\rlap{\textsuperscript{c}} & Recovers all $20$ FB ($+9$ \code{CN10}, $+11$ \code{CN12}; 42 derivs) \\
Type-Level Feedback (CLI, $n{=}3$) & 12/12 & 0 & 18/18 & 0/3 & 0/33 & 3/3\rlap{\textsuperscript{c}} & $46$ calls ($1$--$2$/case), \$0.43; matches completion \\
\textbf{Direct-API 3-Family ($\mathrm{temp}{=}0$, $1\times$)} & \textbf{4/4} & \textbf{0} & \textbf{6/6} & \textbf{0/1}\rlap{\textsuperscript{b}} & \textbf{0/11} & 1/1\rlap{\textsuperscript{f}} & $60/60$ valid, $45.1$k/$17.6$k tok, \textbf{\$0.14} (all 3 prov.), $19.2$\,s p50 \\
\midrule
\multicolumn{8}{@{}l}{\textit{Frozen 36-Case Held-Out Suite $\times 5$ Paired Reps ($540$ execs = $465$ within [$135$ act., $255$ defer, $75$ False Pre.] + $75$ OOA)}} \\
S1 (1$\times$ Sonnet, full tel, full gate) & 43/45 & 2\rlap{\textsuperscript{d}} & 85/85 & 0/25 & 0/155 & 11/25 & $199$ calls, \$5.20, $6.8$\,s p50/case \\
H1 (5$\times$ Sonnet, part.\ tel, full gate) & 45/45 & 0 & 85/85 & 0/25 & 0/155 & 20/25 & $920$ calls, \$22.27, $26.4$\,s p50/case \\
E1 (5$\times$ hetero, part.\ tel, full gate) & 45/45 & 0 & 85/85 & 0/25 & 0/155 & 16/25 & $911$ calls, \$13.81+GPT, $43.0$\,s p50/case \\
Paired Total ($5$ reps) & 133/135 & 2 & 255/255 & 0/75 & 0/465 & 47/75\rlap{\textsuperscript{e}} & $2{,}030$ calls, \$41.28+GPT; $0$ H1-vs-E1 within-disc. \\
\bottomrule
\multicolumn{8}{@{}p{\textwidth}@{}}{\scriptsize \textsuperscript{a}$9$ on \code{CN10} + $11$ on \code{CN12}. \textsuperscript{b}\code{CN09} blocked atomically at Tier~E2 (\code{REJECTED\_COMMIT\_PRECONDITION\_FALSE}; $9/20$ shifted under completion). \textsuperscript{c}\code{CN11} ($E_2$ unsound rule) flips 0/20$\to$20/20 once completion removes baseline hesitation. \textsuperscript{d}Both false blocks on \code{H32}. \textsuperscript{e}OOA controls: \code{H16} ($E_3$), \code{H33} ($E_1$), \code{H34} ($E_2$; deferred 15/15), \code{H35} ($E_5$), \code{H36} ($E_6$). \textsuperscript{f}On \code{CN11}, Bedrock Anthropic agents hesitated while Bedrock GPT and SGLang \code{Qwen/Qwen3-8B} applied the unsound trusted rule (zero completion).}
\end{tabularx}
\end{table*}

\subsection{Synthesis-vs.-Verification Asymmetry, Authoring Footprint, and Online Scalability Stress}
\label{sec:synthesis_scalability}

To empirically ground the architectural trade-offs raised in Section~\ref{sec:llm_role} and quantify PIEQ's online resource caps under load, we executed two dedicated software stress suites (\code{scripts/run\_rev12\_synthesis\_vs\_completion.py} and \code{scripts/run\_rev12\_scalability\_stress.py}; Table~\ref{tab:rev12_scalability}).

\textbf{TCB Model-Authoring Footprint.} Auditing all $100$ benchmark incidents across all ten domains (\code{frozen40}: $3$ domains; \code{heldout36}: $4$ domains; \code{software\_confirm12}: \code{dns\_anycast\_cluster}; \code{live\_confirm12}: \code{cdn\_tls\_edge}, \code{mq\_broker\_cluster}) shows compact per-domain TCB specifications: $|\mathcal{S}_{\mathrm{cand}}| = 2$--$3$ fault worlds, $|\mathcal{M}_{\mathrm{sys}}| = 1$--$41$ Horn rules ($1$--$15$ in core single-shot domains, up to $41$ in probe-augmented held-out suites), $|P_{\mathrm{sys}}| = 0$--$2$ conflict invariants, and $M = 1$--$9$ primitive observations per incident---well inside the $M \le 64$ bitmask cap.

\textbf{Multi-Hop Proof Search vs.\ Bounded TCB Completion.} Holding $M{=}33$ authenticated observations across $4$ domains ($k_a{=}3, f{=}2$) constant, we swept background distractor rules $R_{\mathrm{bg}} \in \{0, 20, 80, 250\}$ and derivation hop depth $d \in \{1, 2, 3, 4\}$ ($16$ regimes; Table~\ref{tab:rev12_scalability}, top). The existing 1-hop TCB completion helper (\code{pieq/completion.py}, cap $128$) admitted only $1/16$ regimes ($R_{\mathrm{bg}}{=}0, d{=}1$, matching \code{CN10}/\code{CN12}), blocking every multi-hop chain ($d \ge 2$) and truncating once $R_{\mathrm{bg}} \ge 20$. Generalizing TCB completion to bounded multi-hop BFS admitted $4/16$ regimes at cap $128$ and $8/16$ at cap $512$ ($2.38$--$2.89\,\mathrm{ms}$), yet truncated fail-closed across $0/8$ regimes once $R_{\mathrm{bg}} \ge 80$ ($3.46$--$3.62\,\mathrm{ms}$) as distractor rules saturated the derivation budget. By contrast, an untrusted goal-directed deterministic software proposer submitting only the relevant multi-hop DAGs to \code{add\_verified\_derivation} achieved $16/16$ admissions in $0.07$--$0.28\,\mathrm{ms}$ p50 with zero TCB expansion and $0$ unsafe admissions, while $4/4$ adversarial hallucination controls (forged unauthenticated log claim, hallucinated rule ID, same-domain multi-context spoofing, and skipped-hop antecedent mismatch) were rejected deterministically. Because this $16$-regime sweep uses a deterministic software proposer rather than live language-model inference, it isolates the computational asymmetry between bounded TCB forward-chaining and targeted DAG verification; measuring live-LLM proof-search recall across deep multi-hop chains amidst hundreds of distractor rules remains open empirical work.

\textbf{Online Scalability and Fail-Closed Truncation Boundaries.} Table~\ref{tab:rev12_scalability} (bottom) characterizes latency and fail-closed truncation across all four resource axes: (1)~\emph{ATMS and complete-proof packing ($M \to 64$):} ATMS registration stays $0.057$--$1.405\,\mathrm{ms}$ p50 up to $M{=}64, d{=}6$, and single-leaf packing stays exact in $0.372\,\mathrm{ms}$ at $M{=}64$. Under adversarial dense intersecting 3-leaf hypergraphs over $18$ shared domains ($M{=}64, k_a{=}3, B_{\mathrm{pack}}{=}10{,}000$), $|\mathcal{W}|{=}32$ finishes exact in $1.43\,\mathrm{ms}$ ($0\%$ false blocks), whereas $|\mathcal{W}| \in \{64, 128, 256\}$ exhausts branch-and-bound ($13.4$--$47.4\,\mathrm{ms}$, $100\%$ truncation) with \emph{zero unsafe admissions} across all actionable and trap instances. Crucially, Q1's sound lower bound $\hat{\nu}_{\mathrm{pack}} \ge k_a$ still certifies valid disjoint packings ($0\%$ Q1 false blocks), whereas Q2's strict conservative \code{packing\_truncated} guard retains worlds ($100\%$ Q2 false blocks), pinpointing witness deduplication or lower-bound refutation pruning as the key lever for dense intersecting refutation sets. (2)~\emph{ATMS label and hitting-set caps:} $L_{\max}{=}64$ flags \code{budget\_exceeded} cleanly at $80$ minimal environments; $B_{\mathrm{HS}}{=}256$ completes exact through $6$ disjoint 2-leaf nogoods ($64$ hitting sets in $0.166\,\mathrm{ms}$) and fails closed cleanly at $\ge 8$ independent conflicts ($0.625\,\mathrm{ms}$). (3)~\emph{Plausible-world shielding and Mode~C:} Shield latency scales linearly from $0.030\,\mathrm{ms}$ ($|\mathcal{S}_{\mathrm{cand}}|{=}4$) to $2.232\,\mathrm{ms}$ ($|\mathcal{S}_{\mathrm{cand}}|{=}256$, $M{=}64, f{=}2$). Mode~C R1 checks $4, 15, 64, 325$ schedules for unordered $|\mathcal{U}_C|{=}1..4$ in $0.012$--$0.556\,\mathrm{ms}$ ($|\mathcal{S}_{\mathrm{base}}|{=}1$) and $0.027$--$3.095\,\mathrm{ms}$ ($|\mathcal{S}_{\mathrm{base}}|{=}16$; DAG ordering cuts $|\mathcal{U}_C|{=}4$ to $95$ schedules in $0.206$--$0.806\,\mathrm{ms}$), failing closed at $|\mathcal{U}_C|{=}5$ in $0.001\,\mathrm{ms}$; Regime~R2 checks $2, 18, 270$ schedules for $|\mathcal{P}|{=}2,4,6$ in $0.008, 0.044, 0.539\,\mathrm{ms}$, rejects write-skew pre-dispatch, and fails closed at $|\mathcal{P}|{=}7$ in $0.001\,\mathrm{ms}$. (4)~\emph{Multi-scope alert storms:} Because $M \le 64$ is scoped per active resource tuple $(\mathrm{res}, \tau, h_{\mathrm{man}})$ rather than cluster-wide, $20$ concurrent scopes $\times 32$ signals ($640$ simultaneous alerts) achieved $20/20$ admissions with $0$ truncations at $0.125\,\mathrm{ms}$/scope p50, whereas unpartitioned single-scope storms ($M \in \{65, 96, 128\}$) rejected overflow signals fail-closed with $0$ unsafe admissions.

\begin{table*}[!t]
\centering
\caption{Revision~12 synthesis-vs.-completion asymmetry ($16$ regimes, $M{=}33, k_a{=}3$; deterministic software proposer isolating verifier-vs.-completion scaling rather than live-LLM multi-hop recall) and online scalability / fail-closed truncation stress benchmarks (\code{scripts/run\_rev12\_synthesis\_vs\_completion.py}, \code{scripts/run\_rev12\_scalability\_stress.py}; Apple M-series, Python~3.13.9). Every saturated or adversarial regime recorded \textbf{0 unsafe admissions}.}
\label{tab:rev12_scalability}
\vspace{0.2mm}
\scriptsize
\setlength{\tabcolsep}{3.4pt}
\renewcommand{\arraystretch}{1.08}
\begin{tabularx}{\textwidth}{@{}L{48mm} c c c c Y@{}}
\toprule
\textbf{Benchmark / Stress Regime} & \textbf{Admitted / Exact} & \textbf{Trunc.\,Rate} & \textbf{Unsafe} & \textbf{p50 (ms)} & \textbf{Key Finding / Fail-Closed Boundary} \\
\midrule
\multicolumn{6}{@{}l}{\textit{Synthesis-vs.-Verification Asymmetry ($R_{\mathrm{bg}} \in \{0,20,80,250\} \times d \in \{1..4\}$, 16 Regimes; 4/4 Adversarial Controls Blocked)}} \\
TCB 1-hop completion (cap $128$, \code{completion.py}) & 1/16 & 12/16 & 0 & 0.06--0.91 & Succeeds only at $R_{\mathrm{bg}}{=}0, d{=}1$; blocks all $d \ge 2$ \\
TCB multi-hop BFS (cap $128$ / cap $512$) & 4/16 \;|\; 8/16 & 12/16 \;|\; 8/16 & 0 & 0.10--3.62 & Exhausts budget ($0/8$) once $R_{\mathrm{bg}} \ge 80$ \\
\textbf{Targeted software proposer + TCB verifier} & \textbf{16/16} & \textbf{0/16} & \textbf{0} & \textbf{0.07--0.28} & Zero TCB growth; $4/4$ hallucination controls blocked \\
\midrule
\multicolumn{6}{@{}l}{\textit{Online Scalability \& Fail-Closed Truncation Stress ($M \to 64$, $|\mathcal{S}_{\mathrm{cand}}| \to 256$, Mode~C R1/R2, Alert Storms)}} \\
Packing realistic ($M{=}64, d{=}1$ / $d{=}6$, $|\mathcal{W}|{=}64$) & $\hat{\nu}{=}64$ / $\hat{\nu}{=}7$ & 0.00 / 1.00 & 0 & 0.37 / 14.38 & ATMS reg.\ $0.60$--$1.41$\,ms; lower bound $\hat{\nu}$ valid \\
Adversarial 3-leaf hypergraph ($|\mathcal{W}|{=}32, B{=}10^4$) & Exact (20/20) & 0.00 & 0 & 1.43 & $0\%$ Q1 \& $0\%$ Q2 false blocks at $M{=}64$ \\
Adversarial 3-leaf ($|\mathcal{W}|{=}64 / 256, B{=}10^4$) & Q1 20/20; Q2 0/20 & 1.00 & 0 & 13.44 / 47.40 & Q1 $\hat{\nu}{\ge}3$ ($0\%$ FB); Q2 strict \code{trunc} retains worlds \\
Hitting-set cap ($B_{\mathrm{HS}}{=}256$, $6$ vs.\ $8$ nogoods) & Exact / Fail-closed & 0.00 / 1.00 & 0 & 0.17 / 0.63 & $64$ HS exact at $6$; truncates safely at $2^8{=}256$ \\
Plausible shield ($|\mathcal{S}_{\mathrm{cand}}|{=}4 \to 256, M{=}64, f{=}2$) & 100\% true kept & 0.00 & 0 & 0.03 $\to$ 2.23 & Linear scaling; forced \code{trunc} retains all $256$ worlds \\
Mode~C R1 ($|\mathcal{U}_C|{=}4$ unord.\ $|\mathcal{S}|{=}1 / 16$; $|\mathcal{U}_C|{=}5$) & 325 scheds / Deny & -- & 0 & 0.56 / 3.10 & DAG cuts to $95$ ($0.21$--$0.81$\,ms); $|\mathcal{U}_C|{=}5$ rejects $0.001$\,ms \\
Mode~C R2 ($|\mathcal{P}|{=}2 / 4 / 6$; $|\mathcal{P}|{=}7$) & 2 / 18 / 270 / Deny & -- & 0 & 0.01--0.54 & Write-skew rejected pre-dispatch; $|\mathcal{P}|{=}7$ rejects $0.001$\,ms \\
Alert storm ($20\times 32{=}640$ scoped vs.\ $M{>}64$ single) & 20/20 vs.\ Overflow & 0.00 / 1.00 & 0 & 0.13/scope & Per-$(\mathrm{res},\tau,h_{\mathrm{man}})$ scoping avoids $M{=}64$ saturation \\
\bottomrule
\end{tabularx}
\end{table*}

\subsection{Execution-Boundary Fault Injection, Physical PostgreSQL/Patroni, and Real-Database Tier~E2 Evaluations}
\label{sec:physical_eval}

To narrow the gap between Section~\ref{sec:execution}'s concurrency theory and physical validation, we evaluate execution-boundary contracts across three distinct real-storage tiers without pooling them: (1)~multi-process OS-level POSIX/SQLite WAL fault injection, (2)~a distributed three-member containerized \texttt{etcd}~v3.7.3 Raft cluster in CI, and (3)~a physical PostgreSQL~16/Patroni cluster.

\textbf{Multi-Process OS-Level Section~5 Fault Suite ($80$ Trials) and Distributed \texttt{etcd} Trials ($140$ Trials).} Extending our $200$ baseline POSIX/SQLite WAL trials, we executed $80$ new forked-process OS fault trials (\code{scripts/run\_rev12\_section5\_fault\_eval.py}, $20$ repetitions across four schedules synchronized over POSIX pipes against on-disk SQLite WAL \texttt{PRAGMA synchronous=FULL}): (i)~in \code{concurrent\_coordinator\_radm\_cas}, two independent coordinator processes verified Mode~C R1 (\code{ModeCCompatibilityVerifier}) at identical $r_{\mathrm{adm}}{=}1$ and raced at \texttt{BEGIN IMMEDIATE} commit, yielding $20/20$ exact single-winner commits ($r_{\mathrm{adm}} \to 2$), deterministic stale-revision rejection of the loser, and clean post-revalidation commit ($6.32$\,ms p50); (ii)~in \code{midflight\_auth\_revocation\_e2}, forked coordinator and revocation processes raced across a pipe barrier against Tier~E2 \code{SQLiteProtectedResourceBoundary}, achieving $20/20$ valid serializations with zero post-revocation commit leaks ($\delta_{\mathrm{fresh}}^{\mathrm{E2}}{=}0$, $8.71$\,ms p50); (iii)~in \code{tier\_e1\_plus\_quiescence\_6step}, an on-disk two-phase quiescence barrier empirically reproduced the exact six-step trace of Corollary~\ref{cor:composition}(c-E1+) across $20/20$ trials ($20/20$ prepared-slot completions during the pre-$\mathsf{FenceAck}$ window, $0/20$ post-$\mathsf{FenceAck}$ revoked commits, and $20/20$ conflict blocks after real \code{SIGKILL} while prepared, $15.18$\,ms p50); and (iv)~in \code{mode\_c\_r2\_phase\_decomposed\_write\_skew}, Regime~R2 rejected write-skew pairs pre-dispatch while admitting disjoint-key overlapping two-phase transactions across forked processes with $0/20$ invariant violations ($6.53$\,ms p50). On the real three-member containerized \texttt{etcd} Raft cluster in GitHub Actions CI (run~\code{45335050195}, commit~\code{80836b20}), $140/140$ trials ($20\times$ across seven schedules: \code{stale\_resume}, \code{simultaneous\_drains}, \code{duplicate\_replay}, \code{predecessor\_frontier}, \code{leader\_failure}, \code{quorum\_loss}, and \code{client\_isolation}) preserved exposure caps, monotonic epochs, and predecessor frontiers across Raft elections and network cuts, and \code{EtcdQuiescenceBoundary} passed the six-step pre-/post-$\mathsf{FenceAck}$ exclusion and crash-blocking check directly on the three-member \texttt{etcd} cluster. In the companion \code{rev12-stress-and-pg-e2} CI job against a live \code{postgres:16} container, \code{rev11\_exp08\_e2\_adapter.py --include-rev12-races} passed all $12$ baseline Tier~E2 contract checks plus barrier-synchronized concurrent-coordinator $\mathbf{r}_{\mathrm{adm}}$ CAS races (single winner advancing $r_{\mathrm{adm}} \to 2$, deterministic loser abort, clean revalidated retry) and $10/10$ barrier-synchronized mid-flight authorization revocation races with zero post-revocation commit leaks.

\textbf{Physical PostgreSQL/Patroni Cluster (Tier~E0+Native) and Real-PostgreSQL Tier~E2 Reservations.} To test how PIEQ behaves on real stateful database infrastructure rather than software simulators alone, we deployed a physical cluster of three Patroni/PostgreSQL~16 nodes (\code{spilo-16:3.3-p1}) backed by a three-member \texttt{etcd}~v3.7.3 consensus cluster inside a container VM under continuous write traffic. An independent 1\,Hz SQL observer queried every node to log roles, timelines, and write-witness rows ($\pm 1$\,s resolution, so sub-second dual-primary states cannot be excluded). Because Patroni's REST promotion endpoint does not re-verify PIEQ certificates at the PostgreSQL commit boundary, we classify this failover adapter strictly as Tier~E0 plus target-native high-availability safeguards and run a tier-limited physical profile reusing PIEQ's epistemic, temporal, and authorization checks for pre-dispatch gating only. Because all six containers share one VM kernel, domain-independence claims apply strictly to injected container and network faults.

We executed five physical trial families on this cluster. In the native baseline, killing the leader triggered Patroni auto-promotion in $9.9$--$10.2$\,s with zero dual-primary intervals at 1\,Hz. In four supervised live-LLM promotion trials (Patroni paused mode), after real leader death Sonnet submitted $5/5$ verified typed derivations spanning three real collector domains (the in-band SQL path, the out-of-band REST bridge, and the \texttt{etcd} lease), achieving $\nu_{\mathrm{pack}}=3$ and a single surviving fault world; REST dispatch then produced observer-recorded promotion in $57$--$67$\,s with zero dual-primary intervals. An event-timeline audit verified causal attribution: no writable primary existed between leader kill and earliest PIEQ dispatch (fault plus $35$\,s lease lapse plus $6.1$--$12.1$\,s live inference), and the new primary appeared $9$--$15$\,s after dispatch. Under an asymmetric network partition blackholing the SQL path while the leader held its lease, PIEQ deferred on plausible-world ambiguity while the leader continued serving ($20/20$ writes), and duplicated same-path telemetry never inflated $\nu_{\mathrm{pack}}$. In policy and revocation refusal trials, a certified promotion ($\nu_{\mathrm{pack}}=3$) was rejected pre-dispatch after authorization revocation, and a commit-time-sensitive policy failed closed on the Tier~E0 target while a snapshot-triggered policy admitted cleanly. Finally, in an out-of-assumption bypass trial, a direct \code{patronictl} switchover bypassing PIEQ executed as modeled outside complete writer mediation and was flagged by reconciliation.

Causal attribution on this testbed must remain precise: Patroni and \texttt{etcd} natively prevented split-brain in every trial, a conventional deterministic supervisor inspecting the same three signals would make the same clean leader-death promotion, and a naive LLM baseline deferred on four partition trials out of model prudence rather than formal enforcement. PIEQ's measured physical increment on Patroni failover is therefore certified pre-dispatch gating (verifying provenance and fault budgets, enforcing authorization and temporal-policy refusals, deferring on partition ambiguity, and detecting unmediated writes) rather than split-brain prevention or faster clean failover.

To validate genuine Tier~E2 enforcement on a real database engine, we additionally built a capacity-reservation adapter enforcing our ten-point Tier~E2 contract inside a single \texttt{SERIALIZABLE} PostgreSQL transaction on the live Patroni primary. In-database \texttt{pgcrypto} HMAC verifies signed operation identity and temporal-policy binding against a server-stored key, while control-row \texttt{SELECT \dots FOR UPDATE} locking serializes aggregate exposure; concurrent over-capacity attempts yielded one commit and one retryable serialization abort, with final $\mathrm{SUM} = 10 \le \mathrm{cap}$ confirmed by an independent SQL read. The same transaction enforces authorization epoch, database-clock TTL, admission-revision CAS, predecessor settlement, idempotency, identity-conflict rejection, and atomic commit-time predicate revalidation, rejecting stale transient predicates with zero state mutation. Crash-before-commit leaves no row, while crash-after-commit reconciles from the authoritative operation table with idempotent re-presentation and no automatic redispatch. At the privilege boundary, an unmediated agent role is denied (\texttt{42501}), whereas a \texttt{postgres} superuser bypass succeeds as an administrator outside mediation assumptions. This adapter covers a bounded reservation vocabulary only and does not upgrade Patroni failover promotion from Tier~E0.

\section{Related Work and Research Positioning}
\label{sec:related}

PIEQ's central contribution is execution-bound epistemic certification: composing classical formal, control-theoretic, and distributed-systems primitives across the boundary between agentic reasoning and physical actuation without claiming originality for any component taken alone.

In symbolic diagnosis and lineage tracking, de~Kleer~\cite{dekleer1986,dekleer1987}, Reiter~\cite{reiter1987}, and Provan~\cite{provan1990} established minimal-environment assumption-based truth maintenance, hitting-set diagnosis, and their worst-case complexity bounds, while Green et al.~\cite{green2007provenance} formalized algebraic provenance semirings. PIEQ builds directly on these foundations by pairing bounded ATMS label maintenance with recursive physical failure-domain footprints, exact complete-proof packing under fail-closed compute truncation, and typed verifier-vocabulary certificate synthesis.

In multi-agent language-model coordination, studies of debate~\cite{du2023debate,choi2025debate,freemad2025,sycophancy2025,failuremodes2025,demystifying2026} and collective reasoning under distributed information~\cite{diverse2026,hiddenbench2026} document conformity collapse, sycophancy, and shared-evidence herding. Rather than trying to engineer debate prompts that resist conformity, PIEQ removes language models from the trusted computing base entirely while retaining their multi-hop diagnostic hypotheses through typed proof claims and read-only diagnostic probing.

In control theory and runtime assurance, PIEQ's plausible-world gating draws on partially observable Markov decision process (POMDP) shielding~\cite{spf2024,melcer2024shield,ipomdp2026}, discrete-event supervisory control and diagnosability~\cite{ramadge1987,sampath1995}, viability theory~\cite{aubin1991}, and Feldbaum's dual control~\cite{feldbaum1960}. PIEQ extends these ideas to fault-prone distributed telemetry through $f$-fault-tolerant refutation packing ($\widehat{\mathcal{B}}_{\tau, v}^{(f)}$), conservative world retention under compute caps, snapshot-audited read-only probing, and explicit separation of compositional state invariants from non-compositional coupled viability kernels.

In security and agent governance, Necula~\cite{necula1997pcc} and Appel and Felten~\cite{appel1999pca} pioneered proof-carrying code and proof-carrying authentication, inspiring recent agent-governance frameworks including CaMeL~\cite{camel2025}, Conformal Policy Control~\cite{conformal2026}, Proof-Carrying Agent Actions~\cite{kadyrbek2026,wang2026pcaa}, and $\mathrm{PC}^2$~\cite{ciliberto2026pc2} that decouple untrusted LLM planning from deterministic policy verification. PIEQ complements single-action policy checking with multi-domain conjunctive provenance isolation, fault-tolerant belief retention, and tier-exact concurrent execution contracts.

Finally, linearizability~\cite{herlihy1990linearizability}, snapshot-isolation anomalies~\cite{berenson1995ansi}, stuttering refinement mappings~\cite{abadi1991refinement}, escrow and demarcation protocols~\cite{oneil1986escrow,barbara1992demarcation}, leases and fencing tokens~\cite{gray1989leases,burrows2006chubby}, and Byzantine fault tolerance~\cite{lamport1982} are classical foundations of distributed systems. Within the KedgeFlow control plane~\cite{he2026kedgeflow,he2026honest}, PIEQ bridges these execution-layer primitives directly to epistemic shielding through contiguous-version snapshot barriers, universal $\mathbf{r}_{\mathrm{adm}}$ CAS, explicit $\mathsf{Refines}_{\mathcal{D}}$ regimes, tier-exact authorization freshness, and trusted commit-time precondition policies.

\section{Residual Limitations and Conclusion}
\label{sec:conclusion}

\subsection{Residual Limitations}
\label{sec:limitations}

PIEQ provides no safety guarantee outside four external trust contracts and two empirical scope boundaries. First, safety depends fundamentally on trusted domain models: omitted fault worlds ($E_1$, illustrated by \code{H33}), unsound deductive rules ($E_2$, illustrated by \code{H34} and \code{CN11}), or missing invariant couplings in $\mathcal{G}_{\mathrm{inv}}$ ($E_8$) permit unsafe admissions even with $k_a$ disjoint proofs. Enabling bounded deterministic completion recovers all $20$ single-shot false blocks on the confirmation suite at the cost of removing accidental LLM-reticence masking against $E_2$ rule defects (\code{CN11} flips from $0/20$ to $20/20$), making completion an explicit operator trade-off rather than an unconditional safeguard. Second, physical fault counting relies on accurate failure-domain classification: misclassified physical dependencies in $h_{\mathrm{man}}$ ($E_3$, \code{H16}), faulty domains exceeding budget $f$ ($E_5$, \code{H35}), or exogenous cross-domain common-cause faults ($E_4$) defeat deterministic quorum guarantees.

Third, verifying a finite abstract phase graph proves safety of the abstract model alone and never discharges physical-to-abstract refinement ($\mathsf{Refines}_{\mathcal{D}}$) automatically; unmodeled internal phases, cached reads, background callbacks, crash-recovery continuations, or non-idempotent retries invalidate R1 or R2 unless formally verified. Fourth, target-native enforcement and durability constrain what PIEQ can promise at the actuator boundary: local WAL deduplication never renders a nontransactional remote actuator idempotent, unmediated external writers ($E_6$, such as direct \code{patronictl} switchovers or PostgreSQL superuser bypasses) bypass $L_{\mathrm{exp}}$, Tier~E1+-quiescence retains a pre-$\mathsf{FenceAck}$ authorization-revocation window and trades liveness for safety via indefinite \texttt{UNKNOWN} blocking, and persistence resides in the SQLite WAL, \texttt{etcd}, or PostgreSQL backends rather than the in-memory ledger core.

Two empirical boundaries also constrain our current evidence. Earlier CLI-tier live runs (\code{heldout36} and repeated studies) used unpinned snapshots with uncontrolled sampling and partial cost reporting (omitting GPT token billing); although our new direct-API three-family \code{confirmation12} run closes controlled sampling ($\mathrm{temperature}{=}0.0$, $\mathrm{seed}{=}17$), server-reported model IDs, and complete billing (\$$0.14$ across all three families), its AWS Bedrock cloud endpoints use cross-region inference-profile aliases rather than immutable date-stamped snapshots, leaving date-pinned cloud API evaluation across \code{heldout36} as future work. Furthermore, every benchmark is author-labeled, our five-repetition agent-count study has limited statistical power because its liveness divergence concentrated on a single incident (\code{H32}), and dense intersecting refutation hypergraphs ($|\mathcal{W}| \ge 64$ 3-leaf witnesses per world at $B_{\mathrm{pack}}{=}10{,}000$) trigger fail-closed truncation under strict Q2 world retention unless witnesses are deduplicated or lower-bound pruned (Section~\ref{sec:synthesis_scalability}). On the physical testbed, the Patroni REST failover adapter remains Tier~E0 plus native safeguards (with Patroni and \texttt{etcd} supplying split-brain protection natively), our real-PostgreSQL Tier~E2 adapter covers a bounded capacity-reservation vocabulary only, all six testbed containers share one VM kernel, 1\,Hz SQL observation cannot rule out sub-second dual-primary states, and while concurrent-coordinator $\mathbf{r}_{\mathrm{adm}}$ CAS, mid-flight revocation, Regime~R2, and Tier~E1+-quiescence races passed $80/80$ multi-process POSIX/SQLite WAL trials as well as CI-verified containerized \texttt{etcd} quiescence and PostgreSQL~16 concurrent/revocation races, live multi-host Patroni failover fault injection remains future work.

\subsection{Conclusion}
\label{sec:concluding_remarks}

Safe autonomous infrastructure mutation requires separating three obligations that generative-agent agreement can never replace: independently sourced evidence certification, uncertainty-aware plausible-world safety gating, and execution-bound target governance. Across conditional formal guarantees, automated CI and distributed \texttt{etcd} fault suites, frozen software benchmarks, live language-model evaluations, supervised physical PostgreSQL/Patroni failovers, and real-PostgreSQL Tier~E2 transactional enforcement, PIEQ demonstrates that autonomous mutations can be governed by deterministic, evidence-backed safety contracts while keeping every language model strictly outside the trusted computing base. Fault tolerance rests on independent physical failure domains rather than agent headcount or model diversity, commit-time semantics are enforced by trusted policy metadata rather than caller discretion, and fail-closed verification converts occasional agent proof omissions into recoverable false blocks rather than within-assumption unsafe mutations. Four concrete directions remain open for future work: immutable date-stamped cloud API evaluation across the full 36-case held-out suite, Tier~E2 physical failover promotion, independent external benchmark annotation, and multi-host hardware fault isolation.

\appendix
\counterwithin{theorem}{section}
\counterwithin{definition}{section}
\counterwithin{remark}{section}
\section{Reproducibility, Artifact Availability, and CI Verification}
\label{app:repro}

\begingroup\sloppy
All implementation code, regression tests, frozen benchmark definitions, preregistrations, experiment scripts, traces, and audit reports are published at \url{https://github.com/openkedge/openkedge} under the authoritative path \code{research/pieq/} at verified Revision~12 commit \code{80836b200ec56d8d4ef845cf86c0c2873314b287} (extending Revision~11 baseline commit \code{525ff84e}).

From \code{research/pieq/}, evaluation dependencies install via \code{python3 -m pip install -r requirements-eval.txt}. The repository workflow \code{.github/workflows/pieq-evaluation.yml} ran against commit \code{80836b20} on GitHub-hosted runners (\texttt{ubuntu-latest}, Python~3.13, \code{pytest}, NumPy) across two container-backed jobs in verified workflow run \url{https://github.com/openkedge/openkedge/actions/runs/45335050195}: (1)~\code{etcd-faults} ($302\,\mathrm{s}$ total) launched an isolated digest-pinned three-member containerized \texttt{etcd} cluster (\code{containers/etcd-compose.yml}) and passed all $249/249$ tests ($0$ failed, $0$ skipped in $169\,\mathrm{s}$) under \code{python -m pytest tests/ -q}---including all $14$ container-backed \texttt{etcd} integration tests (featuring \code{test\_etcd\_quiescence\_6step\_revocation\_and\_crash\_blocking} on \code{EtcdQuiescenceBoundary}) and all $6$ Revision~12 regression tests (\code{tests/test\_revision12\_evals.py})---followed by $140/140$ passing distributed \texttt{etcd} fault trials ($111\,\mathrm{s}$) under \code{python scripts/run\_etcd\_fault\_eval.py --repetitions 20 --output results/ci\_etcd\_evidence.json}, uploaded as CI artifact \code{pieq-etcd-evidence}; and (2)~\code{rev12-stress-and-pg-e2} ($34\,\mathrm{s}$ total) re-executed all three Revision~12 suites on Linux and ran \code{experiments/pg\_patroni\_failover/rev11\_exp08\_e2\_adapter.py 127.0.0.1 --include-rev12-races} against a live containerized \code{postgres:16} service container, passing all $12$ baseline Tier~E2 contract checks plus barrier-synchronized concurrent-coordinator $\mathbf{r}_{\mathrm{adm}}$ CAS and $10$-trial mid-flight authorization revocation races, uploaded as CI artifact \code{pieq-rev12-evidence}. Because pytest's \texttt{etcd} integration tests exercise overlapping schedules with the standalone fault suite, those two suites are reported separately rather than summed. GitHub CI constitutes an independently rerun execution of the authors' published implementation rather than an independent third-party scientific replication.

Frozen hashes and preregistration manifests reside at \code{results/rev11/rev11\_phase0\_freeze.json}, \code{results/rev11/REVISION11\_EVIDENCE\_MANIFEST.json}, \code{results/rev10\_frozen\_hashes.json}, \code{results/rev9\_frozen\_hashes.json}, and \code{results/rev7\_frozen\_hashes.json}. Study outputs and runners reside at \code{results/rev11/exp05/} ($20$ raw traces and \code{exp05\_analysis.json}) for the $20\times$ live reliability study; \code{results/rev11/exp06/c1\_results.jsonl}, \code{c2\_results.jsonl}, and \code{c3\_results.jsonl} (deterministic replay via \code{scripts/run\_rev11\_exp06.py --condition C1|C2}) for bounded deterministic completion; \code{results/rev11/exp07/} and \code{experiments/pg\_patroni\_failover/rev10\_logs/} for physical failover causality; \code{results/rev11/exp08/} and \code{experiments/pg\_patroni\_failover/rev11\_exp08\_e2\_adapter.py} (extended with \code{--include-rev12-races}) for real-PostgreSQL Tier~E2 reservations; \code{results/rev11/exp09/} ($15$ traces and \code{exp09\_analysis.json}) for the $5\times$ paired agent-count study; and \code{results/rev12/} (\code{rev12\_synthesis\_vs\_completion.json}, \code{rev12\_scalability\_stress.json}, \code{rev12\_section5\_multiprocess\_faults.json}, \code{pinned\_conf12\_full\_pieq.jsonl} [SHA-256 \texttt{9964a33f...}], \code{pinned\_conf12\_full\_pieq\_summary.json}, \code{pinned\_conf12\_full\_pieq\_scored.json}, and \code{rev12\_pinned\_confirmation\_preflight.json}) for the Revision~12 synthesis-vs.-completion, scalability/truncation, $80$-trial multi-process Section~5, and direct-API controlled-sampling three-family confirmation suites. All deterministic software suites, frozen held-out evaluations, paired completion replays, scalability stress benchmarks, and multi-process POSIX/SQLite WAL schedules reproduce deterministically from committed traces and scripts, whereas fresh live LLM runs cannot reproduce bit-for-bit across future model versions because model snapshots are unpinned and sampling is uncontrolled. Re-running the physical PostgreSQL/Patroni studies requires Docker Compose with \code{ghcr.io/zalando/spilo-16:3.3-p1} and \code{quay.io/coreos/etcd:v3.7.3} (\code{experiments/pg\_patroni\_failover/docker-compose.rev10.yml}). For replicators with direct provider API keys, \code{scripts/run\_rev8\_pinned\_live\_eval.py --dataset heldout36|confirmation12} provides a pinned multi-provider API runner alongside its preflight manifests (\code{results/rev10\_pinned\_preflight.json}, \code{results/rev12/rev12\_pinned\_confirmation\_preflight.json}). Extended archival tables appear in the companion supplementary document and \code{research/pieq/docs/}.
\endgroup

\section{Conformity Dynamics: Formal Statement, Spectral Proof, and Variance Bound}
\label{app:thm1_proof}

This appendix provides the full formalization of Section~\ref{sec:theory}. Let $N \ge 2$ proposers evaluate a binary proposition $h \in \mathcal{H}$ via unbounded latent score residuals $\varepsilon_i^{(t)} = z_i^{(t)} - z^\star(S_\tau) \in \mathbb{R}$, with probability-valued beliefs produced by a separate nonlinear readout $y_i^{(t)} = \psi(z_i^{(t)}) \in [0,1]$. Let $\Sigma^{(t)} = \operatorname{Cov}(\varepsilon^{(t)} \mid S_\tau) \in \mathbb{R}^{N \times N}$ have equal marginal variance $\Sigma_{ii}^{(t)} = \sigma_t^2 > 0$, pairwise correlations $\rho_{ij}^{(t)} = \Sigma_{ij}^{(t)} / \sigma_t^2$, and mean off-diagonal correlation $\bar{\rho}_t = \frac{1}{N(N-1)}\sum_{i \neq j} \rho_{ij}^{(t)} \ge 0$.

\begin{definition}[Effective Sample Size vs.\ Participation Ratio]
\label{def:metrics}
For $N \ge 2$ equal-variance score residuals with $\bar{\rho}_t \ge 0$, the effective sample size governing $\operatorname{Var}(\bar{\varepsilon}^{(t)} \mid S_\tau) = \sigma_t^2 / N_{\mathrm{eff}}(\Sigma^{(t)})$ is $N_{\mathrm{eff}}(\Sigma^{(t)}) \triangleq N / [1 + (N-1)\bar{\rho}_t] \in [1, N]$, whereas the participation ratio is $r_{\mathrm{PR}}(\Sigma^{(t)}) \triangleq (\operatorname{tr}\Sigma^{(t)})^2 / \|\Sigma^{(t)}\|_F^2 = N^2 / [N + \sum_{i \neq j} (\rho_{ij}^{(t)})^2]$. Under equicorrelation, $r_{\mathrm{PR}} = N/[1+(N-1)\bar{\rho}_t^2] > N_{\mathrm{eff}}$ strictly for every $\bar{\rho}_t \in (0,1)$, so $r_{\mathrm{PR}}$ overstates variance reduction (for example, at $N=5$ and $\bar{\rho}_t = 0.142$, $N_{\mathrm{eff}} = 3.19$ whereas $r_{\mathrm{PR}} = 4.63$).
\end{definition}

In a stylized linear model of shared-context debate~\cite{choi2025debate,sycophancy2025}, each proposer updates its latent score residual toward the peer mean $\bar{\varepsilon}^{(t)}$ with conformity weight $\gamma \in (0, 1]$ while accumulating an additive reasoning innovation $\eta_i^{(t)} \in \mathbb{R}$:
\begin{equation}
  \varepsilon_i^{(t+1)} = (1 - \gamma)\varepsilon_i^{(t)} + \gamma \bar{\varepsilon}^{(t)} + \eta_i^{(t)}.
  \label{eq:delib_update}
\end{equation}

\begin{theorem}[Latent-Score Conformity Covariance Contraction]
\label{thm:conformity}
Let $N \ge 2$ and suppose unbounded latent score residuals $\varepsilon^{(t)} \in \mathbb{R}^N$ evolve under Eq.~\eqref{eq:delib_update} from equicorrelated $\Sigma^{(0)}$ ($\sigma_0^2 > 0, \bar{\rho}_0 \in [0, 1)$), driven by temporally white zero-mean innovations $\{\eta^{(t)}\}_{t \ge 0}$ with $\operatorname{Cov}(\eta^{(t)}, \eta^{(s)}) = 0$ ($t \neq s$), $\operatorname{Cov}(\varepsilon^{(t)}, \eta^{(t)}) = 0$, and equicorrelated covariance $Q_\eta$ ($\sigma_\eta^2 \ge 0, \rho_{\mathrm{model}} \in [0, 1]$). Then for all $t \ge 0$, $\Sigma^{(t)}$ stays equicorrelated on $\operatorname{span}(\mathbf{1}) \oplus \mathbf{1}^\perp$ with consensus eigenvalue $\lambda_\parallel^{(t)} = \sigma_0^2[1+(N-1)\bar{\rho}_0] + t\sigma_\eta^2[1+(N-1)\rho_{\mathrm{model}}]$, disagreement eigenvalue $\lambda_\perp^{(t)} = \sigma_0^2(1-\bar{\rho}_0)(1-\gamma)^{2t} + \frac{\sigma_\eta^2(1-\rho_{\mathrm{model}})}{\gamma(2-\gamma)}[1-(1-\gamma)^{2t}]$ ($0^0 = 1$), and $N_{\mathrm{eff}}(\Sigma^{(t)}) = 1 + (N-1)\lambda_\perp^{(t)}/\lambda_\parallel^{(t)}$. Specifically, (1)~if $\sigma_\eta^2 = 0$, the excess effective sample size $N_{\mathrm{eff}} - 1$ contracts geometrically as $(1-\gamma)^{2t}$; and (2)~if $\sigma_\eta^2 > 0$ and $\rho_{\mathrm{model}} < 1$, $\lambda_\perp^{(t)}$ converges to a positive floor while $\lambda_\parallel^{(t)} = \Theta(t)$, driving $\bar{\rho}_t \to 1$ and $N_{\mathrm{eff}}(\Sigma^{(t)}) = 1 + \Theta(t^{-1}) \to 1$.
\end{theorem}

\begin{proof}
Write Eq.~\eqref{eq:delib_update} as $\varepsilon^{(t+1)} = W \varepsilon^{(t)} + \eta^{(t)}$, where $W = (1-\gamma)I_N + \frac{\gamma}{N}\mathbf{1}\mathbf{1}^\top = P_\parallel + (1-\gamma)P_\perp$, $P_\parallel = \frac{1}{N}\mathbf{1}\mathbf{1}^\top$, and $P_\perp = I_N - P_\parallel$. By orthogonality ($\operatorname{Cov}(\varepsilon^{(t)}, \eta^{(t)}) = 0$) and temporal whiteness ($\operatorname{Cov}(\eta^{(t)}, \eta^{(s)}) = 0$ for $t \neq s$), covariance propagates via $\Sigma^{(t+1)} = W \Sigma^{(t)} W^\top + Q_\eta$. Any equicorrelated matrix $M(\sigma^2, \rho) = \sigma^2[(1-\rho)I_N + \rho \mathbf{1}\mathbf{1}^\top]$ diagonalizes over $\{P_\parallel, P_\perp\}$ with eigenvalues $\lambda_\parallel = \sigma^2[1 + (N-1)\rho]$ (multiplicity~$1$) and $\lambda_\perp = \sigma^2(1-\rho)$ (multiplicity~$N-1$). Since $W P_\parallel = P_\parallel$ and $W P_\perp = (1-\gamma)P_\perp$, induction on $t$ yields
\begin{align*}
  \lambda_\parallel^{(t)} &= \sigma_0^2\bigl[1+(N{-}1)\bar{\rho}_0\bigr] + t\,\sigma_\eta^2\bigl[1+(N{-}1)\rho_{\mathrm{model}}\bigr], \\
  \lambda_\perp^{(t)} &= \sigma_0^2(1{-}\bar{\rho}_0)(1{-}\gamma)^{2t} + \tfrac{\sigma_\eta^2(1-\rho_{\mathrm{model}})}{\gamma(2-\gamma)}\bigl[1-(1{-}\gamma)^{2t}\bigr],
\end{align*}
with $0^0 = 1$ and $N_{\mathrm{eff}}(\Sigma^{(t)}) = 1 + (N-1)\lambda_\perp^{(t)}/\lambda_\parallel^{(t)}$. The two asymptotic regimes follow immediately.
\end{proof}

\textbf{Bounded-Probability Popoviciu Constraint.} For bounded probability beliefs $Y_i^{(t)} \in [0, 1]$ almost surely, Popoviciu's inequality yields $\operatorname{Var}(Y_i^{(t)} \mid S_\tau) \le 1/4$, forcing $\lambda_\parallel^{(t)} = \frac{1}{N}\mathbf{1}^\top \Sigma_y^{(t)} \mathbf{1} \le \operatorname{tr}(\Sigma_y^{(t)}) \le N/4$ for all $t \ge 0$. Additive persistent-innovation dynamics ($\sigma_\eta^2 > 0$) therefore cannot preserve $[0,1]$ over an unbounded horizon; separating unbounded latent scores $z_i^{(t)} \in \mathbb{R}$ from a nonlinear probability readout $y_i^{(t)} = \psi(z_i^{(t)}) \in [0,1]$ restores exact consistency (\code{tests/test\_theorems\_and\_counterexamples.py}).

\begin{remark}[Correlation Collapse vs.\ Single-Agent Error]
\label{rem:counterexample_1c}
Correlation collapse alone does not revert error rates to a single agent: if $N=5$ independent agents ($\bar{\rho}_0 = 0$) draw $y_i^{(0)} \sim \mathrm{Bernoulli}(0.30)$, apply noiseless convex averaging ($\gamma = 1, \sigma_\eta^2 = 0$), and threshold at $0.5$, their votes become perfectly correlated ($\bar{\rho}_1 = 1$), yet unanimous false consensus equals the Round-0 majority tail $\sum_{j=3}^5 \binom{5}{j}0.3^j 0.7^{5-j} = 0.16308 < 0.30$. Symmetric averaging destroys vote diversity by collapsing a $5$-of-$5$ gate to a Round-0 $3$-of-$5$ gate, while empirical debate degrades further under $\bar{\rho}_0 \gg 0$ or asymmetric sycophancy~\cite{hiddenbench2026,sycophancy2025,failuremodes2025}.
\end{remark}

\end{document}